\documentclass[aps,prb,reprint,superscriptaddress]{revtex4-1}

\usepackage{graphicx}
\usepackage{physics}
\usepackage{braket}
\usepackage{amsfonts}
\usepackage{natbib}
\usepackage{hyperref}
\usepackage{amssymb}
\usepackage{xcolor}
\usepackage{enumitem}
\usepackage[normalem]{ulem}
\graphicspath{{"../figures/"}}

\begin{document}

\title{Ideal refocusing of an optically active spin qubit under strong hyperfine interactions}

\author{Leon Zaporski\textsuperscript{1,$\dagger$}}
\author{Noah Shofer\textsuperscript{1,*}}
\author{Jonathan H.\,Bodey\textsuperscript{1,*}}
\author{Santanu Manna\textsuperscript{2,*}}
\author{George Gillard\textsuperscript{3}}
\author{Daniel M.\,Jackson\textsuperscript{1}}
\author{Martin Hayhurst Appel\textsuperscript{1}}
\author{Christian Schimpf\textsuperscript{2}}
\author{Saimon Covre da Silva\textsuperscript{2}}
\author{John Jarman\textsuperscript{1}}
\author{Geoffroy Delamare\textsuperscript{1}}
\author{Gunhee Park\textsuperscript{1}}
\author{Urs Haeusler\textsuperscript{1}}
\author{Evgeny A.\,Chekhovich\textsuperscript{3}}
\author{Armando Rastelli\textsuperscript{2}}
\author{Dorian A.\,Gangloff\textsuperscript{1,4}}
\author{Mete Atat\"ure\textsuperscript{1,$\dagger$}}
\author{Claire Le Gall\textsuperscript{1,$\dagger$}}

\noaffiliation
\affiliation{Cavendish Laboratory, University of Cambridge, JJ Thomson Avenue, Cambridge, CB3 0HE, United Kingdom}
\affiliation{Institute of Semiconductor and Solid State Physics, Johannes Kepler University, Altenbergerstraße 69, Linz 4040, Austria
}\affiliation{Department of Physics and Astronomy, University of Sheffield, Sheffield, S3 7RH, United Kingdom}
\affiliation{Department of Engineering Science, University of Oxford, Parks Road, Oxford, OX1 3PJ
\\ \ \\
\textsuperscript{*}\,These authors contributed equally to this work.
\\
\textsuperscript{$\dagger$}\,Correspondence should be addressed to: lz412@cam.ac.uk; ma424@cam.ac.uk; cl538@cam.ac.uk.}


\begin{abstract}
Combining highly coherent spin control with efficient light-matter coupling offers great opportunities for quantum communication and networks, as well as quantum computing. Optically active semiconductor quantum dots have unparalleled photonic properties, but also modest spin coherence limited by their resident nuclei. Here, we demonstrate that eliminating strain inhomogeneity using lattice-matched GaAs-AlGaAs quantum dot devices prolongs the electron spin coherence by nearly two orders of magnitude, beyond $0.113(3)$\,ms. To do this, we leverage the $99.30(5)\%$ fidelity of our optical $\pi$-pulse gates to implement dynamical decoupling. We vary the number of decoupling pulses up to $N_\pi=81$ and find a coherence time scaling of $ N_\pi^{0.75(2)}$. This scaling manifests an ideal refocusing of strong interactions between the electron and the nuclear-spin ensemble, holding the promise of lifetime-limited spin coherence. Our findings demonstrate that the most punishing material science challenge for such quantum-dot devices has a remedy, and constitute the basis for highly coherent spin-photon interfaces.

%


\end{abstract}

\maketitle

\begin{figure*}[t!]
    \centering
    \includegraphics[width=\textwidth]{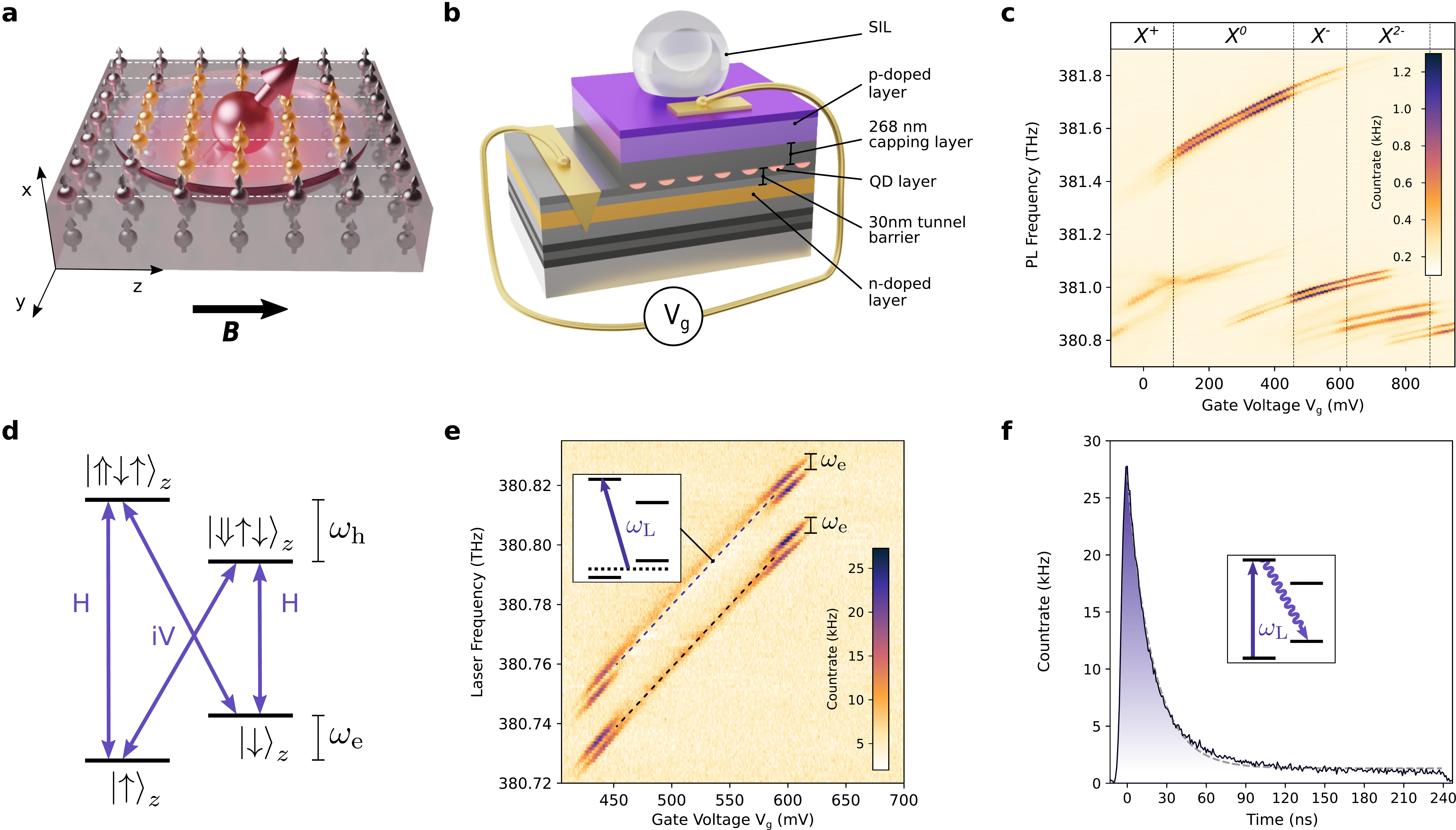}
    \caption{\textbf{Optically addressable spin in a GaAs-AlGaAs QD. a}, The QD electron spin (large arrow) interacts with nuclei (small arrows). White dashed lines highlight the lattice matching between the QD and barrier crystal. The $z$-axis is the external magnetic field $B$ axis; the $x$-axis corresponds to the growth and optical axis. \textbf{b}, n-i-p device structure tuned by applying a DC voltage $V_\text{g}$ -- for controlling the QD charge state -- and equipped with a cubic Zirconia superhemispherical lense (SIL) -- for enhanced light-matter coupling. \textbf{c}, Single QD photoluminescence (PL) spectrum as a function of gate voltage at $B=6.5$\,T, using above-band-gap laser excitation at $475.718$\,THz. The dominant PL lines -- corresponding to an exciton-state emission labelled in the upper strip -- change abruptly as a function of $V_\text{g}$ (dashed lines), defining charge plateaus. The PL overlap between different charge states comes from out-of-equilibrium emission specific to above-band-gap excitation (the QD captures an ill-defined number of photo-created electrons and holes from the barrier). \textbf{d}, Energy levels and selection rules of a negatively-charged QD in a magnetic field, $B$. The states of the electron ($\uparrow$, $\downarrow$) and hole ($\Uparrow$, $\Downarrow$) are Zeeman split by $\omega_e$ and $\omega_h$ respectively.   \textbf{e}, 2D map of resonance fluorescence counts at $B=6.5$\,T as a function of gate voltage ($V_\text{g}$) and laser frequency ($\omega_\text{L}$). The dashed lines highlight single-laser repumping (when the laser frequency excites both electron spin projections off-resonantly and equally, as depicted in the inset). When repumping uses the highest excited state ($\left |\Uparrow\downarrow\uparrow \right \rangle$), the fluorescence signal deviates from the dashed line, due to nuclear spin polarisation effects. \textbf{f}, Time-resolved resonance fluorescence following spin initialization to a $\uparrow$-state. The dashed curve is an exponential fit to the data with a characteristic $1/e$ electron spin pumping time of $19.3(2)$ ns. The inset shows the laser-addressed transition and the decay to the optically-dark spin-state. The laser Rabi frequency is $\Omega_\text{L}^\text{OP}\sim 0.3 \Gamma$, where $\Gamma$ is the excited state linewidth.
    }
\label{fig1}
\end{figure*}

\section{Introduction}
\label{sec_intro}
A pristine spin-photon interface that allows to entangle long-lived matter-based qubits with routable flying qubits, can serve as both a key component of memory-based quantum networks \cite{Nickerson2014, Monroe2014, Cohen2021} and a valuable source of entangled photon streams for measurement-based quantum computing \cite{Gimeno2015}. An ideal spin-photon interface marries long spin coherence with high-fidelity operations, such as single-qubit gates, single-shot read-out, and spin-photon entanglement. Further, near-unity collection efficiencies and coherent coupling to multiple `data qubits' are not only desirable assets but become critical requirements for measurement-based and memory-based quantum computing, respectively. Bringing these attributes together has spurred sustained efforts to develop the performance of candidate systems such as trapped ions \cite{Stephenson2020, Postler2021}, atomic impurities in solids \cite{Abobeih2021, Bergeron2020, Christle2017, Ruskuc2021, Raha2020} and quantum dot (QD) spins \cite{Berezovsky2008, DeGreve2011,Godden2012}, delivering in turn tremendous advances towards quantum networks \cite{Pfaff2014, Pompili2021} and photonic cluster-states \cite{Schwartz2016a,Istrati2020,Cogan2021}. 

Within the pool of promising systems, optically active QD devices made out of III-V materials feature unparalleled optical properties, combining near-unity quantum efficiency, sub-nanosecond photon generation rates and close-to-transform-limited photons with successful integration in photonic structures \cite{Wang2019, Liu2019,Tomm2021, Appel2021,Thomas2021}. The strive for optical performance has brought QD devices to $57\%$ photonic end-to-end efficiency \cite{Tomm2021}, performing well beyond any other physical platform and very close to meeting the stringent requirements of scalable photonic quantum computing \cite{Varnava2008,Pant2019}. The matter qubit implementation for QDs is realised through the spin states of a confined electron or a hole. It can be controlled optically on picosecond timescale\cite{Press2008, Berezovsky2008}, and can be entangled with photons\cite{DeGreve2012, Gao2012, Cogan2021, https://doi.org/10.48550/arxiv.2111.12523} and with other remote spin qubits\cite{Delteil2015,Stockill2017} at superior operational rates.

In parallel, the confined electron couples strongly to an ensemble of $\sim 10^{5}$ nuclear spins residing within the electronic wavefunction. Recent advances in controlling the nuclear ensemble\cite{Gangloff2018} make them desirable candidates for a dedicated multi-mode quantum memory for the optically active spin qubit\cite{Taylor2003, Denning2019}. On the flip side, this ever present interaction with the nuclear environment, if uncontrolled, limits the spin coherence, and dynamical decoupling has thus far yielded modest enhancement constrained to a few microseconds\cite{Stockill2016}. This constraint on spin qubit coherence is fundamentally a materials science challenge arising from local strain and random alloying in QD devices, inherent to a lattice-mismatched growth\cite{doi:10.1063/1.1831564,Rastelli2008}. The nanoscale variation of strain, and disorder at the atomic level, within the QD causes an inhomogeneously broadened nuclear spectrum seen by the electron, which diminishes the effectiveness of the conventional dynamical decoupling protocols\cite{Stockill2016,Bechtold2015}. In particular, this undesirable inhomogeneity within the QD creates electric field gradients that shift the energy levels of high-spin nuclei ($I>1/2$) via the quadrupolar interaction. From a device material perspective, it is highly desirable to have tunable uniform strain instead to enable pristine electron-nuclear coupling for improved qubit performance and for realising a nuclear quantum memory simultaneously\cite{Denning2019}. One way to approach this is to identify a strain-free material platform and introduce strain in a tunable manner via external means. Advances in lattice-matched growth of high-quality III-V QDs\cite{DaSilva2021,Scholl2019, Liu2019,Zhai2020} (Fig. \ref{fig1}a) have precisely struck this chord by bringing the prospect of a dramatic reduction of nuclear spectral inhomogeneities\cite{Chekhovich2020}.

In this work, we bring a materials solution to the so-far limited spin qubit performance by demonstrating the retention of electron spin coherence for at least $0.113(3)$\,ms in lattice-matched GaAs-AlGaAs QD devices.  To do this, we implement all-optical quantum control of the spin qubit and achieve $99.30(5)\%$ fidelity for single-qubit gates. This allows us to apply multi-pulse decoupling sequences and reveal a dramatic improvement of spin coherence over the Hahn-Echo coherence time. Finally, we intersect nuclear magnetic resonance spectroscopy with a microscopic model of our central-spin system to explain the scaling of coherence that we measured in dynamical decoupling. Our work establishes an unprecedented leap towards highly coherent spin dynamics in optically active QDs. 

\begin{figure}[t!]
    \centering
    \includegraphics[width=\linewidth]{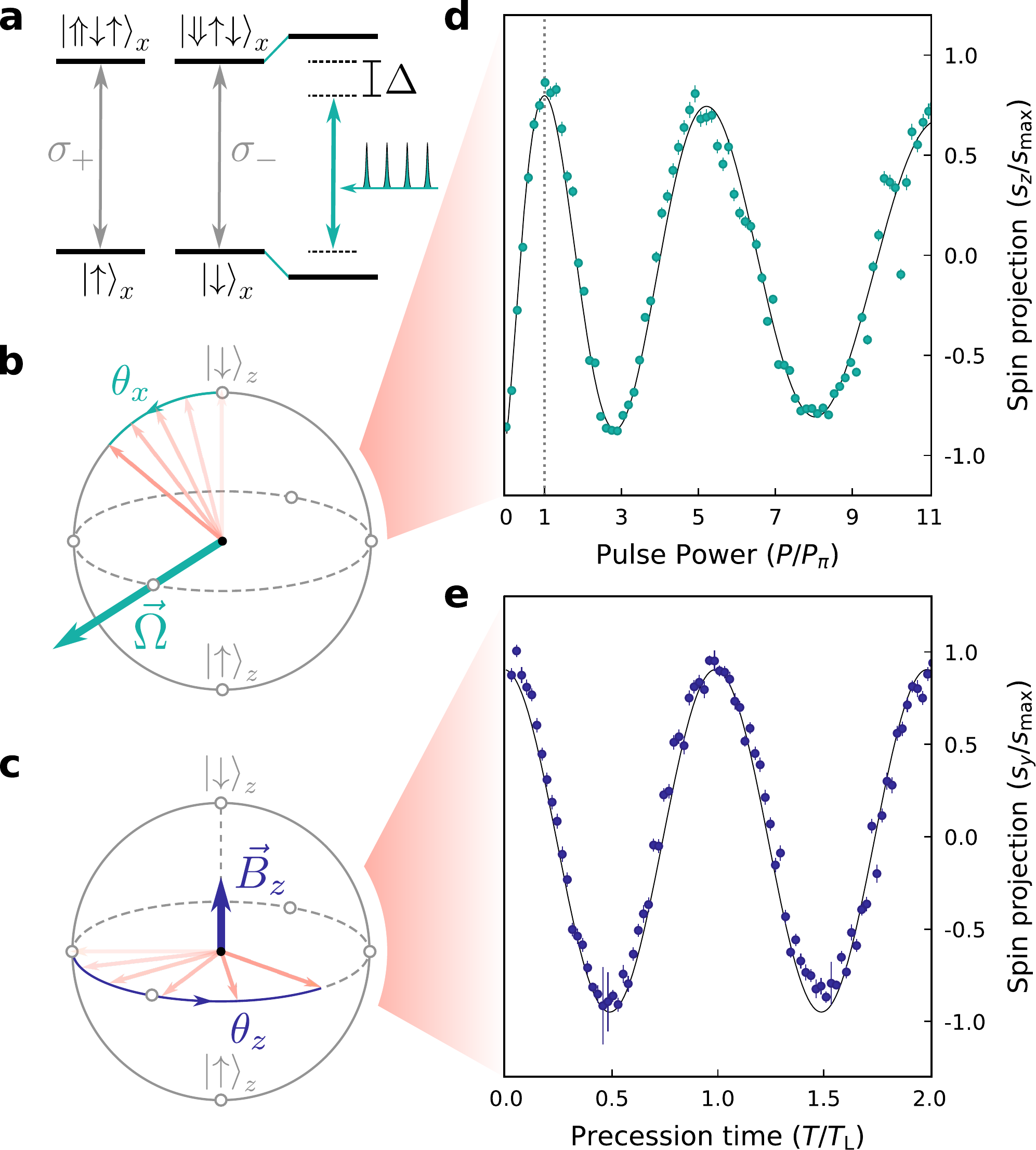}
    \caption{\textbf{All-Optical SU(2) Control. a}, Simple AC Stark-shift picture using the $B=0$\,T QD states and selection rules (the external magnetic field is neglected because the qubit Larmor precession time, $1/\omega_e$, is much longer than the pulse duration, $t_\text{p}$). The $\downarrow_x$-state shifts relative to the $\uparrow_x$-state due to the interaction with a strong $\sigma_-$ pulse detuned by $\Delta$ from the optically excited state, leading to a relative phase-shift $\theta_x \sim \Omega_L^2 t_\text{p}/\Delta$ between the two spin-states (here $x$ is the growth axis, $\Omega_\text{L}\propto \sqrt{P}$ is the optical Rabi frequency of the rotation laser, $\Delta\approx 600$\,GHz and $t_\text{p}\approx 4$\,ps). \textbf{b}, Bloch-sphere representation of the $\theta_x$ spin-rotation induced by the optical pulse. \textbf{c}, Bloch-sphere representation of the $\theta_z$ spin-rotation induced by the external magnetic field, $B$. \textbf{d},  Longitudinal spin projection ($S_z$) of the electron spin as a function of the normalised power of the optical pulse \textbf{e}, Transverse spin projection ($S_y$) of an electron spin as a function of time. In addition to initialisation and read-out via optical pumping, $\left (\frac{\pi}{2}\right)_x$-pulses are used to change basis from $S_z$ to $S_y$.
    } 
\label{fig2}
\end{figure}

\textbf{Device design and characterisation.} 
The QDs used in this work consist of GaAs grown by nanohole infilling of an AlGaAs barrier \cite{Huo2013, DaSilva2021}. To control the QD charge state deterministically, the GaAs-AlGaAs QDs are embedded in an n-i-p diode structure (Fig. \ref{fig1}b) tunnel-coupled to a Fermi sea from the n-doped back-contact \cite{Zhai2020, 10.1117/1.AP.3.6.065001, Supplementary}. Applying gate voltage, $V_\text{g}$, shifts the energy of the QD states relative to the Fermi sea in the Coulomb blockade regime, allowing to step the number of electrons residing in the QD ground state (plateaus in the QD photoluminescence spectral map shown in Fig. \ref{fig1}c) and to tune the optical transition energy via the DC-Stark shift. 
%

To realise an optically active spin qubit, we load the QD with a single electron and we access its corresponding excitonic ($X^-$) optical transitions (Fig. \ref{fig1}d). A magnetic field of $\gtrsim 2$\,T applied perpendicular to the growth axis lifts the spin degeneracy of the ground and excited state energy levels, giving rise to four linearly polarised transitions of equal strength. Figure \ref{fig1}e displays the resonance fluorescence counts under circularly-polarised excitation as a function of gate voltage laser frequency across the $X^-$ plateau. At the center of the plateau, the steady-state resonance fluorescence is vastly reduced, as the frequency-selective addressing of any of the four $X^-$ transitions initializes the electron spin. On the edges of the plateau ($V_G\approx 440$\,mV and $V_G\approx 600$\,mV), where a QD level is resonant with the Fermi energy of the electron reservoir, tunneling-induced relaxation reduces the efficiency of spin initialization, and the four optical transitions of $X^-$ unfold. The splitting between these two line doublets corresponds to two g-factor values. The weak re-pumping conditions in the charge-stable region (Fig. \ref{fig1}e, inset) identify which of these g-factor values corresponds to the ground state. In this way, we find an in-plane g-factor $|g_\text{e}|= 0.049(1)$ and $|g_\text{h}|=0.232(1)$ for the electron and hole spins, respectively. %

Figure \ref{fig1}f displays the time-resolved spin initialization transient, as the electron is illuminated resonantly on the optical transition $\ket{\uparrow}-\ket{\Uparrow\downarrow\uparrow}$ (Fig. \ref{fig1}d). Sequentially addressing the optical transitions that pump into the state $\ket{\downarrow}$ and $\ket{\uparrow}$ yields a robust differential signal that measures the absolute electron polarisation, $S_z$\cite{Supplementary}. A simple exponential fit yields an initialization fidelity of $94.9(1) \%$ into the state $\ket{\downarrow}$, limited by broadening of the optical transition by charge noise\cite{Kuhlmann2013}. Inserting a variable delay between an initialisation and readout pulse, we measure an electron-spin lifetime in excess of $400$\,$\mu$s\cite{Supplementary}, commensurate with previous experiments reporting of order a millisecond \cite{Gillard2021}.

\textbf{Quantum control of the spin qubit.} 
We use far-detuned ($\sim0.6$\,THz) $4$-ps laser pulses to achieve coherent control of the electron spin qubit, as was previously employed for highly strained self-assembled QDs\cite{Berezovsky2008, Press2008}. The effect of such a spectrally broad optical pulse is best captured by a Stark-shift picture (Fig. \ref{fig2}a) and corresponds to a near-instantaneous $x$-rotation on the Bloch sphere, where $x$ denotes the optical axis (Fig. \ref{fig2}b). Arbitrary qubit rotations require a second rotation axis, $z$, which we achieve via the free precession of the electron spin around the applied magnetic field (Fig. \ref{fig2}c). 

Figure \ref{fig2}d shows the spin projection as a function of power of the $x$-rotation laser. Having initialised the qubit in the $\ket{\downarrow}$-state, our $\ket{\uparrow}$-readout displays coherent oscillations due to the laser-induced $\theta_x$-rotation. The $P^{0.67}$ power scaling of the $\theta_x$ rotation, deviating from an ideal linear scaling, typically arises from broad optical absorption in QD devices\cite{Press2008,Berezovsky2008,Stockill2016,Bodey2019}.
This limits our $\pi$-pulse fidelity to $99.30(5)\%$\cite{Supplementary}. Compared to previous reports in conventional InAs-GaAs QD devices\cite{Press2008,Berezovsky2008,Stockill2016, Bodey2019}, this constitutes a two-fold reduction of the $\pi$-gate error.



Figure \ref{fig2}e demonstrates $z$-axis qubit rotations by controlling the time delay between two pulses. A first rotation pulse takes the electron spin, initially in the $\ket{\downarrow}$-state to the $\ket{\uparrow}_y$-state at $t=0$. This state then precesses around the quantisation axis at the electron Larmor frequency. A second $(\frac{\pi}{2})_x$ rotation maps the electron coherence $S_y(t)$ onto our read-out basis, allowing to monitor $\theta_z(\tau)$. A sinusoidal fit (black curve) to the data confirms the quality of the $\theta_z$-rotation. It yields a precise measure of the Larmor precession of our qubit of $2.398(3)$\,GHz, which corresponds to a $209$\,ps calibrated $\pi_z$ gate (at $B=3.5$\,T).

Common to all spin-rich materials, including III-V QDs, slow electron-spin-sensitive drifts of the average nuclear spin polarisation modify the Larmor precession frequency via the Overhauser effect. Maintaining a well-defined $\theta_z$ rotation thus requires preventing these drifts. We implement this correction after every spin readout using a 100-ns high-power laser pulse whose frequency is set to randomize the electron spin (Fig. \ref{fig1}e inset) on a nanosecond timescale \cite{Supplementary}. In this instance, the hyperfine-coupled electron acts as a high-bandwith near-thermal spin reservoir that relaxes the nuclei. 


\begin{figure*}[t!]
    \centering
    \includegraphics[width=\textwidth]{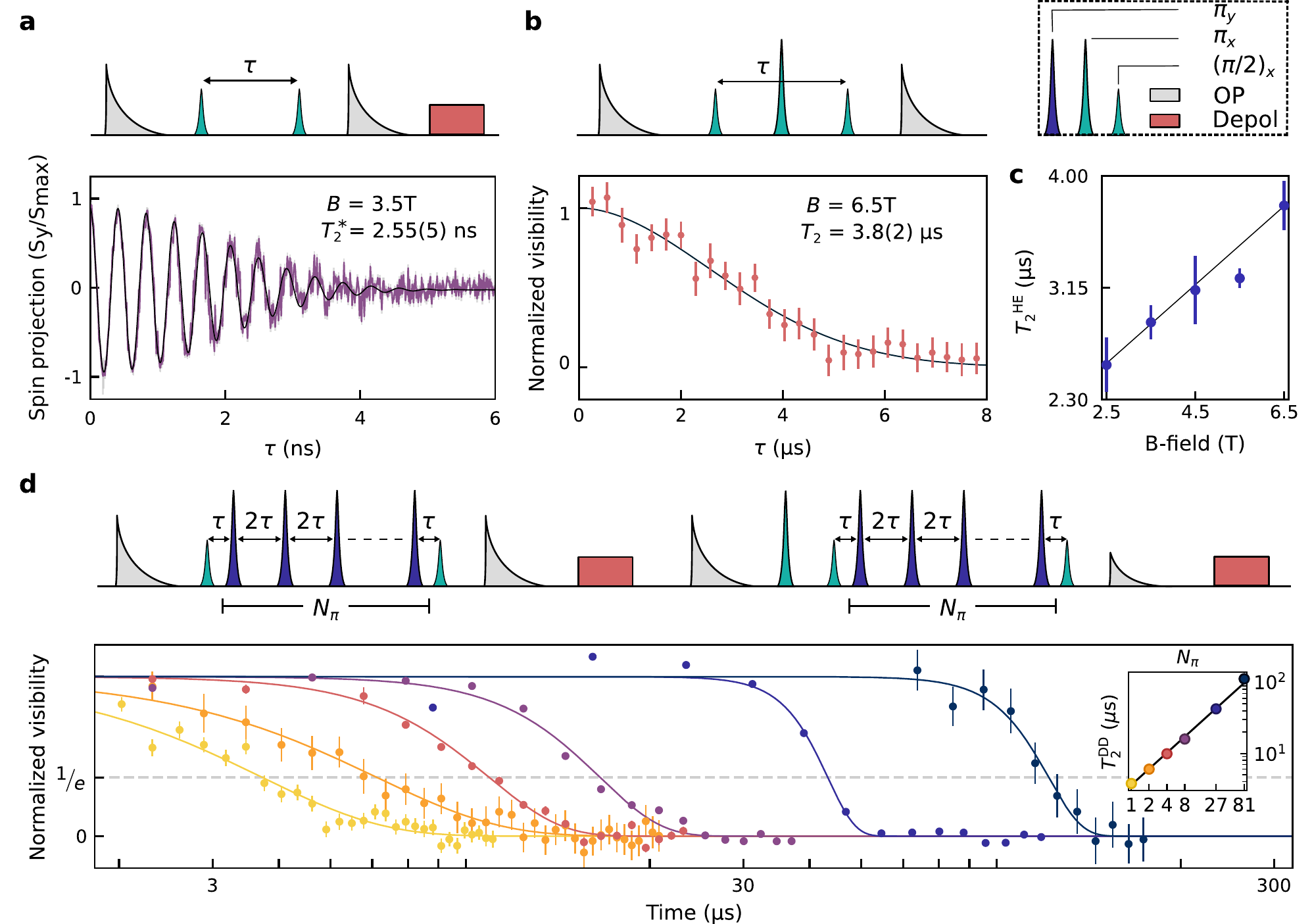}
    \caption{\textbf{Electron spin coherence a}, Electron-spin Ramsey interferometry as a function of the time-delay bewteen the two Ramsey pulses. The solid curve is a damped sinusoidal fit to the data with a $1/e$-decay time of $T_2^*=2.55(5)$\,ns\cite{Supplementary}. \textbf{b}, Hahn-Echo signal as a function of time-delay. The solid curve is a super-exponential fit to the data, $e^{-(T/T_2^\text{HE})^{\alpha}}$ with $\alpha=1.9(3)$ and $T_2^\text{HE}=3.8(2)$\, $\mu$s. \textbf{c}, Magnetic-field dependence of the Hahn-Echo coherence time. The solid black curve is a linear fit. \textbf{d} Time-dependence of the CPMG visibility. The number of decoupling pulses is varied between $1$ and $81$. The solid curves are super-exponential fits to the data, $e^{-(T/T_2^\text{CPMG})^{\alpha}}$ with $\alpha=\left\{1.9(3),1.8(2),3.1(3),3.3(5),6.6(6),7(2)\right\}$ and $T_2^\text{CPMG}=\left\{3.8(2),6.0(3),9.9(3),16.0(5),42.5(4),113(3)\right\}$ $\mu$s for $N_{\pi}=\left\{1,2,4,8,27,81\right\}$. \textbf{Pulse legend} Schematics of the pulse sequences are displayed above the data panel, and the pulse legend box is on the top right. `OP' is the optical pumping pulse for spin initialisation and read-out (frequency as in Figure 1f, inset and Rabi frequency $\Omega_\text{L}^\text{OP}\sim 0.3 \Gamma$, where $\Gamma$ is the excited state linewidth); `Depol' is a high-power depolarisation pulse (frequency as in Figure 1e, inset and Rabi frequency $\Omega_\text{L}^\text{Depol}\sim 5 \Gamma$). 
    } 
\label{fig3}
\end{figure*}

\textbf{Spin qubit coherence.}
The QD spin interacts with $N\sim 10^5$ nuclei via the contact hyperfine interaction, $\hat{H}_{\text{hf}}=\sum_i A_i \mathbf{S}\cdot\mathbf{I_i}$, where $\mathbf{S}$ is the electron spin operator ($S=1/2$), $A_i \sim \mathcal{A}/N$ is the hyperfine coupling to the $i$-th nuclear spin, $\mathbf{I_i}$ ($I=3/2$), and $\mathcal{A}$, the material hyperfine constant. Operating under magnetic field $>1$\,T validates a canonical transformation to arrive at an approximate dephasing Hamiltonian\cite{Cywinski2009} that retains the leading order corrections emanating from the non-secular flip-flop terms ($\propto \hat{S}_{+,i}\hat{I}_{-,i}+ \hat{S}_{-,i}\hat{I}_{+,i}$):

\begin{equation}\label{hf_simple}
\tilde{H}_{\text{hf}}\approx \hat{S}_z\sum_i A_i \hat{I}_{z,i}+\hat{S}_z\sum_{i\neq j}\frac{A_i A_j}{2 \omega_\mathrm{e}}\hat{I}_{+,i}\hat{I}_{-,j}\text{,}
\end{equation}

\noindent and operates in a low-energy electron-nuclear subspace where the electron polarisation, $S_z$, is conserved \cite{Cywinski2009}. In our system, the effective non-collinear hyperfine interaction terms--arising in QDs either from an electron g-factor anisotropy \cite{Botzem2015} or strain\cite{Stockill2016} -- are negligible \cite{Supplementary}.

The first dephasing term in Eq. \ref{hf_simple} ($\sum_i A_i \hat{I}_{z,i}$) is so-called `frozen' as the nuclear polarisation parallel to the external magnetic field changes on slow (millisecond) timescales due to the indirect coupling mediated by the central spin (second term in Eq. \ref{hf_simple}) or intrinsic nuclear dipole interactions \cite{Bluhm2010}. The characteristic amplitude of this term ($\sim \mathcal {A}/\sqrt{N}$) is accessed experimentally through an ensemble-averaged measurement of the electron dephasing in a Ramsey sequence (Fig. \ref{fig2}e) at long delays (Fig. \ref{fig3}a). The Ramsey signal features a quasi-Gaussian envelope, as expected for slow shot-to-shot variations of the nuclear polarisation. The $1/e$-decay time of this envelope, $T_2^*$, is $2.55(5)$\,ns and indicates correspondingly the number of nuclear spins\cite{Supplementary} $N\approx 6.5(3)\times 10^4$. This nuclear ensemble is a factor $\sim 2$ larger than typical self-assembled InAs-GaAs QDs\cite{Stockill2016, Stockill2017, Gangloff2018}, and $\sim10$ smaller than for gate-defined GaAs-AlGaAs QDs\cite{Bluhm2010}.

Next, we use an echo pulse sequence to refocus the slow noise to extend the spin coherence (Fig. \ref{fig3}b). A super-exponential fit, $e^{-(T/T_2^\text{HE})^{\alpha}}$ (solid curve in Fig. \ref{fig3}b), yields exponent $\alpha=1.9(3)$  and a Hahn-Echo coherence time $T_2^\text{HE}=3.8(2)\,\mu$s. Figure \ref{fig3}c displays the magnetic-field dependence of $T_2^\text{HE}$, which increases with magnetic field. This dependence stems from electron-mediated nuclear flip-flops (second term of Eq. \ref{hf_simple}), which, in a semi-classical picture, correspond to a quadratic coupling to transverse nuclear spin fluctuations \cite{Bluhm2010}. As the amplitude of this term is proportional to $1/\omega_\text{e}$, this noise  becomes increasingly suppressed by the external magnetic field. 

Decoherence in this `intermediate' field regime ($\mathcal{A}/\sqrt{N}<\omega_e<\mathcal{A}$), is elucidated theoretically \cite{Cywinski2009}. 
In particular, when the inter-species frequency differences between the $^{75}\mathrm{As}$, $^{69}\mathrm{Ga}$, $^{71}\mathrm{Ga}$ QD nuclei ($(\omega_k-\omega_l)\sim10$\,MHz) are much greater than the single nucleus hyperfine coupling ($\mathcal{A}/N\sim 100$\,kHz), the decay of spin coherence is dominated by homonuclear effects (frequency differences within the same nuclear-spin species) \cite{Cywinski2009, Supplementary}. As such, the monotonic decay we observe in Hahn echo indicates a smooth homonuclear broadening centered around zero frequency -- in contrast to non-zero frequency peaks in the heteronuclear spectrum, as observed at lower magnetic field in gate defined QDs\cite{Bluhm2010, Malinowski2016b, Malinowski2017}). Unlike in highly strained QDs, the Hahn-Echo decay here is super-exponential, due to correlated transverse nuclear-spin noise, suggesting that dynamical decoupling of the electron is possible. 

Figure \ref{fig3}d displays the spin projection visibility (see Methods) as a function of time as we implement Carr-Purcell-Meiboom-Gill (CPMG) pulse sequences with $N=\{1,2,...,81\}$ $\pi$-pulses for $B=6.5$\,T. To perform a $\pi_y$-gate, we use the equivalent \emph{composite} rotation $\left (\frac{\pi}{2}\right )_z\pi_x\left (-\frac{\pi}{2}\right )_z$ accomplished by shifting the $\pi_x$-optical decoupling pulse by a quarter Larmor period ($\delta t= 1/4\omega_\text{e}$) relative to the $\left (\frac{\pi}{2}\right )_x$-rotations. Super-exponential fits in the form of $e^{-(T/T_2^\text{CPMG})^{\alpha}}$ (solid curves in Fig. \ref{fig3}c) reveal the spin coherence time and for $N=81$ decoupling pulses, we measure $T_2^\text{CPMG}=113(3)\,\mu$s. Leveraging the high-fidelity of our optical control pulses, we decouple this qubit from a hyperfine interaction,  that is $10$-times larger ($\sim\mathcal{A}^2/(N\omega_\text{e})$) than in III-V gated-defined singlet-triplet qubits to reach on-par coherence times \cite{Bluhm2010, Malinowski2016b}. 

A power-law fit to the dependence of coherence time on $N_{\pi}$  ($T_2^\text{DD}\propto N_\pi^\gamma$) yields a scaling $\gamma=0.75(2)$. This scaling is similar to the one measured in gate-defined III-V quantum dots where it is set predominantly by the quadratic coupling to transverse nuclear spin fluctuations \cite{Malinowski2017}, and greater than the $\gamma=2/3$ scaling set by the nuclear-spin noise in weakly coupled dipolar platforms \cite{DeLange2010} or the $\gamma\sim 0.35$ scaling set by extrinsic electrical noise \cite{Huthmacher2018}. 

\begin{figure}[t!]
    \centering
    \includegraphics[width=0.48\textwidth]{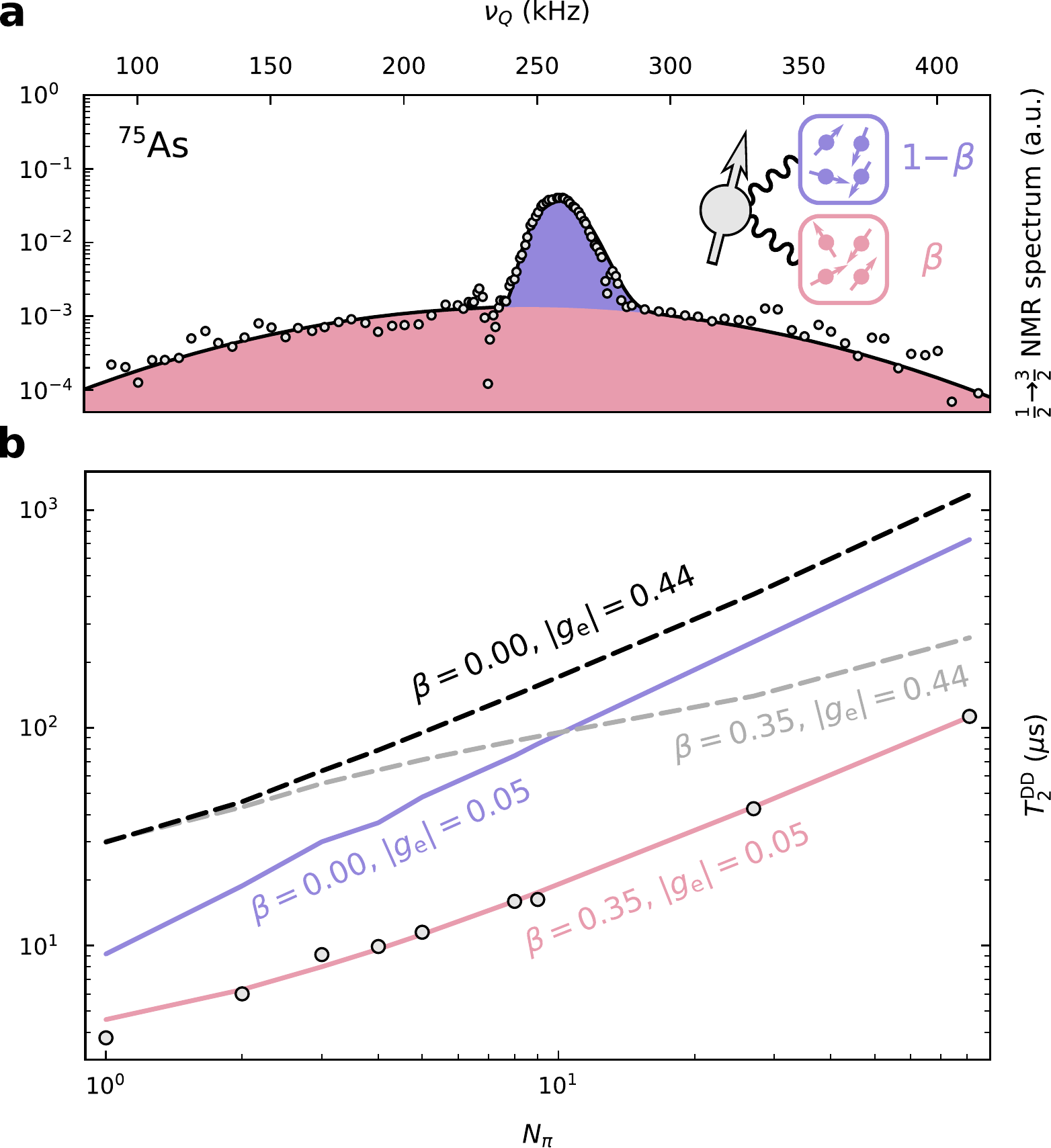}
    \caption{\textbf{Key parameters for device performance. a}, Nuclear spectrum of the arsenic satellite transition ($\big\{+\frac{3}{2},+\frac{1}{2}\big\} $), displaying an arsenic sub-ensemble with low quadrupolar broadening ($\sim 30$\,kHz) and another sub-ensemble with higher quadrupolar broadening ($\sim 200$\,kHz). This spectrum, obtained by differentiating the integral NMR data, is presented for illustrative purposes and the quantitative analysis -- based of the raw integral NMR data \cite{Supplementary} -- shows that the broad arsenic ensemble constitutes $\beta=0.244(7)$ of the total NMR signal. \textbf{Inset}, The electron spin couples to GaAs nuclei (wavefunction overlap, $1-\beta$) as well as AlGaAs nuclei (wavefunction overlap $\beta$). Nuclear spectral inhomogeneities in AlGaAs are stronger than in GaAs because of alloying. \textbf{b}, Electron spin coherence time as a function of the number of decoupling pulses. The data points correspond to the coherence times measured in Figure 3d). The solid and dashed curves correspond to theoretical values, $T_2^\text{DD}$, from the microscopic modelling presented in \cite{Supplementary} for selected values of $\beta$ and $|g_{\mathrm{e}}|$, indicated in the plot.   
    } 
\label{fig4}
\end{figure}

To verify the origin of the measured scaling and gain further insight into the nature of the nuclear ensemble we compare our CPMG results with a nuclear spectrum measured directly via the integral Nuclear Magnetic Resonance (NMR) technique\cite{ragunathan_2019,Ulhaq2016,Supplementary} in a device made from the same wafer. In the studied QD devices, a small amount of residual strain leads to satellite transitions via the nuclear quadrupolar interaction\cite{Chekhovich2020,Supplementary}. By probing the NMR-lineshape of the resolved $\frac{1}{2}\to\frac{3}{2}$ satellite transition, we extract the distribution of residual quadrupolar broadening\cite{Supplementary}. The resulting spectrum features a dominant peak of 30\,kHz width (the purple-shaded area in Fig. \ref{fig4}a) and a weaker 200-kHz-wide pedestal (the pink-shaded area in Fig. \ref{fig4}a), weighing a fraction $\beta=0.244(7)$ of the total integrated area\cite{Supplementary}. We use this distribution to constrain a microscopic theory\cite{Supplementary}, which extends previous work\cite{Cywinski2009,Bluhm2010,Malinowski2016b, Malinowski2017} to capture the effect of quadrupolar-split nuclei on electron spin coherence. This model produces a best fit to our CPMG data for $\beta=0.35$ (the solid pink curve in Fig. \ref{fig4}b) and provides strong evidence that the origin of spin qubit dephasing in this system is quadrupolar-broadened transverse nuclear noise.

\textbf{Outlook for device performance improvement via quantum-dot engineering.} Refocusing of the hyperfine interactions in this optically active spin qubit to longer coherence time is possible. One clear route forward relies on reducing $\beta$ - the fraction of the electronic wavefunction that overlaps with the 200 kHz quadrupolar-broadened nuclear sub-ensemble. The quadrupolar broadening of arsenic nuclei is sensitive to compositional modulations (cation-alloying of gallium and aluminium)\cite{Chekhovich2015,Knijn2010}, which hints to proximity of the broadened sub-ensemble to the AlGaAs alloy. The reported $\beta=0.35$ is likely a result of GaAs-AlGaAs intermixing within the QD, as the wavefunction leakage into the barrier has been estimated to be below $10\%$\cite{Chekhovich2017}. For an equal number of refocusing pulses, $N_\pi$, the solid purple curve in Fig. \ref{fig4}b illustrates the improvement of coherence time expected to follow from a reduction of $\beta$ to zero. Another route forward relies on increasing the electron Zeeman energy, $\omega_{\mathrm{e}}$, via the g-factor\cite{PhysRevB.93.035311} or a higher magnetic field, which -- as shown in Fig. \ref{fig3}c -- would extend the coherence time. The dashed curves in Fig. \ref{fig4}b highlight the added improvement we anticipate if the spin qubit operated at $B=6.5$\,T had the same g-factor as that of an electron in bulk GaAs. Taken together, these improvements would lead to a coherence time exceeding 1 ms (the dashed black curve in Fig. \ref{fig4}b). \newline

\section{Conclusion}
In this work, we demonstrate that the new generation of lattice-matched QD devices, with vanishing strain, enable dramatic prolongation of the electron spin coherence over conventional counterparts. The near two orders of magnitude improvement of electron spin coherence also places these QD devices in a new regime where the spin coherence is much longer than typical electron, nuclear and electron-nuclear gate durations.
This lifts a major road block towards long-lived quantum storage and many-body phenomena in an isolated spin ensemble \cite{Gangloff2018,Jackson2021,Gangloff2021}. This improvement of spin coherence further helps with interfacing QD devices with atomic memories \cite{PhysRevLett.119.060502} or generating photonic cluster states \cite{Istrati2020,doi:10.1126/science.aah4758,PhysRevLett.105.093601}.

\section{Acknowledgements}
We acknowledge support from the Royal Society (EA/181068), the US Office of Naval Research Global (N62909-19-1-2115), EU H2020 FET-Open project QLUSTER (862035), EU H2020 Research and Innovation Programme under the Marie Sklodowska-Curie grant QUDOT-TECH (861097), EPSRC grant EP/V048333/1, and the Austrian Science Fund (FWF): FG 5, P 30459, I 4380, I 4320, and I 3762. The LIT Secure and Correct Systems Lab are funded by the state of Upper Austria and the European Union's Horizon 2020 research and innovation program under Grant Agreement Nos. 899814 (Qurope) and 871130 (ASCENT+). D.A.G. acknowledges a St John's College Fellowship and a Royal Society University Research Fellowship, C.LG. a Dorothy Hodgkin Royal Society Fellowship, and E.A.C. a Royal Society University Research Fellowship. We also thank \L. Cywi\'nski, M. E. Flatt\'e, C. E. Pryor, H. Bluhm, P. Atkinson, A.J. Garcia, G. Undeutsch and P. Klenovsk\'y for fruitful discussions.

\section{Author Contributions}
L.Z., D.A.G., M.A. and C.LG. conceived the spin control experiments; E.A.C. conceived the NMR experiments; E.A.C. and A.R. conceived the QD device. L.Z., J.H.B., N.S., M.H.A., G.D., G.P. and C.L.G. carried out the spin control and resonance fluorescence experiments; L.Z. and C.LG. performed the corresponding data analysis, theory and simulations; G.G. and E.A.C. carried out the NMR experiments; S.M., S.F.C.d.S. performed the MBE growth and C.S. characterized the samples. S.M., J.J. and N.S. processed the QD devices. All authors contributed to the discussion of the analysis and the results. All authors participated in the preparation of the manuscript.

\section{Data Availability}

The datasets generated as part of the current study are available from the corresponding authors upon reasonable request.

\section{Code Availability}

The codes used for the analysis included in the current study are available from the corresponding authors upon reasonable request.


\bibliography{A-GaAs}

\begin{thebibliography}{70}%
\makeatletter
\providecommand \@ifxundefined [1]{%
 \@ifx{#1\undefined}
}%
\providecommand \@ifnum [1]{%
 \ifnum #1\expandafter \@firstoftwo
 \else \expandafter \@secondoftwo
 \fi
}%
\providecommand \@ifx [1]{%
 \ifx #1\expandafter \@firstoftwo
 \else \expandafter \@secondoftwo
 \fi
}%
\providecommand \natexlab [1]{#1}%
\providecommand \enquote  [1]{``#1''}%
\providecommand \bibnamefont  [1]{#1}%
\providecommand \bibfnamefont [1]{#1}%
\providecommand \citenamefont [1]{#1}%
\providecommand \href@noop [0]{\@secondoftwo}%
\providecommand \href [0]{\begingroup \@sanitize@url \@href}%
\providecommand \@href[1]{\@@startlink{#1}\@@href}%
\providecommand \@@href[1]{\endgroup#1\@@endlink}%
\providecommand \@sanitize@url [0]{\catcode `\\12\catcode `\$12\catcode
  `\&12\catcode `\#12\catcode `\^12\catcode `\_12\catcode `\%12\relax}%
\providecommand \@@startlink[1]{}%
\providecommand \@@endlink[0]{}%
\providecommand \url  [0]{\begingroup\@sanitize@url \@url }%
\providecommand \@url [1]{\endgroup\@href {#1}{\urlprefix }}%
\providecommand \urlprefix  [0]{URL }%
\providecommand \Eprint [0]{\href }%
\providecommand \doibase [0]{http://dx.doi.org/}%
\providecommand \selectlanguage [0]{\@gobble}%
\providecommand \bibinfo  [0]{\@secondoftwo}%
\providecommand \bibfield  [0]{\@secondoftwo}%
\providecommand \translation [1]{[#1]}%
\providecommand \BibitemOpen [0]{}%
\providecommand \bibitemStop [0]{}%
\providecommand \bibitemNoStop [0]{.\EOS\space}%
\providecommand \EOS [0]{\spacefactor3000\relax}%
\providecommand \BibitemShut  [1]{\csname bibitem#1\endcsname}%
\let\auto@bib@innerbib\@empty
\bibitem [{\citenamefont {Nickerson}\ \emph {et~al.}(2014)\citenamefont
  {Nickerson}, \citenamefont {Fitzsimons},\ and\ \citenamefont
  {Benjamin}}]{Nickerson2014}%
  \BibitemOpen
  \bibfield  {author} {\bibinfo {author} {\bibfnamefont {N.~H.}\ \bibnamefont
  {Nickerson}}, \bibinfo {author} {\bibfnamefont {J.~F.}\ \bibnamefont
  {Fitzsimons}}, \ and\ \bibinfo {author} {\bibfnamefont {S.~C.}\ \bibnamefont
  {Benjamin}},\ }\href {\doibase 10.1103/PhysRevX.4.041041} {\bibfield
  {journal} {\bibinfo  {journal} {Phys. Rev. X}\ }\textbf {\bibinfo {volume}
  {4}},\ \bibinfo {pages} {041041} (\bibinfo {year} {2014})}\BibitemShut
  {NoStop}%
\bibitem [{\citenamefont {Monroe}\ \emph {et~al.}(2014)\citenamefont {Monroe},
  \citenamefont {Raussendorf}, \citenamefont {Ruthven}, \citenamefont {Brown},
  \citenamefont {Maunz}, \citenamefont {Duan},\ and\ \citenamefont
  {Kim}}]{Monroe2014}%
  \BibitemOpen
  \bibfield  {author} {\bibinfo {author} {\bibfnamefont {C.}~\bibnamefont
  {Monroe}}, \bibinfo {author} {\bibfnamefont {R.}~\bibnamefont {Raussendorf}},
  \bibinfo {author} {\bibfnamefont {A.}~\bibnamefont {Ruthven}}, \bibinfo
  {author} {\bibfnamefont {K.~R.}\ \bibnamefont {Brown}}, \bibinfo {author}
  {\bibfnamefont {P.}~\bibnamefont {Maunz}}, \bibinfo {author} {\bibfnamefont
  {L.-M.}\ \bibnamefont {Duan}}, \ and\ \bibinfo {author} {\bibfnamefont
  {J.}~\bibnamefont {Kim}},\ }\href {\doibase 10.1103/PhysRevA.89.022317}
  {\bibfield  {journal} {\bibinfo  {journal} {Phys. Rev. A}\ }\textbf {\bibinfo
  {volume} {89}},\ \bibinfo {pages} {022317} (\bibinfo {year}
  {2014})}\BibitemShut {NoStop}%
\bibitem [{\citenamefont {Cohen}\ \emph {et~al.}(2021)\citenamefont {Cohen},
  \citenamefont {Kim}, \citenamefont {Bartlett},\ and\ \citenamefont
  {Brown}}]{Cohen2021}%
  \BibitemOpen
  \bibfield  {author} {\bibinfo {author} {\bibfnamefont {L.~Z.}\ \bibnamefont
  {Cohen}}, \bibinfo {author} {\bibfnamefont {I.~H.}\ \bibnamefont {Kim}},
  \bibinfo {author} {\bibfnamefont {S.~D.}\ \bibnamefont {Bartlett}}, \ and\
  \bibinfo {author} {\bibfnamefont {B.~J.}\ \bibnamefont {Brown}},\ }\href
  {http://arxiv.org/abs/2110.10794} {\  (\bibinfo {year} {2021})},\ \Eprint
  {http://arxiv.org/abs/2110.10794} {arXiv:2110.10794} \BibitemShut {NoStop}%
\bibitem [{\citenamefont {Gimeno-Segovia}\ \emph {et~al.}(2015)\citenamefont
  {Gimeno-Segovia}, \citenamefont {Shadbolt}, \citenamefont {Browne},\ and\
  \citenamefont {Rudolph}}]{Gimeno2015}%
  \BibitemOpen
  \bibfield  {author} {\bibinfo {author} {\bibfnamefont {M.}~\bibnamefont
  {Gimeno-Segovia}}, \bibinfo {author} {\bibfnamefont {P.}~\bibnamefont
  {Shadbolt}}, \bibinfo {author} {\bibfnamefont {D.~E.}\ \bibnamefont
  {Browne}}, \ and\ \bibinfo {author} {\bibfnamefont {T.}~\bibnamefont
  {Rudolph}},\ }\href {\doibase 10.1103/PhysRevLett.115.020502} {\bibfield
  {journal} {\bibinfo  {journal} {Phys. Rev. Lett.}\ }\textbf {\bibinfo
  {volume} {115}},\ \bibinfo {pages} {020502} (\bibinfo {year}
  {2015})}\BibitemShut {NoStop}%
\bibitem [{\citenamefont {Stephenson}\ \emph {et~al.}(2020)\citenamefont
  {Stephenson}, \citenamefont {Nadlinger}, \citenamefont {Nichol},
  \citenamefont {An}, \citenamefont {Drmota}, \citenamefont {Ballance},
  \citenamefont {Thirumalai}, \citenamefont {Goodwin}, \citenamefont {Lucas},\
  and\ \citenamefont {Ballance}}]{Stephenson2020}%
  \BibitemOpen
  \bibfield  {author} {\bibinfo {author} {\bibfnamefont {L.~J.}\ \bibnamefont
  {Stephenson}}, \bibinfo {author} {\bibfnamefont {D.~P.}\ \bibnamefont
  {Nadlinger}}, \bibinfo {author} {\bibfnamefont {B.~C.}\ \bibnamefont
  {Nichol}}, \bibinfo {author} {\bibfnamefont {S.}~\bibnamefont {An}}, \bibinfo
  {author} {\bibfnamefont {P.}~\bibnamefont {Drmota}}, \bibinfo {author}
  {\bibfnamefont {T.~G.}\ \bibnamefont {Ballance}}, \bibinfo {author}
  {\bibfnamefont {K.}~\bibnamefont {Thirumalai}}, \bibinfo {author}
  {\bibfnamefont {J.~F.}\ \bibnamefont {Goodwin}}, \bibinfo {author}
  {\bibfnamefont {D.~M.}\ \bibnamefont {Lucas}}, \ and\ \bibinfo {author}
  {\bibfnamefont {C.~J.}\ \bibnamefont {Ballance}},\ }\href {\doibase
  10.1103/PhysRevLett.124.110501} {\bibfield  {journal} {\bibinfo  {journal}
  {Phys. Rev. Lett.}\ }\textbf {\bibinfo {volume} {124}},\ \bibinfo {pages}
  {110501} (\bibinfo {year} {2020})}\BibitemShut {NoStop}%
\bibitem [{\citenamefont {Postler}\ \emph {et~al.}(2021)\citenamefont
  {Postler}, \citenamefont {Heu{\ss}en}, \citenamefont {Pogorelov},
  \citenamefont {Rispler}, \citenamefont {Feldker}, \citenamefont {Meth},
  \citenamefont {Marciniak}, \citenamefont {Stricker}, \citenamefont
  {Ringbauer}, \citenamefont {Blatt}, \citenamefont {Schindler}, \citenamefont
  {M{\"{u}}ller},\ and\ \citenamefont {Monz}}]{Postler2021}%
  \BibitemOpen
  \bibfield  {author} {\bibinfo {author} {\bibfnamefont {L.}~\bibnamefont
  {Postler}}, \bibinfo {author} {\bibfnamefont {S.}~\bibnamefont {Heu{\ss}en}},
  \bibinfo {author} {\bibfnamefont {I.}~\bibnamefont {Pogorelov}}, \bibinfo
  {author} {\bibfnamefont {M.}~\bibnamefont {Rispler}}, \bibinfo {author}
  {\bibfnamefont {T.}~\bibnamefont {Feldker}}, \bibinfo {author} {\bibfnamefont
  {M.}~\bibnamefont {Meth}}, \bibinfo {author} {\bibfnamefont {C.~D.}\
  \bibnamefont {Marciniak}}, \bibinfo {author} {\bibfnamefont {R.}~\bibnamefont
  {Stricker}}, \bibinfo {author} {\bibfnamefont {M.}~\bibnamefont {Ringbauer}},
  \bibinfo {author} {\bibfnamefont {R.}~\bibnamefont {Blatt}}, \bibinfo
  {author} {\bibfnamefont {P.}~\bibnamefont {Schindler}}, \bibinfo {author}
  {\bibfnamefont {M.}~\bibnamefont {M{\"{u}}ller}}, \ and\ \bibinfo {author}
  {\bibfnamefont {T.}~\bibnamefont {Monz}},\ }\href
  {http://arxiv.org/abs/2111.12654
  https://journals.aps.org/prl/pdf/10.1103/PhysRevLett.124.110501} {\
  (\bibinfo {year} {2021})},\ \Eprint {http://arxiv.org/abs/2111.12654}
  {arXiv:2111.12654} \BibitemShut {NoStop}%
\bibitem [{\citenamefont {Abobeih}\ \emph {et~al.}(2022)\citenamefont
  {Abobeih}, \citenamefont {Wang}, \citenamefont {Randall}, \citenamefont
  {Loenen}, \citenamefont {Bradley}, \citenamefont {Markham}, \citenamefont
  {Twitchen}, \citenamefont {Terhal},\ and\ \citenamefont
  {Taminiau}}]{Abobeih2021}%
  \BibitemOpen
  \bibfield  {author} {\bibinfo {author} {\bibfnamefont {M.~H.}\ \bibnamefont
  {Abobeih}}, \bibinfo {author} {\bibfnamefont {Y.}~\bibnamefont {Wang}},
  \bibinfo {author} {\bibfnamefont {J.}~\bibnamefont {Randall}}, \bibinfo
  {author} {\bibfnamefont {S.~J.~H.}\ \bibnamefont {Loenen}}, \bibinfo {author}
  {\bibfnamefont {C.~E.}\ \bibnamefont {Bradley}}, \bibinfo {author}
  {\bibfnamefont {M.}~\bibnamefont {Markham}}, \bibinfo {author} {\bibfnamefont
  {D.~J.}\ \bibnamefont {Twitchen}}, \bibinfo {author} {\bibfnamefont {B.~M.}\
  \bibnamefont {Terhal}}, \ and\ \bibinfo {author} {\bibfnamefont {T.~H.}\
  \bibnamefont {Taminiau}},\ }\href {\doibase 10.1038/s41586-022-04819-6}
  {\bibfield  {journal} {\bibinfo  {journal} {Nature}\ } (\bibinfo {year}
  {2022}),\ 10.1038/s41586-022-04819-6}\BibitemShut {NoStop}%
\bibitem [{\citenamefont {Bergeron}\ \emph {et~al.}(2020)\citenamefont
  {Bergeron}, \citenamefont {Chartrand}, \citenamefont {Kurkjian},
  \citenamefont {Morse}, \citenamefont {Riemann}, \citenamefont {Abrosimov},
  \citenamefont {Becker}, \citenamefont {Pohl}, \citenamefont {Thewalt},\ and\
  \citenamefont {Simmons}}]{Bergeron2020}%
  \BibitemOpen
  \bibfield  {author} {\bibinfo {author} {\bibfnamefont {L.}~\bibnamefont
  {Bergeron}}, \bibinfo {author} {\bibfnamefont {C.}~\bibnamefont {Chartrand}},
  \bibinfo {author} {\bibfnamefont {A.~T.~K.}\ \bibnamefont {Kurkjian}},
  \bibinfo {author} {\bibfnamefont {K.~J.}\ \bibnamefont {Morse}}, \bibinfo
  {author} {\bibfnamefont {H.}~\bibnamefont {Riemann}}, \bibinfo {author}
  {\bibfnamefont {N.~V.}\ \bibnamefont {Abrosimov}}, \bibinfo {author}
  {\bibfnamefont {P.}~\bibnamefont {Becker}}, \bibinfo {author} {\bibfnamefont
  {H.-J.}\ \bibnamefont {Pohl}}, \bibinfo {author} {\bibfnamefont {M.~L.~W.}\
  \bibnamefont {Thewalt}}, \ and\ \bibinfo {author} {\bibfnamefont
  {S.}~\bibnamefont {Simmons}},\ }\href {\doibase 10.1103/PRXQuantum.1.020301}
  {\bibfield  {journal} {\bibinfo  {journal} {PRX Quantum}\ }\textbf {\bibinfo
  {volume} {1}},\ \bibinfo {pages} {020301} (\bibinfo {year}
  {2020})}\BibitemShut {NoStop}%
\bibitem [{\citenamefont {Christle}\ \emph {et~al.}(2017)\citenamefont
  {Christle}, \citenamefont {Klimov}, \citenamefont {de~las Casas},
  \citenamefont {Sz{\'{a}}sz}, \citenamefont {Iv{\'{a}}dy}, \citenamefont
  {Jokubavicius}, \citenamefont {{Ul Hassan}}, \citenamefont
  {Syv{\"{a}}j{\"{a}}rvi}, \citenamefont {Koehl}, \citenamefont {Ohshima},
  \citenamefont {Son}, \citenamefont {Janz{\'{e}}n}, \citenamefont {Gali},\
  and\ \citenamefont {Awschalom}}]{Christle2017}%
  \BibitemOpen
  \bibfield  {author} {\bibinfo {author} {\bibfnamefont {D.~J.}\ \bibnamefont
  {Christle}}, \bibinfo {author} {\bibfnamefont {P.~V.}\ \bibnamefont
  {Klimov}}, \bibinfo {author} {\bibfnamefont {C.~F.}\ \bibnamefont {de~las
  Casas}}, \bibinfo {author} {\bibfnamefont {K.}~\bibnamefont {Sz{\'{a}}sz}},
  \bibinfo {author} {\bibfnamefont {V.}~\bibnamefont {Iv{\'{a}}dy}}, \bibinfo
  {author} {\bibfnamefont {V.}~\bibnamefont {Jokubavicius}}, \bibinfo {author}
  {\bibfnamefont {J.}~\bibnamefont {{Ul Hassan}}}, \bibinfo {author}
  {\bibfnamefont {M.}~\bibnamefont {Syv{\"{a}}j{\"{a}}rvi}}, \bibinfo {author}
  {\bibfnamefont {W.~F.}\ \bibnamefont {Koehl}}, \bibinfo {author}
  {\bibfnamefont {T.}~\bibnamefont {Ohshima}}, \bibinfo {author} {\bibfnamefont
  {N.~T.}\ \bibnamefont {Son}}, \bibinfo {author} {\bibfnamefont
  {E.}~\bibnamefont {Janz{\'{e}}n}}, \bibinfo {author} {\bibfnamefont
  {{\'{A}}.}~\bibnamefont {Gali}}, \ and\ \bibinfo {author} {\bibfnamefont
  {D.~D.}\ \bibnamefont {Awschalom}},\ }\href {\doibase
  10.1103/PhysRevX.7.021046} {\bibfield  {journal} {\bibinfo  {journal} {Phys.
  Rev. X}\ }\textbf {\bibinfo {volume} {7}},\ \bibinfo {pages} {021046}
  (\bibinfo {year} {2017})}\BibitemShut {NoStop}%
\bibitem [{\citenamefont {Ruskuc}\ \emph {et~al.}(2022)\citenamefont {Ruskuc},
  \citenamefont {Wu}, \citenamefont {Rochman}, \citenamefont {Choi},\ and\
  \citenamefont {Faraon}}]{Ruskuc2021}%
  \BibitemOpen
  \bibfield  {author} {\bibinfo {author} {\bibfnamefont {A.}~\bibnamefont
  {Ruskuc}}, \bibinfo {author} {\bibfnamefont {C.-J.}\ \bibnamefont {Wu}},
  \bibinfo {author} {\bibfnamefont {J.}~\bibnamefont {Rochman}}, \bibinfo
  {author} {\bibfnamefont {J.}~\bibnamefont {Choi}}, \ and\ \bibinfo {author}
  {\bibfnamefont {A.}~\bibnamefont {Faraon}},\ }\href {\doibase
  10.1038/s41586-021-04293-6} {\bibfield  {journal} {\bibinfo  {journal}
  {Nature}\ }\textbf {\bibinfo {volume} {602}},\ \bibinfo {pages} {408}
  (\bibinfo {year} {2022})}\BibitemShut {NoStop}%
\bibitem [{\citenamefont {Raha}\ \emph {et~al.}(2020)\citenamefont {Raha},
  \citenamefont {Chen}, \citenamefont {Phenicie}, \citenamefont {Ourari},
  \citenamefont {Dibos},\ and\ \citenamefont {Thompson}}]{Raha2020}%
  \BibitemOpen
  \bibfield  {author} {\bibinfo {author} {\bibfnamefont {M.}~\bibnamefont
  {Raha}}, \bibinfo {author} {\bibfnamefont {S.}~\bibnamefont {Chen}}, \bibinfo
  {author} {\bibfnamefont {C.~M.}\ \bibnamefont {Phenicie}}, \bibinfo {author}
  {\bibfnamefont {S.}~\bibnamefont {Ourari}}, \bibinfo {author} {\bibfnamefont
  {A.~M.}\ \bibnamefont {Dibos}}, \ and\ \bibinfo {author} {\bibfnamefont
  {J.~D.}\ \bibnamefont {Thompson}},\ }\href {\doibase
  10.1038/s41467-020-15138-7} {\bibfield  {journal} {\bibinfo  {journal} {Nat.
  Commun.}\ }\textbf {\bibinfo {volume} {11}},\ \bibinfo {pages} {1605}
  (\bibinfo {year} {2020})}\BibitemShut {NoStop}%
\bibitem [{\citenamefont {Berezovsky}\ \emph {et~al.}(2008)\citenamefont
  {Berezovsky}, \citenamefont {Mikkelsen}, \citenamefont {Stoltz},
  \citenamefont {Coldren},\ and\ \citenamefont {Awschalom}}]{Berezovsky2008}%
  \BibitemOpen
  \bibfield  {author} {\bibinfo {author} {\bibfnamefont {J.}~\bibnamefont
  {Berezovsky}}, \bibinfo {author} {\bibfnamefont {M.~H.}\ \bibnamefont
  {Mikkelsen}}, \bibinfo {author} {\bibfnamefont {N.~G.}\ \bibnamefont
  {Stoltz}}, \bibinfo {author} {\bibfnamefont {L.~A.}\ \bibnamefont {Coldren}},
  \ and\ \bibinfo {author} {\bibfnamefont {D.~D.}\ \bibnamefont {Awschalom}},\
  }\href {\doibase 10.1126/science.1154798} {\bibfield  {journal} {\bibinfo
  {journal} {Science}\ }\textbf {\bibinfo {volume} {320}},\ \bibinfo {pages}
  {349} (\bibinfo {year} {2008})},\ \Eprint
  {http://arxiv.org/abs/https://www.science.org/doi/pdf/10.1126/science.1154798}
  {https://www.science.org/doi/pdf/10.1126/science.1154798} \BibitemShut
  {NoStop}%
\bibitem [{\citenamefont {{De Greve}}\ \emph {et~al.}(2011)\citenamefont {{De
  Greve}}, \citenamefont {McMahon}, \citenamefont {Press}, \citenamefont
  {Ladd}, \citenamefont {Bisping}, \citenamefont {Schneider}, \citenamefont
  {Kamp}, \citenamefont {Worschech}, \citenamefont {H{\"{o}}fling},
  \citenamefont {Forchel},\ and\ \citenamefont {Yamamoto}}]{DeGreve2011}%
  \BibitemOpen
  \bibfield  {author} {\bibinfo {author} {\bibfnamefont {K.}~\bibnamefont {{De
  Greve}}}, \bibinfo {author} {\bibfnamefont {P.~L.}\ \bibnamefont {McMahon}},
  \bibinfo {author} {\bibfnamefont {D.}~\bibnamefont {Press}}, \bibinfo
  {author} {\bibfnamefont {T.~D.}\ \bibnamefont {Ladd}}, \bibinfo {author}
  {\bibfnamefont {D.}~\bibnamefont {Bisping}}, \bibinfo {author} {\bibfnamefont
  {C.}~\bibnamefont {Schneider}}, \bibinfo {author} {\bibfnamefont
  {M.}~\bibnamefont {Kamp}}, \bibinfo {author} {\bibfnamefont {L.}~\bibnamefont
  {Worschech}}, \bibinfo {author} {\bibfnamefont {S.}~\bibnamefont
  {H{\"{o}}fling}}, \bibinfo {author} {\bibfnamefont {A.}~\bibnamefont
  {Forchel}}, \ and\ \bibinfo {author} {\bibfnamefont {Y.}~\bibnamefont
  {Yamamoto}},\ }\href {\doibase 10.1038/nphys2078} {\bibfield  {journal}
  {\bibinfo  {journal} {Nat. Phys.}\ }\textbf {\bibinfo {volume} {7}},\
  \bibinfo {pages} {872} (\bibinfo {year} {2011})}\BibitemShut {NoStop}%
\bibitem [{\citenamefont {Godden}\ \emph {et~al.}(2012)\citenamefont {Godden},
  \citenamefont {Quilter}, \citenamefont {Ramsay}, \citenamefont {Wu},
  \citenamefont {Brereton}, \citenamefont {Boyle}, \citenamefont {Luxmoore},
  \citenamefont {Puebla-Nunez}, \citenamefont {Fox},\ and\ \citenamefont
  {Skolnick}}]{Godden2012}%
  \BibitemOpen
  \bibfield  {author} {\bibinfo {author} {\bibfnamefont {T.~M.}\ \bibnamefont
  {Godden}}, \bibinfo {author} {\bibfnamefont {J.~H.}\ \bibnamefont {Quilter}},
  \bibinfo {author} {\bibfnamefont {A.~J.}\ \bibnamefont {Ramsay}}, \bibinfo
  {author} {\bibfnamefont {Y.}~\bibnamefont {Wu}}, \bibinfo {author}
  {\bibfnamefont {P.}~\bibnamefont {Brereton}}, \bibinfo {author}
  {\bibfnamefont {S.~J.}\ \bibnamefont {Boyle}}, \bibinfo {author}
  {\bibfnamefont {I.~J.}\ \bibnamefont {Luxmoore}}, \bibinfo {author}
  {\bibfnamefont {J.}~\bibnamefont {Puebla-Nunez}}, \bibinfo {author}
  {\bibfnamefont {A.~M.}\ \bibnamefont {Fox}}, \ and\ \bibinfo {author}
  {\bibfnamefont {M.~S.}\ \bibnamefont {Skolnick}},\ }\href {\doibase
  10.1103/PHYSREVLETT.108.017402/FIGURES/3/MEDIUM} {\bibfield  {journal}
  {\bibinfo  {journal} {Phys. Rev. Lett.}\ }\textbf {\bibinfo {volume} {108}},\
  \bibinfo {pages} {017402} (\bibinfo {year} {2012})}\BibitemShut {NoStop}%
\bibitem [{\citenamefont {Pfaff}\ \emph {et~al.}(2014)\citenamefont {Pfaff},
  \citenamefont {Hensen}, \citenamefont {Bernien}, \citenamefont {van Dam},
  \citenamefont {Blok}, \citenamefont {Taminiau}, \citenamefont {Tiggelman},
  \citenamefont {Schouten}, \citenamefont {Markham}, \citenamefont {Twitchen},\
  and\ \citenamefont {Hanson}}]{Pfaff2014}%
  \BibitemOpen
  \bibfield  {author} {\bibinfo {author} {\bibfnamefont {W.}~\bibnamefont
  {Pfaff}}, \bibinfo {author} {\bibfnamefont {B.~J.}\ \bibnamefont {Hensen}},
  \bibinfo {author} {\bibfnamefont {H.}~\bibnamefont {Bernien}}, \bibinfo
  {author} {\bibfnamefont {S.~B.}\ \bibnamefont {van Dam}}, \bibinfo {author}
  {\bibfnamefont {M.~S.}\ \bibnamefont {Blok}}, \bibinfo {author}
  {\bibfnamefont {T.~H.}\ \bibnamefont {Taminiau}}, \bibinfo {author}
  {\bibfnamefont {M.~J.}\ \bibnamefont {Tiggelman}}, \bibinfo {author}
  {\bibfnamefont {R.~N.}\ \bibnamefont {Schouten}}, \bibinfo {author}
  {\bibfnamefont {M.}~\bibnamefont {Markham}}, \bibinfo {author} {\bibfnamefont
  {D.~J.}\ \bibnamefont {Twitchen}}, \ and\ \bibinfo {author} {\bibfnamefont
  {R.}~\bibnamefont {Hanson}},\ }\href {\doibase 10.1126/science.1253512}
  {\bibfield  {journal} {\bibinfo  {journal} {Science}\ }\textbf {\bibinfo
  {volume} {345}},\ \bibinfo {pages} {532} (\bibinfo {year} {2014})},\ \Eprint
  {http://arxiv.org/abs/https://www.science.org/doi/pdf/10.1126/science.1253512}
  {https://www.science.org/doi/pdf/10.1126/science.1253512} \BibitemShut
  {NoStop}%
\bibitem [{\citenamefont {Pompili}\ \emph {et~al.}(2021)\citenamefont
  {Pompili}, \citenamefont {Hermans}, \citenamefont {Baier}, \citenamefont
  {Beukers}, \citenamefont {Humphreys}, \citenamefont {Schouten}, \citenamefont
  {Vermeulen}, \citenamefont {Tiggelman}, \citenamefont {dos Santos~Martins},
  \citenamefont {Dirkse}, \citenamefont {Wehner},\ and\ \citenamefont
  {Hanson}}]{Pompili2021}%
  \BibitemOpen
  \bibfield  {author} {\bibinfo {author} {\bibfnamefont {M.}~\bibnamefont
  {Pompili}}, \bibinfo {author} {\bibfnamefont {S.~L.~N.}\ \bibnamefont
  {Hermans}}, \bibinfo {author} {\bibfnamefont {S.}~\bibnamefont {Baier}},
  \bibinfo {author} {\bibfnamefont {H.~K.~C.}\ \bibnamefont {Beukers}},
  \bibinfo {author} {\bibfnamefont {P.~C.}\ \bibnamefont {Humphreys}}, \bibinfo
  {author} {\bibfnamefont {R.~N.}\ \bibnamefont {Schouten}}, \bibinfo {author}
  {\bibfnamefont {R.~F.~L.}\ \bibnamefont {Vermeulen}}, \bibinfo {author}
  {\bibfnamefont {M.~J.}\ \bibnamefont {Tiggelman}}, \bibinfo {author}
  {\bibfnamefont {L.}~\bibnamefont {dos Santos~Martins}}, \bibinfo {author}
  {\bibfnamefont {B.}~\bibnamefont {Dirkse}}, \bibinfo {author} {\bibfnamefont
  {S.}~\bibnamefont {Wehner}}, \ and\ \bibinfo {author} {\bibfnamefont
  {R.}~\bibnamefont {Hanson}},\ }\href {\doibase 10.1126/science.abg1919}
  {\bibfield  {journal} {\bibinfo  {journal} {Science}\ }\textbf {\bibinfo
  {volume} {372}},\ \bibinfo {pages} {259} (\bibinfo {year} {2021})},\ \Eprint
  {http://arxiv.org/abs/https://www.science.org/doi/pdf/10.1126/science.abg1919}
  {https://www.science.org/doi/pdf/10.1126/science.abg1919} \BibitemShut
  {NoStop}%
\bibitem [{\citenamefont {Schwartz}\ \emph
  {et~al.}(2016{\natexlab{a}})\citenamefont {Schwartz}, \citenamefont {Cogan},
  \citenamefont {Schmidgall}, \citenamefont {Don}, \citenamefont {Gantz},
  \citenamefont {Kenneth}, \citenamefont {Lindner},\ and\ \citenamefont
  {Gershoni}}]{Schwartz2016a}%
  \BibitemOpen
  \bibfield  {author} {\bibinfo {author} {\bibfnamefont {I.}~\bibnamefont
  {Schwartz}}, \bibinfo {author} {\bibfnamefont {D.}~\bibnamefont {Cogan}},
  \bibinfo {author} {\bibfnamefont {E.~R.}\ \bibnamefont {Schmidgall}},
  \bibinfo {author} {\bibfnamefont {Y.}~\bibnamefont {Don}}, \bibinfo {author}
  {\bibfnamefont {L.}~\bibnamefont {Gantz}}, \bibinfo {author} {\bibfnamefont
  {O.}~\bibnamefont {Kenneth}}, \bibinfo {author} {\bibfnamefont {N.~H.}\
  \bibnamefont {Lindner}}, \ and\ \bibinfo {author} {\bibfnamefont
  {D.}~\bibnamefont {Gershoni}},\ }\href {\doibase 10.1126/science.aah4758}
  {\bibfield  {journal} {\bibinfo  {journal} {Science}\ }\textbf {\bibinfo
  {volume} {354}},\ \bibinfo {pages} {434} (\bibinfo {year}
  {2016}{\natexlab{a}})}\BibitemShut {NoStop}%
\bibitem [{\citenamefont {Istrati}\ \emph {et~al.}(2020)\citenamefont
  {Istrati}, \citenamefont {Pilnyak}, \citenamefont {Loredo}, \citenamefont
  {Ant{\'{o}}n}, \citenamefont {Somaschi}, \citenamefont {Hilaire},
  \citenamefont {Ollivier}, \citenamefont {Esmann}, \citenamefont {Cohen},
  \citenamefont {Vidro}, \citenamefont {Millet}, \citenamefont
  {Lema{\^{i}}tre}, \citenamefont {Sagnes}, \citenamefont {Harouri},
  \citenamefont {Lanco}, \citenamefont {Senellart},\ and\ \citenamefont
  {Eisenberg}}]{Istrati2020}%
  \BibitemOpen
  \bibfield  {author} {\bibinfo {author} {\bibfnamefont {D.}~\bibnamefont
  {Istrati}}, \bibinfo {author} {\bibfnamefont {Y.}~\bibnamefont {Pilnyak}},
  \bibinfo {author} {\bibfnamefont {J.~C.}\ \bibnamefont {Loredo}}, \bibinfo
  {author} {\bibfnamefont {C.}~\bibnamefont {Ant{\'{o}}n}}, \bibinfo {author}
  {\bibfnamefont {N.}~\bibnamefont {Somaschi}}, \bibinfo {author}
  {\bibfnamefont {P.}~\bibnamefont {Hilaire}}, \bibinfo {author} {\bibfnamefont
  {H.}~\bibnamefont {Ollivier}}, \bibinfo {author} {\bibfnamefont
  {M.}~\bibnamefont {Esmann}}, \bibinfo {author} {\bibfnamefont
  {L.}~\bibnamefont {Cohen}}, \bibinfo {author} {\bibfnamefont
  {L.}~\bibnamefont {Vidro}}, \bibinfo {author} {\bibfnamefont
  {C.}~\bibnamefont {Millet}}, \bibinfo {author} {\bibfnamefont
  {A.}~\bibnamefont {Lema{\^{i}}tre}}, \bibinfo {author} {\bibfnamefont
  {I.}~\bibnamefont {Sagnes}}, \bibinfo {author} {\bibfnamefont
  {A.}~\bibnamefont {Harouri}}, \bibinfo {author} {\bibfnamefont
  {L.}~\bibnamefont {Lanco}}, \bibinfo {author} {\bibfnamefont
  {P.}~\bibnamefont {Senellart}}, \ and\ \bibinfo {author} {\bibfnamefont
  {H.~S.}\ \bibnamefont {Eisenberg}},\ }\href {\doibase
  10.1038/s41467-020-19341-4} {\bibfield  {journal} {\bibinfo  {journal} {Nat.
  Commun.}\ }\textbf {\bibinfo {volume} {11}},\ \bibinfo {pages} {5501}
  (\bibinfo {year} {2020})},\ \Eprint {http://arxiv.org/abs/1912.04375}
  {arXiv:1912.04375} \BibitemShut {NoStop}%
\bibitem [{\citenamefont {Cogan}\ \emph {et~al.}(2021)\citenamefont {Cogan},
  \citenamefont {Su}, \citenamefont {Kenneth},\ and\ \citenamefont
  {Gershoni}}]{Cogan2021}%
  \BibitemOpen
  \bibfield  {author} {\bibinfo {author} {\bibfnamefont {D.}~\bibnamefont
  {Cogan}}, \bibinfo {author} {\bibfnamefont {Z.-E.}\ \bibnamefont {Su}},
  \bibinfo {author} {\bibfnamefont {O.}~\bibnamefont {Kenneth}}, \ and\
  \bibinfo {author} {\bibfnamefont {D.}~\bibnamefont {Gershoni}},\ }\href
  {https://arxiv.org/abs/2110.05908v1 http://arxiv.org/abs/2110.05908} {\
  (\bibinfo {year} {2021})},\ \Eprint {http://arxiv.org/abs/2110.05908}
  {arXiv:2110.05908} \BibitemShut {NoStop}%
\bibitem [{\citenamefont {Wang}\ \emph {et~al.}(2019)\citenamefont {Wang},
  \citenamefont {He}, \citenamefont {Chung}, \citenamefont {Hu}, \citenamefont
  {Yu}, \citenamefont {Chen}, \citenamefont {Ding}, \citenamefont {Chen},
  \citenamefont {Qin}, \citenamefont {Yang}, \citenamefont {Liu}, \citenamefont
  {Duan}, \citenamefont {Li}, \citenamefont {Gerhardt}, \citenamefont
  {Winkler}, \citenamefont {Jurkat}, \citenamefont {Wang}, \citenamefont
  {Gregersen}, \citenamefont {Huo}, \citenamefont {Dai}, \citenamefont {Yu},
  \citenamefont {H{\"{o}}fling}, \citenamefont {Lu},\ and\ \citenamefont
  {Pan}}]{Wang2019}%
  \BibitemOpen
  \bibfield  {author} {\bibinfo {author} {\bibfnamefont {H.}~\bibnamefont
  {Wang}}, \bibinfo {author} {\bibfnamefont {Y.-M.}\ \bibnamefont {He}},
  \bibinfo {author} {\bibfnamefont {T.-H.}\ \bibnamefont {Chung}}, \bibinfo
  {author} {\bibfnamefont {H.}~\bibnamefont {Hu}}, \bibinfo {author}
  {\bibfnamefont {Y.}~\bibnamefont {Yu}}, \bibinfo {author} {\bibfnamefont
  {S.}~\bibnamefont {Chen}}, \bibinfo {author} {\bibfnamefont {X.}~\bibnamefont
  {Ding}}, \bibinfo {author} {\bibfnamefont {M.-C.}\ \bibnamefont {Chen}},
  \bibinfo {author} {\bibfnamefont {J.}~\bibnamefont {Qin}}, \bibinfo {author}
  {\bibfnamefont {X.}~\bibnamefont {Yang}}, \bibinfo {author} {\bibfnamefont
  {R.-Z.}\ \bibnamefont {Liu}}, \bibinfo {author} {\bibfnamefont {Z.-C.}\
  \bibnamefont {Duan}}, \bibinfo {author} {\bibfnamefont {J.-P.}\ \bibnamefont
  {Li}}, \bibinfo {author} {\bibfnamefont {S.}~\bibnamefont {Gerhardt}},
  \bibinfo {author} {\bibfnamefont {K.}~\bibnamefont {Winkler}}, \bibinfo
  {author} {\bibfnamefont {J.}~\bibnamefont {Jurkat}}, \bibinfo {author}
  {\bibfnamefont {L.-J.}\ \bibnamefont {Wang}}, \bibinfo {author}
  {\bibfnamefont {N.}~\bibnamefont {Gregersen}}, \bibinfo {author}
  {\bibfnamefont {Y.-H.}\ \bibnamefont {Huo}}, \bibinfo {author} {\bibfnamefont
  {Q.}~\bibnamefont {Dai}}, \bibinfo {author} {\bibfnamefont {S.}~\bibnamefont
  {Yu}}, \bibinfo {author} {\bibfnamefont {S.}~\bibnamefont {H{\"{o}}fling}},
  \bibinfo {author} {\bibfnamefont {C.-Y.}\ \bibnamefont {Lu}}, \ and\ \bibinfo
  {author} {\bibfnamefont {J.-W.}\ \bibnamefont {Pan}},\ }\href {\doibase
  10.1038/s41566-019-0494-3} {\bibfield  {journal} {\bibinfo  {journal} {Nat.
  Photonics}\ }\textbf {\bibinfo {volume} {13}},\ \bibinfo {pages} {770}
  (\bibinfo {year} {2019})}\BibitemShut {NoStop}%
\bibitem [{\citenamefont {Liu}\ \emph {et~al.}(2019)\citenamefont {Liu},
  \citenamefont {Su}, \citenamefont {Wei}, \citenamefont {Yao}, \citenamefont
  {da~Silva}, \citenamefont {Yu}, \citenamefont {Iles-Smith}, \citenamefont
  {Srinivasan}, \citenamefont {Rastelli}, \citenamefont {Li},\ and\
  \citenamefont {Wang}}]{Liu2019}%
  \BibitemOpen
  \bibfield  {author} {\bibinfo {author} {\bibfnamefont {J.}~\bibnamefont
  {Liu}}, \bibinfo {author} {\bibfnamefont {R.}~\bibnamefont {Su}}, \bibinfo
  {author} {\bibfnamefont {Y.}~\bibnamefont {Wei}}, \bibinfo {author}
  {\bibfnamefont {B.}~\bibnamefont {Yao}}, \bibinfo {author} {\bibfnamefont
  {S.~F.~C.}\ \bibnamefont {da~Silva}}, \bibinfo {author} {\bibfnamefont
  {Y.}~\bibnamefont {Yu}}, \bibinfo {author} {\bibfnamefont {J.}~\bibnamefont
  {Iles-Smith}}, \bibinfo {author} {\bibfnamefont {K.}~\bibnamefont
  {Srinivasan}}, \bibinfo {author} {\bibfnamefont {A.}~\bibnamefont
  {Rastelli}}, \bibinfo {author} {\bibfnamefont {J.}~\bibnamefont {Li}}, \ and\
  \bibinfo {author} {\bibfnamefont {X.}~\bibnamefont {Wang}},\ }\href {\doibase
  10.1038/s41565-019-0435-9} {\bibfield  {journal} {\bibinfo  {journal} {Nat.
  Nanotechnol.}\ }\textbf {\bibinfo {volume} {14}},\ \bibinfo {pages} {586}
  (\bibinfo {year} {2019})}\BibitemShut {NoStop}%
\bibitem [{\citenamefont {Tomm}\ \emph {et~al.}(2021)\citenamefont {Tomm},
  \citenamefont {Javadi}, \citenamefont {Antoniadis}, \citenamefont {Najer},
  \citenamefont {L{\"{o}}bl}, \citenamefont {Korsch}, \citenamefont {Schott},
  \citenamefont {Valentin}, \citenamefont {Wieck}, \citenamefont {Ludwig},\
  and\ \citenamefont {Warburton}}]{Tomm2021}%
  \BibitemOpen
  \bibfield  {author} {\bibinfo {author} {\bibfnamefont {N.}~\bibnamefont
  {Tomm}}, \bibinfo {author} {\bibfnamefont {A.}~\bibnamefont {Javadi}},
  \bibinfo {author} {\bibfnamefont {N.~O.}\ \bibnamefont {Antoniadis}},
  \bibinfo {author} {\bibfnamefont {D.}~\bibnamefont {Najer}}, \bibinfo
  {author} {\bibfnamefont {M.~C.}\ \bibnamefont {L{\"{o}}bl}}, \bibinfo
  {author} {\bibfnamefont {A.~R.}\ \bibnamefont {Korsch}}, \bibinfo {author}
  {\bibfnamefont {R.}~\bibnamefont {Schott}}, \bibinfo {author} {\bibfnamefont
  {S.~R.}\ \bibnamefont {Valentin}}, \bibinfo {author} {\bibfnamefont {A.~D.}\
  \bibnamefont {Wieck}}, \bibinfo {author} {\bibfnamefont {A.}~\bibnamefont
  {Ludwig}}, \ and\ \bibinfo {author} {\bibfnamefont {R.~J.}\ \bibnamefont
  {Warburton}},\ }\href {\doibase 10.1038/s41565-020-00831-x} {\bibfield
  {journal} {\bibinfo  {journal} {Nat. Nanotechnol.}\ }\textbf {\bibinfo
  {volume} {16}},\ \bibinfo {pages} {399} (\bibinfo {year} {2021})},\ \Eprint
  {http://arxiv.org/abs/2007.12654} {arXiv:2007.12654} \BibitemShut {NoStop}%
\bibitem [{\citenamefont {Appel}\ \emph
  {et~al.}(2021{\natexlab{a}})\citenamefont {Appel}, \citenamefont {Tiranov},
  \citenamefont {Javadi}, \citenamefont {L{\"{o}}bl}, \citenamefont {Wang},
  \citenamefont {Scholz}, \citenamefont {Wieck}, \citenamefont {Ludwig},
  \citenamefont {Warburton},\ and\ \citenamefont {Lodahl}}]{Appel2021}%
  \BibitemOpen
  \bibfield  {author} {\bibinfo {author} {\bibfnamefont {M.~H.}\ \bibnamefont
  {Appel}}, \bibinfo {author} {\bibfnamefont {A.}~\bibnamefont {Tiranov}},
  \bibinfo {author} {\bibfnamefont {A.}~\bibnamefont {Javadi}}, \bibinfo
  {author} {\bibfnamefont {M.~C.}\ \bibnamefont {L{\"{o}}bl}}, \bibinfo
  {author} {\bibfnamefont {Y.}~\bibnamefont {Wang}}, \bibinfo {author}
  {\bibfnamefont {S.}~\bibnamefont {Scholz}}, \bibinfo {author} {\bibfnamefont
  {A.~D.}\ \bibnamefont {Wieck}}, \bibinfo {author} {\bibfnamefont
  {A.}~\bibnamefont {Ludwig}}, \bibinfo {author} {\bibfnamefont {R.~J.}\
  \bibnamefont {Warburton}}, \ and\ \bibinfo {author} {\bibfnamefont
  {P.}~\bibnamefont {Lodahl}},\ }\href {\doibase
  10.1103/PhysRevLett.126.013602} {\bibfield  {journal} {\bibinfo  {journal}
  {Phys. Rev. Lett.}\ }\textbf {\bibinfo {volume} {126}},\ \bibinfo {pages}
  {013602} (\bibinfo {year} {2021}{\natexlab{a}})}\BibitemShut {NoStop}%
\bibitem [{\citenamefont {Thomas}\ \emph {et~al.}(2021)\citenamefont {Thomas},
  \citenamefont {Billard}, \citenamefont {Coste}, \citenamefont {Wein},
  \citenamefont {Priya}, \citenamefont {Ollivier}, \citenamefont {Krebs},
  \citenamefont {Taza{\"{i}}rt}, \citenamefont {Harouri}, \citenamefont
  {Lemaitre}, \citenamefont {Sagnes}, \citenamefont {Anton}, \citenamefont
  {Lanco}, \citenamefont {Somaschi}, \citenamefont {Loredo},\ and\
  \citenamefont {Senellart}}]{Thomas2021}%
  \BibitemOpen
  \bibfield  {author} {\bibinfo {author} {\bibfnamefont {S.~E.}\ \bibnamefont
  {Thomas}}, \bibinfo {author} {\bibfnamefont {M.}~\bibnamefont {Billard}},
  \bibinfo {author} {\bibfnamefont {N.}~\bibnamefont {Coste}}, \bibinfo
  {author} {\bibfnamefont {S.~C.}\ \bibnamefont {Wein}}, \bibinfo {author}
  {\bibnamefont {Priya}}, \bibinfo {author} {\bibfnamefont {H.}~\bibnamefont
  {Ollivier}}, \bibinfo {author} {\bibfnamefont {O.}~\bibnamefont {Krebs}},
  \bibinfo {author} {\bibfnamefont {L.}~\bibnamefont {Taza{\"{i}}rt}}, \bibinfo
  {author} {\bibfnamefont {A.}~\bibnamefont {Harouri}}, \bibinfo {author}
  {\bibfnamefont {A.}~\bibnamefont {Lemaitre}}, \bibinfo {author}
  {\bibfnamefont {I.}~\bibnamefont {Sagnes}}, \bibinfo {author} {\bibfnamefont
  {C.}~\bibnamefont {Anton}}, \bibinfo {author} {\bibfnamefont
  {L.}~\bibnamefont {Lanco}}, \bibinfo {author} {\bibfnamefont
  {N.}~\bibnamefont {Somaschi}}, \bibinfo {author} {\bibfnamefont {J.~C.}\
  \bibnamefont {Loredo}}, \ and\ \bibinfo {author} {\bibfnamefont
  {P.}~\bibnamefont {Senellart}},\ }\href {\doibase
  10.1103/PhysRevLett.126.233601} {\bibfield  {journal} {\bibinfo  {journal}
  {Phys. Rev. Lett.}\ }\textbf {\bibinfo {volume} {126}},\ \bibinfo {pages}
  {233601} (\bibinfo {year} {2021})},\ \Eprint
  {http://arxiv.org/abs/2007.04330} {arXiv:2007.04330} \BibitemShut {NoStop}%
\bibitem [{\citenamefont {Varnava}\ \emph {et~al.}(2008)\citenamefont
  {Varnava}, \citenamefont {Browne},\ and\ \citenamefont
  {Rudolph}}]{Varnava2008}%
  \BibitemOpen
  \bibfield  {author} {\bibinfo {author} {\bibfnamefont {M.}~\bibnamefont
  {Varnava}}, \bibinfo {author} {\bibfnamefont {D.~E.}\ \bibnamefont {Browne}},
  \ and\ \bibinfo {author} {\bibfnamefont {T.}~\bibnamefont {Rudolph}},\ }\href
  {\doibase 10.1103/PhysRevLett.100.060502} {\bibfield  {journal} {\bibinfo
  {journal} {Phys. Rev. Lett.}\ }\textbf {\bibinfo {volume} {100}},\ \bibinfo
  {pages} {060502} (\bibinfo {year} {2008})}\BibitemShut {NoStop}%
\bibitem [{\citenamefont {Pant}\ \emph {et~al.}(2019)\citenamefont {Pant},
  \citenamefont {Towsley}, \citenamefont {Englund},\ and\ \citenamefont
  {Guha}}]{Pant2019}%
  \BibitemOpen
  \bibfield  {author} {\bibinfo {author} {\bibfnamefont {M.}~\bibnamefont
  {Pant}}, \bibinfo {author} {\bibfnamefont {D.}~\bibnamefont {Towsley}},
  \bibinfo {author} {\bibfnamefont {D.}~\bibnamefont {Englund}}, \ and\
  \bibinfo {author} {\bibfnamefont {S.}~\bibnamefont {Guha}},\ }\href {\doibase
  10.1038/s41467-019-08948-x} {\bibfield  {journal} {\bibinfo  {journal} {Nat.
  Commun.}\ }\textbf {\bibinfo {volume} {10}},\ \bibinfo {pages} {1070}
  (\bibinfo {year} {2019})},\ \Eprint {http://arxiv.org/abs/1701.03775}
  {arXiv:1701.03775} \BibitemShut {NoStop}%
\bibitem [{\citenamefont {Press}\ \emph {et~al.}(2008)\citenamefont {Press},
  \citenamefont {Ladd}, \citenamefont {Zhang},\ and\ \citenamefont
  {Yamamoto}}]{Press2008}%
  \BibitemOpen
  \bibfield  {author} {\bibinfo {author} {\bibfnamefont {D.}~\bibnamefont
  {Press}}, \bibinfo {author} {\bibfnamefont {T.~D.}\ \bibnamefont {Ladd}},
  \bibinfo {author} {\bibfnamefont {B.}~\bibnamefont {Zhang}}, \ and\ \bibinfo
  {author} {\bibfnamefont {Y.}~\bibnamefont {Yamamoto}},\ }\href {\doibase
  10.1038/nature07530} {\bibfield  {journal} {\bibinfo  {journal} {Nature}\
  }\textbf {\bibinfo {volume} {456}},\ \bibinfo {pages} {218} (\bibinfo {year}
  {2008})}\BibitemShut {NoStop}%
\bibitem [{\citenamefont {{De Greve}}\ \emph {et~al.}(2012)\citenamefont {{De
  Greve}}, \citenamefont {Yu}, \citenamefont {McMahon}, \citenamefont {Pelc},
  \citenamefont {Natarajan}, \citenamefont {Kim}, \citenamefont {Abe},
  \citenamefont {Maier}, \citenamefont {Schneider}, \citenamefont {Kamp},
  \citenamefont {H{\"{o}}fling}, \citenamefont {Hadfield}, \citenamefont
  {Forchel}, \citenamefont {Fejer},\ and\ \citenamefont
  {Yamamoto}}]{DeGreve2012}%
  \BibitemOpen
  \bibfield  {author} {\bibinfo {author} {\bibfnamefont {K.}~\bibnamefont {{De
  Greve}}}, \bibinfo {author} {\bibfnamefont {L.}~\bibnamefont {Yu}}, \bibinfo
  {author} {\bibfnamefont {P.~L.}\ \bibnamefont {McMahon}}, \bibinfo {author}
  {\bibfnamefont {J.~S.}\ \bibnamefont {Pelc}}, \bibinfo {author}
  {\bibfnamefont {C.~M.}\ \bibnamefont {Natarajan}}, \bibinfo {author}
  {\bibfnamefont {N.~Y.}\ \bibnamefont {Kim}}, \bibinfo {author} {\bibfnamefont
  {E.}~\bibnamefont {Abe}}, \bibinfo {author} {\bibfnamefont {S.}~\bibnamefont
  {Maier}}, \bibinfo {author} {\bibfnamefont {C.}~\bibnamefont {Schneider}},
  \bibinfo {author} {\bibfnamefont {M.}~\bibnamefont {Kamp}}, \bibinfo {author}
  {\bibfnamefont {S.}~\bibnamefont {H{\"{o}}fling}}, \bibinfo {author}
  {\bibfnamefont {R.~H.}\ \bibnamefont {Hadfield}}, \bibinfo {author}
  {\bibfnamefont {A.}~\bibnamefont {Forchel}}, \bibinfo {author} {\bibfnamefont
  {M.~M.}\ \bibnamefont {Fejer}}, \ and\ \bibinfo {author} {\bibfnamefont
  {Y.}~\bibnamefont {Yamamoto}},\ }\href {\doibase 10.1038/nature11577}
  {\bibfield  {journal} {\bibinfo  {journal} {Nature}\ }\textbf {\bibinfo
  {volume} {491}},\ \bibinfo {pages} {421} (\bibinfo {year}
  {2012})}\BibitemShut {NoStop}%
\bibitem [{\citenamefont {Gao}\ \emph {et~al.}(2012)\citenamefont {Gao},
  \citenamefont {Fallahi}, \citenamefont {Togan}, \citenamefont
  {Miguel-Sanchez},\ and\ \citenamefont {Imamoglu}}]{Gao2012}%
  \BibitemOpen
  \bibfield  {author} {\bibinfo {author} {\bibfnamefont {W.~B.}\ \bibnamefont
  {Gao}}, \bibinfo {author} {\bibfnamefont {P.}~\bibnamefont {Fallahi}},
  \bibinfo {author} {\bibfnamefont {E.}~\bibnamefont {Togan}}, \bibinfo
  {author} {\bibfnamefont {J.}~\bibnamefont {Miguel-Sanchez}}, \ and\ \bibinfo
  {author} {\bibfnamefont {A.}~\bibnamefont {Imamoglu}},\ }\href {\doibase
  10.1038/nature11573} {\bibfield  {journal} {\bibinfo  {journal} {Nature}\
  }\textbf {\bibinfo {volume} {491}},\ \bibinfo {pages} {426} (\bibinfo {year}
  {2012})}\BibitemShut {NoStop}%
\bibitem [{\citenamefont {Appel}\ \emph
  {et~al.}(2021{\natexlab{b}})\citenamefont {Appel}, \citenamefont {Tiranov},
  \citenamefont {Pabst}, \citenamefont {Chan}, \citenamefont {Starup},
  \citenamefont {Wang}, \citenamefont {Midolo}, \citenamefont {Tiurev},
  \citenamefont {Scholz}, \citenamefont {Wieck}, \citenamefont {Ludwig},
  \citenamefont {Sørensen},\ and\ \citenamefont
  {Lodahl}}]{https://doi.org/10.48550/arxiv.2111.12523}%
  \BibitemOpen
  \bibfield  {author} {\bibinfo {author} {\bibfnamefont {M.~H.}\ \bibnamefont
  {Appel}}, \bibinfo {author} {\bibfnamefont {A.}~\bibnamefont {Tiranov}},
  \bibinfo {author} {\bibfnamefont {S.}~\bibnamefont {Pabst}}, \bibinfo
  {author} {\bibfnamefont {M.~L.}\ \bibnamefont {Chan}}, \bibinfo {author}
  {\bibfnamefont {C.}~\bibnamefont {Starup}}, \bibinfo {author} {\bibfnamefont
  {Y.}~\bibnamefont {Wang}}, \bibinfo {author} {\bibfnamefont {L.}~\bibnamefont
  {Midolo}}, \bibinfo {author} {\bibfnamefont {K.}~\bibnamefont {Tiurev}},
  \bibinfo {author} {\bibfnamefont {S.}~\bibnamefont {Scholz}}, \bibinfo
  {author} {\bibfnamefont {A.~D.}\ \bibnamefont {Wieck}}, \bibinfo {author}
  {\bibfnamefont {A.}~\bibnamefont {Ludwig}}, \bibinfo {author} {\bibfnamefont
  {A.~S.}\ \bibnamefont {Sørensen}}, \ and\ \bibinfo {author} {\bibfnamefont
  {P.}~\bibnamefont {Lodahl}},\ }\href {\doibase 10.48550/ARXIV.2111.12523}
  {\enquote {\bibinfo {title} {A source of indistinguishable time-bin entangled
  photons from a waveguide-embedded quantum dot},}\ } (\bibinfo {year}
  {2021}{\natexlab{b}})\BibitemShut {NoStop}%
\bibitem [{\citenamefont {Delteil}\ \emph {et~al.}(2016)\citenamefont
  {Delteil}, \citenamefont {Sun}, \citenamefont {Gao}, \citenamefont {Togan},
  \citenamefont {Faelt},\ and\ \citenamefont {Imamoğlu}}]{Delteil2015}%
  \BibitemOpen
  \bibfield  {author} {\bibinfo {author} {\bibfnamefont {A.}~\bibnamefont
  {Delteil}}, \bibinfo {author} {\bibfnamefont {Z.}~\bibnamefont {Sun}},
  \bibinfo {author} {\bibfnamefont {W.-b.}\ \bibnamefont {Gao}}, \bibinfo
  {author} {\bibfnamefont {E.}~\bibnamefont {Togan}}, \bibinfo {author}
  {\bibfnamefont {S.}~\bibnamefont {Faelt}}, \ and\ \bibinfo {author}
  {\bibfnamefont {A.}~\bibnamefont {Imamoğlu}},\ }\href {\doibase
  10.1038/nphys3605} {\bibfield  {journal} {\bibinfo  {journal} {Nat. Phys.}\
  }\textbf {\bibinfo {volume} {12}},\ \bibinfo {pages} {218} (\bibinfo {year}
  {2016})},\ \Eprint {http://arxiv.org/abs/arXiv:1507.00465v1}
  {arXiv:arXiv:1507.00465v1} \BibitemShut {NoStop}%
\bibitem [{\citenamefont {Stockill}\ \emph {et~al.}(2017)\citenamefont
  {Stockill}, \citenamefont {Stanley}, \citenamefont {Huthmacher},
  \citenamefont {Clarke}, \citenamefont {Hugues}, \citenamefont {Miller},
  \citenamefont {Matthiesen}, \citenamefont {{Le Gall}},\ and\ \citenamefont
  {Atat{\"{u}}re}}]{Stockill2017}%
  \BibitemOpen
  \bibfield  {author} {\bibinfo {author} {\bibfnamefont {R.}~\bibnamefont
  {Stockill}}, \bibinfo {author} {\bibfnamefont {M.~J.}\ \bibnamefont
  {Stanley}}, \bibinfo {author} {\bibfnamefont {L.}~\bibnamefont {Huthmacher}},
  \bibinfo {author} {\bibfnamefont {E.}~\bibnamefont {Clarke}}, \bibinfo
  {author} {\bibfnamefont {M.}~\bibnamefont {Hugues}}, \bibinfo {author}
  {\bibfnamefont {A.~J.}\ \bibnamefont {Miller}}, \bibinfo {author}
  {\bibfnamefont {C.}~\bibnamefont {Matthiesen}}, \bibinfo {author}
  {\bibfnamefont {C.}~\bibnamefont {{Le Gall}}}, \ and\ \bibinfo {author}
  {\bibfnamefont {M.}~\bibnamefont {Atat{\"{u}}re}},\ }\href {\doibase
  10.1103/PhysRevLett.119.010503} {\bibfield  {journal} {\bibinfo  {journal}
  {Phys. Rev. Lett.}\ }\textbf {\bibinfo {volume} {119}},\ \bibinfo {pages}
  {010503} (\bibinfo {year} {2017})}\BibitemShut {NoStop}%
\bibitem [{\citenamefont {Gangloff}\ \emph {et~al.}(2019)\citenamefont
  {Gangloff}, \citenamefont {Éthier Majcher}, \citenamefont {Lang},
  \citenamefont {Denning}, \citenamefont {Bodey}, \citenamefont {Jackson},
  \citenamefont {Clarke}, \citenamefont {Hugues}, \citenamefont {Gall},\ and\
  \citenamefont {Atatüre}}]{Gangloff2018}%
  \BibitemOpen
  \bibfield  {author} {\bibinfo {author} {\bibfnamefont {D.~A.}\ \bibnamefont
  {Gangloff}}, \bibinfo {author} {\bibfnamefont {G.}~\bibnamefont {Éthier
  Majcher}}, \bibinfo {author} {\bibfnamefont {C.}~\bibnamefont {Lang}},
  \bibinfo {author} {\bibfnamefont {E.~V.}\ \bibnamefont {Denning}}, \bibinfo
  {author} {\bibfnamefont {J.~H.}\ \bibnamefont {Bodey}}, \bibinfo {author}
  {\bibfnamefont {D.~M.}\ \bibnamefont {Jackson}}, \bibinfo {author}
  {\bibfnamefont {E.}~\bibnamefont {Clarke}}, \bibinfo {author} {\bibfnamefont
  {M.}~\bibnamefont {Hugues}}, \bibinfo {author} {\bibfnamefont {C.~L.}\
  \bibnamefont {Gall}}, \ and\ \bibinfo {author} {\bibfnamefont
  {M.}~\bibnamefont {Atatüre}},\ }\href {\doibase 10.1126/science.aaw2906}
  {\bibfield  {journal} {\bibinfo  {journal} {Science}\ }\textbf {\bibinfo
  {volume} {364}},\ \bibinfo {pages} {62} (\bibinfo {year} {2019})},\ \Eprint
  {http://arxiv.org/abs/https://www.science.org/doi/pdf/10.1126/science.aaw2906}
  {https://www.science.org/doi/pdf/10.1126/science.aaw2906} \BibitemShut
  {NoStop}%
\bibitem [{\citenamefont {Taylor}\ \emph {et~al.}(2003)\citenamefont {Taylor},
  \citenamefont {Marcus},\ and\ \citenamefont {Lukin}}]{Taylor2003}%
  \BibitemOpen
  \bibfield  {author} {\bibinfo {author} {\bibfnamefont {J.~M.}\ \bibnamefont
  {Taylor}}, \bibinfo {author} {\bibfnamefont {C.~M.}\ \bibnamefont {Marcus}},
  \ and\ \bibinfo {author} {\bibfnamefont {M.~D.}\ \bibnamefont {Lukin}},\
  }\href {\doibase 10.1103/PhysRevLett.90.206803} {\bibfield  {journal}
  {\bibinfo  {journal} {Phys. Rev. Lett.}\ }\textbf {\bibinfo {volume} {90}},\
  \bibinfo {pages} {206803} (\bibinfo {year} {2003})},\ \Eprint
  {http://arxiv.org/abs/0301323} {arXiv:0301323 [cond-mat]} \BibitemShut
  {NoStop}%
\bibitem [{\citenamefont {Denning}\ \emph {et~al.}(2019)\citenamefont
  {Denning}, \citenamefont {Gangloff}, \citenamefont {Atat{\"{u}}re},
  \citenamefont {M{\o}rk},\ and\ \citenamefont {{Le Gall}}}]{Denning2019}%
  \BibitemOpen
  \bibfield  {author} {\bibinfo {author} {\bibfnamefont {E.~V.}\ \bibnamefont
  {Denning}}, \bibinfo {author} {\bibfnamefont {D.~A.}\ \bibnamefont
  {Gangloff}}, \bibinfo {author} {\bibfnamefont {M.}~\bibnamefont
  {Atat{\"{u}}re}}, \bibinfo {author} {\bibfnamefont {J.}~\bibnamefont
  {M{\o}rk}}, \ and\ \bibinfo {author} {\bibfnamefont {C.}~\bibnamefont {{Le
  Gall}}},\ }\href {\doibase 10.1103/PhysRevLett.123.140502} {\bibfield
  {journal} {\bibinfo  {journal} {Phys. Rev. Lett.}\ }\textbf {\bibinfo
  {volume} {123}},\ \bibinfo {pages} {140502} (\bibinfo {year} {2019})},\
  \Eprint {http://arxiv.org/abs/1904.11180} {arXiv:1904.11180} \BibitemShut
  {NoStop}%
\bibitem [{\citenamefont {Stockill}\ \emph {et~al.}(2016)\citenamefont
  {Stockill}, \citenamefont {{Le Gall}}, \citenamefont {Matthiesen},
  \citenamefont {Huthmacher}, \citenamefont {Clarke}, \citenamefont {Hugues},\
  and\ \citenamefont {Atat{\"{u}}re}}]{Stockill2016}%
  \BibitemOpen
  \bibfield  {author} {\bibinfo {author} {\bibfnamefont {R.}~\bibnamefont
  {Stockill}}, \bibinfo {author} {\bibfnamefont {C.}~\bibnamefont {{Le Gall}}},
  \bibinfo {author} {\bibfnamefont {C.}~\bibnamefont {Matthiesen}}, \bibinfo
  {author} {\bibfnamefont {L.}~\bibnamefont {Huthmacher}}, \bibinfo {author}
  {\bibfnamefont {E.}~\bibnamefont {Clarke}}, \bibinfo {author} {\bibfnamefont
  {M.}~\bibnamefont {Hugues}}, \ and\ \bibinfo {author} {\bibfnamefont
  {M.}~\bibnamefont {Atat{\"{u}}re}},\ }\href {\doibase 10.1038/ncomms12745}
  {\bibfield  {journal} {\bibinfo  {journal} {Nat. Commun.}\ }\textbf {\bibinfo
  {volume} {7}},\ \bibinfo {pages} {12745} (\bibinfo {year}
  {2016})}\BibitemShut {NoStop}%
\bibitem [{\citenamefont {Gong}\ \emph {et~al.}(2004)\citenamefont {Gong},
  \citenamefont {Offermans}, \citenamefont {Nötzel}, \citenamefont
  {Koenraad},\ and\ \citenamefont {Wolter}}]{doi:10.1063/1.1831564}%
  \BibitemOpen
  \bibfield  {author} {\bibinfo {author} {\bibfnamefont {Q.}~\bibnamefont
  {Gong}}, \bibinfo {author} {\bibfnamefont {P.}~\bibnamefont {Offermans}},
  \bibinfo {author} {\bibfnamefont {R.}~\bibnamefont {Nötzel}}, \bibinfo
  {author} {\bibfnamefont {P.~M.}\ \bibnamefont {Koenraad}}, \ and\ \bibinfo
  {author} {\bibfnamefont {J.~H.}\ \bibnamefont {Wolter}},\ }\href {\doibase
  10.1063/1.1831564} {\bibfield  {journal} {\bibinfo  {journal} {Applied
  Physics Letters}\ }\textbf {\bibinfo {volume} {85}},\ \bibinfo {pages} {5697}
  (\bibinfo {year} {2004})},\ \Eprint
  {http://arxiv.org/abs/https://doi.org/10.1063/1.1831564}
  {https://doi.org/10.1063/1.1831564} \BibitemShut {NoStop}%
\bibitem [{\citenamefont {Rastelli}\ \emph {et~al.}(2008)\citenamefont
  {Rastelli}, \citenamefont {Stoffel}, \citenamefont {Malachias}, \citenamefont
  {Merdzhanova}, \citenamefont {Katsaros}, \citenamefont {Kern}, \citenamefont
  {Metzger},\ and\ \citenamefont {Schmidt}}]{Rastelli2008}%
  \BibitemOpen
  \bibfield  {author} {\bibinfo {author} {\bibfnamefont {A.}~\bibnamefont
  {Rastelli}}, \bibinfo {author} {\bibfnamefont {M.}~\bibnamefont {Stoffel}},
  \bibinfo {author} {\bibfnamefont {A.}~\bibnamefont {Malachias}}, \bibinfo
  {author} {\bibfnamefont {T.}~\bibnamefont {Merdzhanova}}, \bibinfo {author}
  {\bibfnamefont {G.}~\bibnamefont {Katsaros}}, \bibinfo {author}
  {\bibfnamefont {K.}~\bibnamefont {Kern}}, \bibinfo {author} {\bibfnamefont
  {T.~H.}\ \bibnamefont {Metzger}}, \ and\ \bibinfo {author} {\bibfnamefont
  {O.~G.}\ \bibnamefont {Schmidt}},\ }\href {\doibase 10.1021/nl080290y}
  {\bibfield  {journal} {\bibinfo  {journal} {Nano Letters}\ }\textbf {\bibinfo
  {volume} {8}},\ \bibinfo {pages} {1404} (\bibinfo {year} {2008})}\BibitemShut
  {NoStop}%
\bibitem [{\citenamefont {Bechtold}\ \emph {et~al.}(2015)\citenamefont
  {Bechtold}, \citenamefont {Rauch}, \citenamefont {Li}, \citenamefont
  {Simmet}, \citenamefont {Ardelt}, \citenamefont {Regler}, \citenamefont
  {M{\"u}ller}, \citenamefont {Sinitsyn},\ and\ \citenamefont
  {Finley}}]{Bechtold2015}%
  \BibitemOpen
  \bibfield  {author} {\bibinfo {author} {\bibfnamefont {A.}~\bibnamefont
  {Bechtold}}, \bibinfo {author} {\bibfnamefont {D.}~\bibnamefont {Rauch}},
  \bibinfo {author} {\bibfnamefont {F.}~\bibnamefont {Li}}, \bibinfo {author}
  {\bibfnamefont {T.}~\bibnamefont {Simmet}}, \bibinfo {author} {\bibfnamefont
  {P.-L.}\ \bibnamefont {Ardelt}}, \bibinfo {author} {\bibfnamefont
  {A.}~\bibnamefont {Regler}}, \bibinfo {author} {\bibfnamefont
  {K.}~\bibnamefont {M{\"u}ller}}, \bibinfo {author} {\bibfnamefont {N.~A.}\
  \bibnamefont {Sinitsyn}}, \ and\ \bibinfo {author} {\bibfnamefont {J.~J.}\
  \bibnamefont {Finley}},\ }\href {\doibase 10.1038/nphys3470} {\bibfield
  {journal} {\bibinfo  {journal} {Nature Physics}\ }\textbf {\bibinfo {volume}
  {11}},\ \bibinfo {pages} {1005} (\bibinfo {year} {2015})}\BibitemShut
  {NoStop}%
\bibitem [{\citenamefont {da~Silva}\ \emph {et~al.}(2021)\citenamefont
  {da~Silva}, \citenamefont {Undeutsch}, \citenamefont {Lehner}, \citenamefont
  {Manna}, \citenamefont {Krieger}, \citenamefont {Reindl}, \citenamefont
  {Schimpf}, \citenamefont {Trotta},\ and\ \citenamefont
  {Rastelli}}]{DaSilva2021}%
  \BibitemOpen
  \bibfield  {author} {\bibinfo {author} {\bibfnamefont {S.~F.~C.}\
  \bibnamefont {da~Silva}}, \bibinfo {author} {\bibfnamefont {G.}~\bibnamefont
  {Undeutsch}}, \bibinfo {author} {\bibfnamefont {B.}~\bibnamefont {Lehner}},
  \bibinfo {author} {\bibfnamefont {S.}~\bibnamefont {Manna}}, \bibinfo
  {author} {\bibfnamefont {T.~M.}\ \bibnamefont {Krieger}}, \bibinfo {author}
  {\bibfnamefont {M.}~\bibnamefont {Reindl}}, \bibinfo {author} {\bibfnamefont
  {C.}~\bibnamefont {Schimpf}}, \bibinfo {author} {\bibfnamefont
  {R.}~\bibnamefont {Trotta}}, \ and\ \bibinfo {author} {\bibfnamefont
  {A.}~\bibnamefont {Rastelli}},\ }\href {\doibase 10.1063/5.0057070}
  {\bibfield  {journal} {\bibinfo  {journal} {Appl. Phys. Lett.}\ }\textbf
  {\bibinfo {volume} {119}},\ \bibinfo {pages} {120502} (\bibinfo {year}
  {2021})},\ \Eprint {http://arxiv.org/abs/2109.01507} {arXiv:2109.01507}
  \BibitemShut {NoStop}%
\bibitem [{\citenamefont {Sch{\"{o}}ll}\ \emph {et~al.}(2019)\citenamefont
  {Sch{\"{o}}ll}, \citenamefont {Hanschke}, \citenamefont {Schweickert},
  \citenamefont {Zeuner}, \citenamefont {Reindl}, \citenamefont {{Covre Da
  Silva}}, \citenamefont {Lettner}, \citenamefont {Trotta}, \citenamefont
  {Finley}, \citenamefont {M{\"{u}}ller}, \citenamefont {Rastelli},
  \citenamefont {Zwiller},\ and\ \citenamefont {J{\"{o}}ns}}]{Scholl2019}%
  \BibitemOpen
  \bibfield  {author} {\bibinfo {author} {\bibfnamefont {E.}~\bibnamefont
  {Sch{\"{o}}ll}}, \bibinfo {author} {\bibfnamefont {L.}~\bibnamefont
  {Hanschke}}, \bibinfo {author} {\bibfnamefont {L.}~\bibnamefont
  {Schweickert}}, \bibinfo {author} {\bibfnamefont {K.~D.}\ \bibnamefont
  {Zeuner}}, \bibinfo {author} {\bibfnamefont {M.}~\bibnamefont {Reindl}},
  \bibinfo {author} {\bibfnamefont {S.~F.}\ \bibnamefont {{Covre Da Silva}}},
  \bibinfo {author} {\bibfnamefont {T.}~\bibnamefont {Lettner}}, \bibinfo
  {author} {\bibfnamefont {R.}~\bibnamefont {Trotta}}, \bibinfo {author}
  {\bibfnamefont {J.~J.}\ \bibnamefont {Finley}}, \bibinfo {author}
  {\bibfnamefont {K.}~\bibnamefont {M{\"{u}}ller}}, \bibinfo {author}
  {\bibfnamefont {A.}~\bibnamefont {Rastelli}}, \bibinfo {author}
  {\bibfnamefont {V.}~\bibnamefont {Zwiller}}, \ and\ \bibinfo {author}
  {\bibfnamefont {K.~D.}\ \bibnamefont {J{\"{o}}ns}},\ }\href {\doibase
  10.1021/ACS.NANOLETT.8B05132/SUPPL_FILE/NL8B05132_SI_002.PDF} {\bibfield
  {journal} {\bibinfo  {journal} {Nano Lett.}\ }\textbf {\bibinfo {volume}
  {19}},\ \bibinfo {pages} {2404} (\bibinfo {year} {2019})},\ \Eprint
  {http://arxiv.org/abs/1901.09721} {arXiv:1901.09721} \BibitemShut {NoStop}%
\bibitem [{\citenamefont {Zhai}\ \emph {et~al.}(2020)\citenamefont {Zhai},
  \citenamefont {L{\"{o}}bl}, \citenamefont {Nguyen}, \citenamefont {Ritzmann},
  \citenamefont {Javadi}, \citenamefont {Spinnler}, \citenamefont {Wieck},
  \citenamefont {Ludwig},\ and\ \citenamefont {Warburton}}]{Zhai2020}%
  \BibitemOpen
  \bibfield  {author} {\bibinfo {author} {\bibfnamefont {L.}~\bibnamefont
  {Zhai}}, \bibinfo {author} {\bibfnamefont {M.~C.}\ \bibnamefont
  {L{\"{o}}bl}}, \bibinfo {author} {\bibfnamefont {G.~N.}\ \bibnamefont
  {Nguyen}}, \bibinfo {author} {\bibfnamefont {J.}~\bibnamefont {Ritzmann}},
  \bibinfo {author} {\bibfnamefont {A.}~\bibnamefont {Javadi}}, \bibinfo
  {author} {\bibfnamefont {C.}~\bibnamefont {Spinnler}}, \bibinfo {author}
  {\bibfnamefont {A.~D.}\ \bibnamefont {Wieck}}, \bibinfo {author}
  {\bibfnamefont {A.}~\bibnamefont {Ludwig}}, \ and\ \bibinfo {author}
  {\bibfnamefont {R.~J.}\ \bibnamefont {Warburton}},\ }\href {\doibase
  10.1038/s41467-020-18625-z} {\bibfield  {journal} {\bibinfo  {journal} {Nat.
  Commun.}\ }\textbf {\bibinfo {volume} {11}},\ \bibinfo {pages} {4745}
  (\bibinfo {year} {2020})}\BibitemShut {NoStop}%
\bibitem [{\citenamefont {Chekhovich}\ \emph {et~al.}(2020)\citenamefont
  {Chekhovich}, \citenamefont {da~Silva},\ and\ \citenamefont
  {Rastelli}}]{Chekhovich2020}%
  \BibitemOpen
  \bibfield  {author} {\bibinfo {author} {\bibfnamefont {E.~A.}\ \bibnamefont
  {Chekhovich}}, \bibinfo {author} {\bibfnamefont {S.~F.~C.}\ \bibnamefont
  {da~Silva}}, \ and\ \bibinfo {author} {\bibfnamefont {A.}~\bibnamefont
  {Rastelli}},\ }\href {\doibase 10.1038/s41565-020-0769-3} {\bibfield
  {journal} {\bibinfo  {journal} {Nat. Nanotechnol.}\ }\textbf {\bibinfo
  {volume} {15}},\ \bibinfo {pages} {999} (\bibinfo {year} {2020})}\BibitemShut
  {NoStop}%
\bibitem [{\citenamefont {Huo}\ \emph {et~al.}(2013)\citenamefont {Huo},
  \citenamefont {Rastelli},\ and\ \citenamefont {Schmidt}}]{Huo2013}%
  \BibitemOpen
  \bibfield  {author} {\bibinfo {author} {\bibfnamefont {Y.~H.}\ \bibnamefont
  {Huo}}, \bibinfo {author} {\bibfnamefont {A.}~\bibnamefont {Rastelli}}, \
  and\ \bibinfo {author} {\bibfnamefont {O.~G.}\ \bibnamefont {Schmidt}},\
  }\href {\doibase 10.1063/1.4802088} {\bibfield  {journal} {\bibinfo
  {journal} {Appl. Phys. Lett.}\ }\textbf {\bibinfo {volume} {102}},\ \bibinfo
  {pages} {152105} (\bibinfo {year} {2013})}\BibitemShut {NoStop}%
\bibitem [{\citenamefont {Schimpf}\ \emph {et~al.}(2021)\citenamefont
  {Schimpf}, \citenamefont {Manna}, \citenamefont {da~Silva}, \citenamefont
  {Aigner},\ and\ \citenamefont {Rastelli}}]{10.1117/1.AP.3.6.065001}%
  \BibitemOpen
  \bibfield  {author} {\bibinfo {author} {\bibfnamefont {C.}~\bibnamefont
  {Schimpf}}, \bibinfo {author} {\bibfnamefont {S.}~\bibnamefont {Manna}},
  \bibinfo {author} {\bibfnamefont {S.~F.~C.}\ \bibnamefont {da~Silva}},
  \bibinfo {author} {\bibfnamefont {M.}~\bibnamefont {Aigner}}, \ and\ \bibinfo
  {author} {\bibfnamefont {A.}~\bibnamefont {Rastelli}},\ }\href {\doibase
  10.1117/1.AP.3.6.065001} {\bibfield  {journal} {\bibinfo  {journal} {Advanced
  Photonics}\ }\textbf {\bibinfo {volume} {3}},\ \bibinfo {pages} {1 }
  (\bibinfo {year} {2021})}\BibitemShut {NoStop}%
\bibitem [{Sup()}]{Supplementary}%
  \BibitemOpen
  \href@noop {} {\bibinfo  {journal} {Supplementary materials}\ }\BibitemShut
  {NoStop}%
\bibitem [{\citenamefont {Kuhlmann}\ \emph {et~al.}(2013)\citenamefont
  {Kuhlmann}, \citenamefont {Houel}, \citenamefont {Ludwig}, \citenamefont
  {Greuter}, \citenamefont {Reuter}, \citenamefont {Wieck}, \citenamefont
  {Poggio},\ and\ \citenamefont {Warburton}}]{Kuhlmann2013}%
  \BibitemOpen
\bibfield  {journal} {  }\bibfield  {author} {\bibinfo {author} {\bibfnamefont
  {A.~V.}\ \bibnamefont {Kuhlmann}}, \bibinfo {author} {\bibfnamefont
  {J.}~\bibnamefont {Houel}}, \bibinfo {author} {\bibfnamefont
  {A.}~\bibnamefont {Ludwig}}, \bibinfo {author} {\bibfnamefont
  {L.}~\bibnamefont {Greuter}}, \bibinfo {author} {\bibfnamefont
  {D.}~\bibnamefont {Reuter}}, \bibinfo {author} {\bibfnamefont {A.~D.}\
  \bibnamefont {Wieck}}, \bibinfo {author} {\bibfnamefont {M.}~\bibnamefont
  {Poggio}}, \ and\ \bibinfo {author} {\bibfnamefont {R.~J.}\ \bibnamefont
  {Warburton}},\ }\href {\doibase 10.1038/nphys2688} {\bibfield  {journal}
  {\bibinfo  {journal} {Nature Physics}\ }\textbf {\bibinfo {volume} {9}},\
  \bibinfo {pages} {570} (\bibinfo {year} {2013})}\BibitemShut {NoStop}%
\bibitem [{\citenamefont {Gillard}\ \emph {et~al.}(2021)\citenamefont
  {Gillard}, \citenamefont {Griffiths}, \citenamefont {Ragunathan},
  \citenamefont {Ulhaq}, \citenamefont {McEwan}, \citenamefont {Clarke},\ and\
  \citenamefont {Chekhovich}}]{Gillard2021}%
  \BibitemOpen
  \bibfield  {author} {\bibinfo {author} {\bibfnamefont {G.}~\bibnamefont
  {Gillard}}, \bibinfo {author} {\bibfnamefont {I.~M.}\ \bibnamefont
  {Griffiths}}, \bibinfo {author} {\bibfnamefont {G.}~\bibnamefont
  {Ragunathan}}, \bibinfo {author} {\bibfnamefont {A.}~\bibnamefont {Ulhaq}},
  \bibinfo {author} {\bibfnamefont {C.}~\bibnamefont {McEwan}}, \bibinfo
  {author} {\bibfnamefont {E.}~\bibnamefont {Clarke}}, \ and\ \bibinfo {author}
  {\bibfnamefont {E.~A.}\ \bibnamefont {Chekhovich}},\ }\href {\doibase
  10.1038/s41534-021-00378-2} {\bibfield  {journal} {\bibinfo  {journal} {npj
  Quantum Information}\ }\textbf {\bibinfo {volume} {7}},\ \bibinfo {pages}
  {43} (\bibinfo {year} {2021})}\BibitemShut {NoStop}%
\bibitem [{\citenamefont {Bodey}\ \emph {et~al.}(2019)\citenamefont {Bodey},
  \citenamefont {Stockill}, \citenamefont {Denning}, \citenamefont {Gangloff},
  \citenamefont {{\'{E}}thier-Majcher}, \citenamefont {Jackson}, \citenamefont
  {Clarke}, \citenamefont {Hugues}, \citenamefont {Gall},\ and\ \citenamefont
  {Atat{\"{u}}re}}]{Bodey2019}%
  \BibitemOpen
  \bibfield  {author} {\bibinfo {author} {\bibfnamefont {J.~H.}\ \bibnamefont
  {Bodey}}, \bibinfo {author} {\bibfnamefont {R.}~\bibnamefont {Stockill}},
  \bibinfo {author} {\bibfnamefont {E.~V.}\ \bibnamefont {Denning}}, \bibinfo
  {author} {\bibfnamefont {D.~A.}\ \bibnamefont {Gangloff}}, \bibinfo {author}
  {\bibfnamefont {G.}~\bibnamefont {{\'{E}}thier-Majcher}}, \bibinfo {author}
  {\bibfnamefont {D.~M.}\ \bibnamefont {Jackson}}, \bibinfo {author}
  {\bibfnamefont {E.}~\bibnamefont {Clarke}}, \bibinfo {author} {\bibfnamefont
  {M.}~\bibnamefont {Hugues}}, \bibinfo {author} {\bibfnamefont {C.~L.}\
  \bibnamefont {Gall}}, \ and\ \bibinfo {author} {\bibfnamefont
  {M.}~\bibnamefont {Atat{\"{u}}re}},\ }\href {\doibase
  10.1038/s41534-019-0206-3} {\bibfield  {journal} {\bibinfo  {journal} {npj
  Quantum Inf.}\ }\textbf {\bibinfo {volume} {5}},\ \bibinfo {pages} {95}
  (\bibinfo {year} {2019})}\BibitemShut {NoStop}%
\bibitem [{\citenamefont {Cywi{\'{n}}ski}\ \emph {et~al.}(2009)\citenamefont
  {Cywi{\'{n}}ski}, \citenamefont {Witzel},\ and\ \citenamefont {{Das
  Sarma}}}]{Cywinski2009}%
  \BibitemOpen
  \bibfield  {author} {\bibinfo {author} {\bibfnamefont {{\L}.}~\bibnamefont
  {Cywi{\'{n}}ski}}, \bibinfo {author} {\bibfnamefont {W.~M.}\ \bibnamefont
  {Witzel}}, \ and\ \bibinfo {author} {\bibfnamefont {S.}~\bibnamefont {{Das
  Sarma}}},\ }\href {\doibase 10.1103/PhysRevB.79.245314} {\bibfield  {journal}
  {\bibinfo  {journal} {Phys. Rev. B}\ }\textbf {\bibinfo {volume} {79}},\
  \bibinfo {pages} {245314} (\bibinfo {year} {2009})}\BibitemShut {NoStop}%
\bibitem [{\citenamefont {Botzem}\ \emph {et~al.}(2016)\citenamefont {Botzem},
  \citenamefont {McNeil}, \citenamefont {Mol}, \citenamefont {Schuh},
  \citenamefont {Bougeard},\ and\ \citenamefont {Bluhm}}]{Botzem2015}%
  \BibitemOpen
  \bibfield  {author} {\bibinfo {author} {\bibfnamefont {T.}~\bibnamefont
  {Botzem}}, \bibinfo {author} {\bibfnamefont {R.~P.~G.}\ \bibnamefont
  {McNeil}}, \bibinfo {author} {\bibfnamefont {J.-M.}\ \bibnamefont {Mol}},
  \bibinfo {author} {\bibfnamefont {D.}~\bibnamefont {Schuh}}, \bibinfo
  {author} {\bibfnamefont {D.}~\bibnamefont {Bougeard}}, \ and\ \bibinfo
  {author} {\bibfnamefont {H.}~\bibnamefont {Bluhm}},\ }\href {\doibase
  10.1038/ncomms11170} {\bibfield  {journal} {\bibinfo  {journal} {Nat.
  Commun.}\ }\textbf {\bibinfo {volume} {7}},\ \bibinfo {pages} {11170}
  (\bibinfo {year} {2016})},\ \Eprint {http://arxiv.org/abs/1508.05136}
  {arXiv:1508.05136} \BibitemShut {NoStop}%
\bibitem [{\citenamefont {Bluhm}\ \emph {et~al.}(2011)\citenamefont {Bluhm},
  \citenamefont {Foletti}, \citenamefont {Neder}, \citenamefont {Rudner},
  \citenamefont {Mahalu}, \citenamefont {Umansky},\ and\ \citenamefont
  {Yacoby}}]{Bluhm2010}%
  \BibitemOpen
  \bibfield  {author} {\bibinfo {author} {\bibfnamefont {H.}~\bibnamefont
  {Bluhm}}, \bibinfo {author} {\bibfnamefont {S.}~\bibnamefont {Foletti}},
  \bibinfo {author} {\bibfnamefont {I.}~\bibnamefont {Neder}}, \bibinfo
  {author} {\bibfnamefont {M.}~\bibnamefont {Rudner}}, \bibinfo {author}
  {\bibfnamefont {D.}~\bibnamefont {Mahalu}}, \bibinfo {author} {\bibfnamefont
  {V.}~\bibnamefont {Umansky}}, \ and\ \bibinfo {author} {\bibfnamefont
  {A.}~\bibnamefont {Yacoby}},\ }\href {\doibase 10.1038/nphys1856} {\bibfield
  {journal} {\bibinfo  {journal} {Nat. Phys.}\ }\textbf {\bibinfo {volume}
  {7}},\ \bibinfo {pages} {109} (\bibinfo {year} {2011})},\ \Eprint
  {http://arxiv.org/abs/1005.2995} {arXiv:1005.2995} \BibitemShut {NoStop}%
\bibitem [{\citenamefont {Malinowski}\ \emph {et~al.}(2016)\citenamefont
  {Malinowski}, \citenamefont {Martins}, \citenamefont {Nissen}, \citenamefont
  {Barnes}, \citenamefont {Cywi{\'{n}}ski}, \citenamefont {Rudner},
  \citenamefont {Fallahi}, \citenamefont {Gardner}, \citenamefont {Manfra},
  \citenamefont {Marcus},\ and\ \citenamefont {Kuemmeth}}]{Malinowski2016b}%
  \BibitemOpen
  \bibfield  {author} {\bibinfo {author} {\bibfnamefont {F.~K.}\ \bibnamefont
  {Malinowski}}, \bibinfo {author} {\bibfnamefont {F.}~\bibnamefont {Martins}},
  \bibinfo {author} {\bibfnamefont {P.~D.}\ \bibnamefont {Nissen}}, \bibinfo
  {author} {\bibfnamefont {E.}~\bibnamefont {Barnes}}, \bibinfo {author}
  {\bibfnamefont {{\L}.}~\bibnamefont {Cywi{\'{n}}ski}}, \bibinfo {author}
  {\bibfnamefont {M.~S.}\ \bibnamefont {Rudner}}, \bibinfo {author}
  {\bibfnamefont {S.}~\bibnamefont {Fallahi}}, \bibinfo {author} {\bibfnamefont
  {G.~C.}\ \bibnamefont {Gardner}}, \bibinfo {author} {\bibfnamefont {M.~J.}\
  \bibnamefont {Manfra}}, \bibinfo {author} {\bibfnamefont {C.~M.}\
  \bibnamefont {Marcus}}, \ and\ \bibinfo {author} {\bibfnamefont
  {F.}~\bibnamefont {Kuemmeth}},\ }\href {\doibase 10.1038/nnano.2016.170}
  {\bibfield  {journal} {\bibinfo  {journal} {Nat. Nanotechnol. 2016 121}\
  }\textbf {\bibinfo {volume} {12}},\ \bibinfo {pages} {16} (\bibinfo {year}
  {2016})}\BibitemShut {NoStop}%
\bibitem [{\citenamefont {Malinowski}\ \emph {et~al.}(2017)\citenamefont
  {Malinowski}, \citenamefont {Martins}, \citenamefont {Cywi{\'{n}}ski},
  \citenamefont {Rudner}, \citenamefont {Nissen}, \citenamefont {Fallahi},
  \citenamefont {Gardner}, \citenamefont {Manfra}, \citenamefont {Marcus},\
  and\ \citenamefont {Kuemmeth}}]{Malinowski2017}%
  \BibitemOpen
  \bibfield  {author} {\bibinfo {author} {\bibfnamefont {F.~K.}\ \bibnamefont
  {Malinowski}}, \bibinfo {author} {\bibfnamefont {F.}~\bibnamefont {Martins}},
  \bibinfo {author} {\bibfnamefont {{\L}.}~\bibnamefont {Cywi{\'{n}}ski}},
  \bibinfo {author} {\bibfnamefont {M.~S.}\ \bibnamefont {Rudner}}, \bibinfo
  {author} {\bibfnamefont {P.~D.}\ \bibnamefont {Nissen}}, \bibinfo {author}
  {\bibfnamefont {S.}~\bibnamefont {Fallahi}}, \bibinfo {author} {\bibfnamefont
  {G.~C.}\ \bibnamefont {Gardner}}, \bibinfo {author} {\bibfnamefont {M.~J.}\
  \bibnamefont {Manfra}}, \bibinfo {author} {\bibfnamefont {C.~M.}\
  \bibnamefont {Marcus}}, \ and\ \bibinfo {author} {\bibfnamefont
  {F.}~\bibnamefont {Kuemmeth}},\ }\href {\doibase
  10.1103/PhysRevLett.118.177702} {\bibfield  {journal} {\bibinfo  {journal}
  {Phys. Rev. Lett.}\ }\textbf {\bibinfo {volume} {118}},\ \bibinfo {pages}
  {177702} (\bibinfo {year} {2017})}\BibitemShut {NoStop}%
\bibitem [{\citenamefont {de~Lange}\ \emph {et~al.}(2010)\citenamefont
  {de~Lange}, \citenamefont {Wang}, \citenamefont {Ristè}, \citenamefont
  {Dobrovitski},\ and\ \citenamefont {Hanson}}]{DeLange2010}%
  \BibitemOpen
  \bibfield  {author} {\bibinfo {author} {\bibfnamefont {G.}~\bibnamefont
  {de~Lange}}, \bibinfo {author} {\bibfnamefont {Z.~H.}\ \bibnamefont {Wang}},
  \bibinfo {author} {\bibfnamefont {D.}~\bibnamefont {Ristè}}, \bibinfo
  {author} {\bibfnamefont {V.~V.}\ \bibnamefont {Dobrovitski}}, \ and\ \bibinfo
  {author} {\bibfnamefont {R.}~\bibnamefont {Hanson}},\ }\href {\doibase
  10.1126/science.1192739} {\bibfield  {journal} {\bibinfo  {journal}
  {Science}\ }\textbf {\bibinfo {volume} {330}},\ \bibinfo {pages} {60}
  (\bibinfo {year} {2010})},\ \Eprint
  {http://arxiv.org/abs/https://www.science.org/doi/pdf/10.1126/science.1192739}
  {https://www.science.org/doi/pdf/10.1126/science.1192739} \BibitemShut
  {NoStop}%
\bibitem [{\citenamefont {Huthmacher}\ \emph {et~al.}(2018)\citenamefont
  {Huthmacher}, \citenamefont {Stockill}, \citenamefont {Clarke}, \citenamefont
  {Hugues}, \citenamefont {{Le Gall}},\ and\ \citenamefont
  {Atat{\"{u}}re}}]{Huthmacher2018}%
  \BibitemOpen
  \bibfield  {author} {\bibinfo {author} {\bibfnamefont {L.}~\bibnamefont
  {Huthmacher}}, \bibinfo {author} {\bibfnamefont {R.}~\bibnamefont
  {Stockill}}, \bibinfo {author} {\bibfnamefont {E.}~\bibnamefont {Clarke}},
  \bibinfo {author} {\bibfnamefont {M.}~\bibnamefont {Hugues}}, \bibinfo
  {author} {\bibfnamefont {C.}~\bibnamefont {{Le Gall}}}, \ and\ \bibinfo
  {author} {\bibfnamefont {M.}~\bibnamefont {Atat{\"{u}}re}},\ }\href {\doibase
  10.1103/PhysRevB.97.241413} {\bibfield  {journal} {\bibinfo  {journal} {Phys.
  Rev. B}\ }\textbf {\bibinfo {volume} {97}},\ \bibinfo {pages} {241413}
  (\bibinfo {year} {2018})},\ \Eprint {http://arxiv.org/abs/1711.09169}
  {arXiv:1711.09169} \BibitemShut {NoStop}%
\bibitem [{\citenamefont {Ragunathan}(2019)}]{ragunathan_2019}%
  \BibitemOpen
  \bibfield  {author} {\bibinfo {author} {\bibfnamefont {G.}~\bibnamefont
  {Ragunathan}},\ }\href {https://etheses.whiterose.ac.uk/25466/} {Ph.D.
  thesis} (\bibinfo {year} {2019})\BibitemShut {NoStop}%
\bibitem [{\citenamefont {Ulhaq}\ \emph {et~al.}(2016)\citenamefont {Ulhaq},
  \citenamefont {Duan}, \citenamefont {Zallo}, \citenamefont {Ding},
  \citenamefont {Schmidt}, \citenamefont {Tartakovskii}, \citenamefont
  {Skolnick},\ and\ \citenamefont {Chekhovich}}]{Ulhaq2016}%
  \BibitemOpen
  \bibfield  {author} {\bibinfo {author} {\bibfnamefont {A.}~\bibnamefont
  {Ulhaq}}, \bibinfo {author} {\bibfnamefont {Q.}~\bibnamefont {Duan}},
  \bibinfo {author} {\bibfnamefont {E.}~\bibnamefont {Zallo}}, \bibinfo
  {author} {\bibfnamefont {F.}~\bibnamefont {Ding}}, \bibinfo {author}
  {\bibfnamefont {O.~G.}\ \bibnamefont {Schmidt}}, \bibinfo {author}
  {\bibfnamefont {A.~I.}\ \bibnamefont {Tartakovskii}}, \bibinfo {author}
  {\bibfnamefont {M.~S.}\ \bibnamefont {Skolnick}}, \ and\ \bibinfo {author}
  {\bibfnamefont {E.~A.}\ \bibnamefont {Chekhovich}},\ }\href {\doibase
  10.1103/PhysRevB.93.165306} {\bibfield  {journal} {\bibinfo  {journal} {Phys.
  Rev. B}\ }\textbf {\bibinfo {volume} {93}},\ \bibinfo {pages} {165306}
  (\bibinfo {year} {2016})},\ \Eprint {http://arxiv.org/abs/1507.06553}
  {arXiv:1507.06553} \BibitemShut {NoStop}%
\bibitem [{\citenamefont {Chekhovich}\ \emph {et~al.}(2015)\citenamefont
  {Chekhovich}, \citenamefont {Hopkinson}, \citenamefont {Skolnick},\ and\
  \citenamefont {Tartakovskii}}]{Chekhovich2015}%
  \BibitemOpen
  \bibfield  {author} {\bibinfo {author} {\bibfnamefont {E.~A.}\ \bibnamefont
  {Chekhovich}}, \bibinfo {author} {\bibfnamefont {M.}~\bibnamefont
  {Hopkinson}}, \bibinfo {author} {\bibfnamefont {M.~S.}\ \bibnamefont
  {Skolnick}}, \ and\ \bibinfo {author} {\bibfnamefont {A.~I.}\ \bibnamefont
  {Tartakovskii}},\ }\href {\doibase 10.1038/ncomms7348} {\bibfield  {journal}
  {\bibinfo  {journal} {Nature Communications}\ }\textbf {\bibinfo {volume}
  {6}},\ \bibinfo {pages} {6348} (\bibinfo {year} {2015})}\BibitemShut
  {NoStop}%
\bibitem [{\citenamefont {Knijn}\ \emph {et~al.}(2010)\citenamefont {Knijn},
  \citenamefont {van Bentum}, \citenamefont {van Eck}, \citenamefont {Fang},
  \citenamefont {Grimminck}, \citenamefont {de~Groot}, \citenamefont
  {Havenith}, \citenamefont {Marsman}, \citenamefont {Meerts}, \citenamefont
  {de~Wijs},\ and\ \citenamefont {Kentgens}}]{Knijn2010}%
  \BibitemOpen
  \bibfield  {author} {\bibinfo {author} {\bibfnamefont {P.~J.}\ \bibnamefont
  {Knijn}}, \bibinfo {author} {\bibfnamefont {P.~J.~M.}\ \bibnamefont {van
  Bentum}}, \bibinfo {author} {\bibfnamefont {E.~R.~H.}\ \bibnamefont {van
  Eck}}, \bibinfo {author} {\bibfnamefont {C.}~\bibnamefont {Fang}}, \bibinfo
  {author} {\bibfnamefont {D.~L. A.~G.}\ \bibnamefont {Grimminck}}, \bibinfo
  {author} {\bibfnamefont {R.~A.}\ \bibnamefont {de~Groot}}, \bibinfo {author}
  {\bibfnamefont {R.~W.~A.}\ \bibnamefont {Havenith}}, \bibinfo {author}
  {\bibfnamefont {M.}~\bibnamefont {Marsman}}, \bibinfo {author} {\bibfnamefont
  {W.~L.}\ \bibnamefont {Meerts}}, \bibinfo {author} {\bibfnamefont {G.~A.}\
  \bibnamefont {de~Wijs}}, \ and\ \bibinfo {author} {\bibfnamefont {A.~P.~M.}\
  \bibnamefont {Kentgens}},\ }\href {\doibase 10.1039/c003624b} {\bibfield
  {journal} {\bibinfo  {journal} {Phys. Chem. Chem. Phys.}\ }\textbf {\bibinfo
  {volume} {12}},\ \bibinfo {pages} {11517} (\bibinfo {year}
  {2010})}\BibitemShut {NoStop}%
\bibitem [{\citenamefont {Chekhovich}\ \emph {et~al.}(2017)\citenamefont
  {Chekhovich}, \citenamefont {Ulhaq}, \citenamefont {Zallo}, \citenamefont
  {Ding}, \citenamefont {Schmidt},\ and\ \citenamefont
  {Skolnick}}]{Chekhovich2017}%
  \BibitemOpen
  \bibfield  {author} {\bibinfo {author} {\bibfnamefont {E.~A.}\ \bibnamefont
  {Chekhovich}}, \bibinfo {author} {\bibfnamefont {A.}~\bibnamefont {Ulhaq}},
  \bibinfo {author} {\bibfnamefont {E.}~\bibnamefont {Zallo}}, \bibinfo
  {author} {\bibfnamefont {F.}~\bibnamefont {Ding}}, \bibinfo {author}
  {\bibfnamefont {O.~G.}\ \bibnamefont {Schmidt}}, \ and\ \bibinfo {author}
  {\bibfnamefont {M.~S.}\ \bibnamefont {Skolnick}},\ }\href {\doibase
  10.1038/nmat4959} {\bibfield  {journal} {\bibinfo  {journal} {Nat. Mater.}\
  }\textbf {\bibinfo {volume} {16}},\ \bibinfo {pages} {982} (\bibinfo {year}
  {2017})},\ \Eprint {http://arxiv.org/abs/1701.02759} {arXiv:1701.02759}
  \BibitemShut {NoStop}%
\bibitem [{\citenamefont {van Bree}\ \emph {et~al.}(2016)\citenamefont {van
  Bree}, \citenamefont {Silov}, \citenamefont {van Maasakkers}, \citenamefont
  {Pryor}, \citenamefont {Flatt\'e},\ and\ \citenamefont
  {Koenraad}}]{PhysRevB.93.035311}%
  \BibitemOpen
  \bibfield  {author} {\bibinfo {author} {\bibfnamefont {J.}~\bibnamefont {van
  Bree}}, \bibinfo {author} {\bibfnamefont {A.~Y.}\ \bibnamefont {Silov}},
  \bibinfo {author} {\bibfnamefont {M.~L.}\ \bibnamefont {van Maasakkers}},
  \bibinfo {author} {\bibfnamefont {C.~E.}\ \bibnamefont {Pryor}}, \bibinfo
  {author} {\bibfnamefont {M.~E.}\ \bibnamefont {Flatt\'e}}, \ and\ \bibinfo
  {author} {\bibfnamefont {P.~M.}\ \bibnamefont {Koenraad}},\ }\href {\doibase
  10.1103/PhysRevB.93.035311} {\bibfield  {journal} {\bibinfo  {journal} {Phys.
  Rev. B}\ }\textbf {\bibinfo {volume} {93}},\ \bibinfo {pages} {035311}
  (\bibinfo {year} {2016})}\BibitemShut {NoStop}%
\bibitem [{\citenamefont {Jackson}\ \emph {et~al.}(2021)\citenamefont
  {Jackson}, \citenamefont {Gangloff}, \citenamefont {Bodey}, \citenamefont
  {Zaporski}, \citenamefont {Bachorz}, \citenamefont {Clarke}, \citenamefont
  {Hugues}, \citenamefont {{Le Gall}},\ and\ \citenamefont
  {Atat{\"{u}}re}}]{Jackson2021}%
  \BibitemOpen
  \bibfield  {author} {\bibinfo {author} {\bibfnamefont {D.~M.}\ \bibnamefont
  {Jackson}}, \bibinfo {author} {\bibfnamefont {D.~A.}\ \bibnamefont
  {Gangloff}}, \bibinfo {author} {\bibfnamefont {J.~H.}\ \bibnamefont {Bodey}},
  \bibinfo {author} {\bibfnamefont {L.}~\bibnamefont {Zaporski}}, \bibinfo
  {author} {\bibfnamefont {C.}~\bibnamefont {Bachorz}}, \bibinfo {author}
  {\bibfnamefont {E.}~\bibnamefont {Clarke}}, \bibinfo {author} {\bibfnamefont
  {M.}~\bibnamefont {Hugues}}, \bibinfo {author} {\bibfnamefont
  {C.}~\bibnamefont {{Le Gall}}}, \ and\ \bibinfo {author} {\bibfnamefont
  {M.}~\bibnamefont {Atat{\"{u}}re}},\ }\href {\doibase
  10.1038/s41567-020-01161-4} {\bibfield  {journal} {\bibinfo  {journal} {Nat.
  Phys.}\ }\textbf {\bibinfo {volume} {17}},\ \bibinfo {pages} {585} (\bibinfo
  {year} {2021})},\ \Eprint {http://arxiv.org/abs/2008.09541}
  {arXiv:2008.09541} \BibitemShut {NoStop}%
\bibitem [{\citenamefont {Gangloff}\ \emph {et~al.}(2021)\citenamefont
  {Gangloff}, \citenamefont {Zaporski}, \citenamefont {Bodey}, \citenamefont
  {Bachorz}, \citenamefont {Jackson}, \citenamefont {{\'E}thier-Majcher},
  \citenamefont {Lang}, \citenamefont {Clarke}, \citenamefont {Hugues},
  \citenamefont {Le~Gall},\ and\ \citenamefont {Atat{\"u}re}}]{Gangloff2021}%
  \BibitemOpen
  \bibfield  {author} {\bibinfo {author} {\bibfnamefont {D.~A.}\ \bibnamefont
  {Gangloff}}, \bibinfo {author} {\bibfnamefont {L.}~\bibnamefont {Zaporski}},
  \bibinfo {author} {\bibfnamefont {J.~H.}\ \bibnamefont {Bodey}}, \bibinfo
  {author} {\bibfnamefont {C.}~\bibnamefont {Bachorz}}, \bibinfo {author}
  {\bibfnamefont {D.~M.}\ \bibnamefont {Jackson}}, \bibinfo {author}
  {\bibfnamefont {G.}~\bibnamefont {{\'E}thier-Majcher}}, \bibinfo {author}
  {\bibfnamefont {C.}~\bibnamefont {Lang}}, \bibinfo {author} {\bibfnamefont
  {E.}~\bibnamefont {Clarke}}, \bibinfo {author} {\bibfnamefont
  {M.}~\bibnamefont {Hugues}}, \bibinfo {author} {\bibfnamefont
  {C.}~\bibnamefont {Le~Gall}}, \ and\ \bibinfo {author} {\bibfnamefont
  {M.}~\bibnamefont {Atat{\"u}re}},\ }\href {\doibase
  10.1038/s41567-021-01344-7} {\bibfield  {journal} {\bibinfo  {journal}
  {Nature Physics}\ }\textbf {\bibinfo {volume} {17}},\ \bibinfo {pages} {1247}
  (\bibinfo {year} {2021})}\BibitemShut {NoStop}%
\bibitem [{\citenamefont {Wolters}\ \emph {et~al.}(2017)\citenamefont
  {Wolters}, \citenamefont {Buser}, \citenamefont {Horsley}, \citenamefont
  {B\'eguin}, \citenamefont {J\"ockel}, \citenamefont {Jahn}, \citenamefont
  {Warburton},\ and\ \citenamefont {Treutlein}}]{PhysRevLett.119.060502}%
  \BibitemOpen
  \bibfield  {author} {\bibinfo {author} {\bibfnamefont {J.}~\bibnamefont
  {Wolters}}, \bibinfo {author} {\bibfnamefont {G.}~\bibnamefont {Buser}},
  \bibinfo {author} {\bibfnamefont {A.}~\bibnamefont {Horsley}}, \bibinfo
  {author} {\bibfnamefont {L.}~\bibnamefont {B\'eguin}}, \bibinfo {author}
  {\bibfnamefont {A.}~\bibnamefont {J\"ockel}}, \bibinfo {author}
  {\bibfnamefont {J.-P.}\ \bibnamefont {Jahn}}, \bibinfo {author}
  {\bibfnamefont {R.~J.}\ \bibnamefont {Warburton}}, \ and\ \bibinfo {author}
  {\bibfnamefont {P.}~\bibnamefont {Treutlein}},\ }\href {\doibase
  10.1103/PhysRevLett.119.060502} {\bibfield  {journal} {\bibinfo  {journal}
  {Phys. Rev. Lett.}\ }\textbf {\bibinfo {volume} {119}},\ \bibinfo {pages}
  {060502} (\bibinfo {year} {2017})}\BibitemShut {NoStop}%
\bibitem [{\citenamefont {Schwartz}\ \emph
  {et~al.}(2016{\natexlab{b}})\citenamefont {Schwartz}, \citenamefont {Cogan},
  \citenamefont {Schmidgall}, \citenamefont {Don}, \citenamefont {Gantz},
  \citenamefont {Kenneth}, \citenamefont {Lindner},\ and\ \citenamefont
  {Gershoni}}]{doi:10.1126/science.aah4758}%
  \BibitemOpen
  \bibfield  {author} {\bibinfo {author} {\bibfnamefont {I.}~\bibnamefont
  {Schwartz}}, \bibinfo {author} {\bibfnamefont {D.}~\bibnamefont {Cogan}},
  \bibinfo {author} {\bibfnamefont {E.~R.}\ \bibnamefont {Schmidgall}},
  \bibinfo {author} {\bibfnamefont {Y.}~\bibnamefont {Don}}, \bibinfo {author}
  {\bibfnamefont {L.}~\bibnamefont {Gantz}}, \bibinfo {author} {\bibfnamefont
  {O.}~\bibnamefont {Kenneth}}, \bibinfo {author} {\bibfnamefont {N.~H.}\
  \bibnamefont {Lindner}}, \ and\ \bibinfo {author} {\bibfnamefont
  {D.}~\bibnamefont {Gershoni}},\ }\href {\doibase 10.1126/science.aah4758}
  {\bibfield  {journal} {\bibinfo  {journal} {Science}\ }\textbf {\bibinfo
  {volume} {354}},\ \bibinfo {pages} {434} (\bibinfo {year}
  {2016}{\natexlab{b}})},\ \Eprint
  {http://arxiv.org/abs/https://www.science.org/doi/pdf/10.1126/science.aah4758}
  {https://www.science.org/doi/pdf/10.1126/science.aah4758} \BibitemShut
  {NoStop}%
\bibitem [{\citenamefont {Economou}\ \emph {et~al.}(2010)\citenamefont
  {Economou}, \citenamefont {Lindner},\ and\ \citenamefont
  {Rudolph}}]{PhysRevLett.105.093601}%
  \BibitemOpen
  \bibfield  {author} {\bibinfo {author} {\bibfnamefont {S.~E.}\ \bibnamefont
  {Economou}}, \bibinfo {author} {\bibfnamefont {N.}~\bibnamefont {Lindner}}, \
  and\ \bibinfo {author} {\bibfnamefont {T.}~\bibnamefont {Rudolph}},\ }\href
  {\doibase 10.1103/PhysRevLett.105.093601} {\bibfield  {journal} {\bibinfo
  {journal} {Phys. Rev. Lett.}\ }\textbf {\bibinfo {volume} {105}},\ \bibinfo
  {pages} {093601} (\bibinfo {year} {2010})}\BibitemShut {NoStop}%
\bibitem [{\citenamefont {Heyn}\ \emph {et~al.}(2009)\citenamefont {Heyn},
  \citenamefont {Stemmann}, \citenamefont {K\"oppen}, \citenamefont {Strelow},
  \citenamefont {Kipp}, \citenamefont {Grave}, \citenamefont {Mendach},\ and\
  \citenamefont {Hansen}}]{Heyn2009}%
  \BibitemOpen
  \bibfield  {author} {\bibinfo {author} {\bibfnamefont {C.}~\bibnamefont
  {Heyn}}, \bibinfo {author} {\bibfnamefont {A.}~\bibnamefont {Stemmann}},
  \bibinfo {author} {\bibfnamefont {T.}~\bibnamefont {K\"oppen}}, \bibinfo
  {author} {\bibfnamefont {C.}~\bibnamefont {Strelow}}, \bibinfo {author}
  {\bibfnamefont {T.}~\bibnamefont {Kipp}}, \bibinfo {author} {\bibfnamefont
  {M.}~\bibnamefont {Grave}}, \bibinfo {author} {\bibfnamefont
  {S.}~\bibnamefont {Mendach}}, \ and\ \bibinfo {author} {\bibfnamefont
  {W.}~\bibnamefont {Hansen}},\ }\href {\doibase 10.1063/1.3133338} {\bibfield
  {journal} {\bibinfo  {journal} {Applied Physics Letters}\ }\textbf {\bibinfo
  {volume} {94}},\ \bibinfo {pages} {183113} (\bibinfo {year}
  {2009})}\BibitemShut {NoStop}%
\bibitem [{\citenamefont {Atkinson}\ \emph {et~al.}(2012)\citenamefont
  {Atkinson}, \citenamefont {Zallo},\ and\ \citenamefont
  {Schmidt}}]{Atkinson2012}%
  \BibitemOpen
  \bibfield  {author} {\bibinfo {author} {\bibfnamefont {P.}~\bibnamefont
  {Atkinson}}, \bibinfo {author} {\bibfnamefont {E.}~\bibnamefont {Zallo}}, \
  and\ \bibinfo {author} {\bibfnamefont {O.~G.}\ \bibnamefont {Schmidt}},\
  }\href {\doibase 10.1063/1.4748183} {\bibfield  {journal} {\bibinfo
  {journal} {Journal of Applied Physics}\ }\textbf {\bibinfo {volume} {112}},\
  \bibinfo {pages} {054303} (\bibinfo {year} {2012})}\BibitemShut {NoStop}%
\bibitem [{\citenamefont {Neder}\ \emph {et~al.}(2011)\citenamefont {Neder},
  \citenamefont {Rudner}, \citenamefont {Bluhm}, \citenamefont {Foletti},
  \citenamefont {Halperin},\ and\ \citenamefont {Yacoby}}]{Neder2011a}%
  \BibitemOpen
  \bibfield  {author} {\bibinfo {author} {\bibfnamefont {I.}~\bibnamefont
  {Neder}}, \bibinfo {author} {\bibfnamefont {M.~S.}\ \bibnamefont {Rudner}},
  \bibinfo {author} {\bibfnamefont {H.}~\bibnamefont {Bluhm}}, \bibinfo
  {author} {\bibfnamefont {S.}~\bibnamefont {Foletti}}, \bibinfo {author}
  {\bibfnamefont {B.~I.}\ \bibnamefont {Halperin}}, \ and\ \bibinfo {author}
  {\bibfnamefont {A.}~\bibnamefont {Yacoby}},\ }\href {\doibase
  10.1103/PhysRevB.84.035441} {\bibfield  {journal} {\bibinfo  {journal} {Phys.
  Rev. B}\ }\textbf {\bibinfo {volume} {84}},\ \bibinfo {pages} {035441}
  (\bibinfo {year} {2011})}\BibitemShut {NoStop}%
\end{thebibliography}%


\begin{thebibliography}{22}%
\makeatletter
\providecommand \@ifxundefined [1]{%
 \@ifx{#1\undefined}
}%
\providecommand \@ifnum [1]{%
 \ifnum #1\expandafter \@firstoftwo
 \else \expandafter \@secondoftwo
 \fi
}%
\providecommand \@ifx [1]{%
 \ifx #1\expandafter \@firstoftwo
 \else \expandafter \@secondoftwo
 \fi
}%
\providecommand \natexlab [1]{#1}%
\providecommand \enquote  [1]{``#1''}%
\providecommand \bibnamefont  [1]{#1}%
\providecommand \bibfnamefont [1]{#1}%
\providecommand \citenamefont [1]{#1}%
\providecommand \href@noop [0]{\@secondoftwo}%
\providecommand \href [0]{\begingroup \@sanitize@url \@href}%
\providecommand \@href[1]{\@@startlink{#1}\@@href}%
\providecommand \@@href[1]{\endgroup#1\@@endlink}%
\providecommand \@sanitize@url [0]{\catcode `\\12\catcode `\$12\catcode
  `\&12\catcode `\#12\catcode `\^12\catcode `\_12\catcode `\%12\relax}%
\providecommand \@@startlink[1]{}%
\providecommand \@@endlink[0]{}%
\providecommand \url  [0]{\begingroup\@sanitize@url \@url }%
\providecommand \@url [1]{\endgroup\@href {#1}{\urlprefix }}%
\providecommand \urlprefix  [0]{URL }%
\providecommand \Eprint [0]{\href }%
\providecommand \doibase [0]{http://dx.doi.org/}%
\providecommand \selectlanguage [0]{\@gobble}%
\providecommand \bibinfo  [0]{\@secondoftwo}%
\providecommand \bibfield  [0]{\@secondoftwo}%
\providecommand \translation [1]{[#1]}%
\providecommand \BibitemOpen [0]{}%
\providecommand \bibitemStop [0]{}%
\providecommand \bibitemNoStop [0]{.\EOS\space}%
\providecommand \EOS [0]{\spacefactor3000\relax}%
\providecommand \BibitemShut  [1]{\csname bibitem#1\endcsname}%
\let\auto@bib@innerbib\@empty
\bibitem [{\citenamefont {da~Silva}\ \emph {et~al.}(2021)\citenamefont
  {da~Silva}, \citenamefont {Undeutsch}, \citenamefont {Lehner}, \citenamefont
  {Manna}, \citenamefont {Krieger}, \citenamefont {Reindl}, \citenamefont
  {Schimpf}, \citenamefont {Trotta},\ and\ \citenamefont
  {Rastelli}}]{doi:10.1063/5.0057070}%
  \BibitemOpen
  \bibfield  {author} {\bibinfo {author} {\bibfnamefont {S.~F.~C.}\
  \bibnamefont {da~Silva}}, \bibinfo {author} {\bibfnamefont {G.}~\bibnamefont
  {Undeutsch}}, \bibinfo {author} {\bibfnamefont {B.}~\bibnamefont {Lehner}},
  \bibinfo {author} {\bibfnamefont {S.}~\bibnamefont {Manna}}, \bibinfo
  {author} {\bibfnamefont {T.~M.}\ \bibnamefont {Krieger}}, \bibinfo {author}
  {\bibfnamefont {M.}~\bibnamefont {Reindl}}, \bibinfo {author} {\bibfnamefont
  {C.}~\bibnamefont {Schimpf}}, \bibinfo {author} {\bibfnamefont
  {R.}~\bibnamefont {Trotta}}, \ and\ \bibinfo {author} {\bibfnamefont
  {A.}~\bibnamefont {Rastelli}},\ }\href {\doibase 10.1063/5.0057070}
  {\bibfield  {journal} {\bibinfo  {journal} {Applied Physics Letters}\
  }\textbf {\bibinfo {volume} {119}},\ \bibinfo {pages} {120502} (\bibinfo
  {year} {2021})},\ \Eprint
  {http://arxiv.org/abs/https://doi.org/10.1063/5.0057070}
  {https://doi.org/10.1063/5.0057070} \BibitemShut {NoStop}%
\bibitem [{\citenamefont {Heyn}\ \emph {et~al.}(2009)\citenamefont {Heyn},
  \citenamefont {Stemmann}, \citenamefont {K\"oppen}, \citenamefont {Strelow},
  \citenamefont {Kipp}, \citenamefont {Grave}, \citenamefont {Mendach},\ and\
  \citenamefont {Hansen}}]{Heyn2009}%
  \BibitemOpen
  \bibfield  {author} {\bibinfo {author} {\bibfnamefont {C.}~\bibnamefont
  {Heyn}}, \bibinfo {author} {\bibfnamefont {A.}~\bibnamefont {Stemmann}},
  \bibinfo {author} {\bibfnamefont {T.}~\bibnamefont {K\"oppen}}, \bibinfo
  {author} {\bibfnamefont {C.}~\bibnamefont {Strelow}}, \bibinfo {author}
  {\bibfnamefont {T.}~\bibnamefont {Kipp}}, \bibinfo {author} {\bibfnamefont
  {M.}~\bibnamefont {Grave}}, \bibinfo {author} {\bibfnamefont
  {S.}~\bibnamefont {Mendach}}, \ and\ \bibinfo {author} {\bibfnamefont
  {W.}~\bibnamefont {Hansen}},\ }\href {\doibase 10.1063/1.3133338} {\bibfield
  {journal} {\bibinfo  {journal} {Applied Physics Letters}\ }\textbf {\bibinfo
  {volume} {94}},\ \bibinfo {pages} {183113} (\bibinfo {year}
  {2009})}\BibitemShut {NoStop}%
\bibitem [{\citenamefont {Atkinson}\ \emph {et~al.}(2012)\citenamefont
  {Atkinson}, \citenamefont {Zallo},\ and\ \citenamefont
  {Schmidt}}]{Atkinson2012}%
  \BibitemOpen
  \bibfield  {author} {\bibinfo {author} {\bibfnamefont {P.}~\bibnamefont
  {Atkinson}}, \bibinfo {author} {\bibfnamefont {E.}~\bibnamefont {Zallo}}, \
  and\ \bibinfo {author} {\bibfnamefont {O.~G.}\ \bibnamefont {Schmidt}},\
  }\href {\doibase 10.1063/1.4748183} {\bibfield  {journal} {\bibinfo
  {journal} {Journal of Applied Physics}\ }\textbf {\bibinfo {volume} {112}},\
  \bibinfo {pages} {054303} (\bibinfo {year} {2012})}\BibitemShut {NoStop}%
\bibitem [{\citenamefont {Berezovsky}\ \emph {et~al.}(2008)\citenamefont
  {Berezovsky}, \citenamefont {Mikkelsen}, \citenamefont {Stoltz},
  \citenamefont {Coldren},\ and\ \citenamefont {Awschalom}}]{Berezovsky2008}%
  \BibitemOpen
  \bibfield  {author} {\bibinfo {author} {\bibfnamefont {J.}~\bibnamefont
  {Berezovsky}}, \bibinfo {author} {\bibfnamefont {M.~H.}\ \bibnamefont
  {Mikkelsen}}, \bibinfo {author} {\bibfnamefont {N.~G.}\ \bibnamefont
  {Stoltz}}, \bibinfo {author} {\bibfnamefont {L.~A.}\ \bibnamefont {Coldren}},
  \ and\ \bibinfo {author} {\bibfnamefont {D.~D.}\ \bibnamefont {Awschalom}},\
  }\href {\doibase 10.1126/science.1154798} {\bibfield  {journal} {\bibinfo
  {journal} {Science}\ }\textbf {\bibinfo {volume} {320}},\ \bibinfo {pages}
  {349} (\bibinfo {year} {2008})},\ \Eprint
  {http://arxiv.org/abs/https://www.science.org/doi/pdf/10.1126/science.1154798}
  {https://www.science.org/doi/pdf/10.1126/science.1154798} \BibitemShut
  {NoStop}%
\bibitem [{\citenamefont {Press}\ \emph {et~al.}(2008)\citenamefont {Press},
  \citenamefont {Ladd}, \citenamefont {Zhang},\ and\ \citenamefont
  {Yamamoto}}]{Press2008}%
  \BibitemOpen
  \bibfield  {author} {\bibinfo {author} {\bibfnamefont {D.}~\bibnamefont
  {Press}}, \bibinfo {author} {\bibfnamefont {T.~D.}\ \bibnamefont {Ladd}},
  \bibinfo {author} {\bibfnamefont {B.}~\bibnamefont {Zhang}}, \ and\ \bibinfo
  {author} {\bibfnamefont {Y.}~\bibnamefont {Yamamoto}},\ }\href {\doibase
  10.1038/nature07530} {\bibfield  {journal} {\bibinfo  {journal} {Nature}\
  }\textbf {\bibinfo {volume} {456}},\ \bibinfo {pages} {218} (\bibinfo {year}
  {2008})}\BibitemShut {NoStop}%
\bibitem [{\citenamefont {Bodey}\ \emph {et~al.}(2019)\citenamefont {Bodey},
  \citenamefont {Stockill}, \citenamefont {Denning}, \citenamefont {Gangloff},
  \citenamefont {{\'E}thier-Majcher}, \citenamefont {Jackson}, \citenamefont
  {Clarke}, \citenamefont {Hugues}, \citenamefont {Gall},\ and\ \citenamefont
  {Atat{\"u}re}}]{Bodey2019}%
  \BibitemOpen
  \bibfield  {author} {\bibinfo {author} {\bibfnamefont {J.~H.}\ \bibnamefont
  {Bodey}}, \bibinfo {author} {\bibfnamefont {R.}~\bibnamefont {Stockill}},
  \bibinfo {author} {\bibfnamefont {E.~V.}\ \bibnamefont {Denning}}, \bibinfo
  {author} {\bibfnamefont {D.~A.}\ \bibnamefont {Gangloff}}, \bibinfo {author}
  {\bibfnamefont {G.}~\bibnamefont {{\'E}thier-Majcher}}, \bibinfo {author}
  {\bibfnamefont {D.~M.}\ \bibnamefont {Jackson}}, \bibinfo {author}
  {\bibfnamefont {E.}~\bibnamefont {Clarke}}, \bibinfo {author} {\bibfnamefont
  {M.}~\bibnamefont {Hugues}}, \bibinfo {author} {\bibfnamefont {C.~L.}\
  \bibnamefont {Gall}}, \ and\ \bibinfo {author} {\bibfnamefont
  {M.}~\bibnamefont {Atat{\"u}re}},\ }\href {\doibase
  10.1038/s41534-019-0206-3} {\bibfield  {journal} {\bibinfo  {journal} {npj
  Quantum Information}\ }\textbf {\bibinfo {volume} {5}},\ \bibinfo {pages}
  {95} (\bibinfo {year} {2019})}\BibitemShut {NoStop}%
\bibitem [{\citenamefont {Houel}\ \emph {et~al.}(2012)\citenamefont {Houel},
  \citenamefont {Kuhlmann}, \citenamefont {Greuter}, \citenamefont {Xue},
  \citenamefont {Poggio}, \citenamefont {Gerardot}, \citenamefont {Dalgarno},
  \citenamefont {Badolato}, \citenamefont {Petroff}, \citenamefont {Ludwig},
  \citenamefont {Reuter}, \citenamefont {Wieck},\ and\ \citenamefont
  {Warburton}}]{Houel2012}%
  \BibitemOpen
  \bibfield  {author} {\bibinfo {author} {\bibfnamefont {J.}~\bibnamefont
  {Houel}}, \bibinfo {author} {\bibfnamefont {A.~V.}\ \bibnamefont {Kuhlmann}},
  \bibinfo {author} {\bibfnamefont {L.}~\bibnamefont {Greuter}}, \bibinfo
  {author} {\bibfnamefont {F.}~\bibnamefont {Xue}}, \bibinfo {author}
  {\bibfnamefont {M.}~\bibnamefont {Poggio}}, \bibinfo {author} {\bibfnamefont
  {B.~D.}\ \bibnamefont {Gerardot}}, \bibinfo {author} {\bibfnamefont {P.~A.}\
  \bibnamefont {Dalgarno}}, \bibinfo {author} {\bibfnamefont {A.}~\bibnamefont
  {Badolato}}, \bibinfo {author} {\bibfnamefont {P.~M.}\ \bibnamefont
  {Petroff}}, \bibinfo {author} {\bibfnamefont {A.}~\bibnamefont {Ludwig}},
  \bibinfo {author} {\bibfnamefont {D.}~\bibnamefont {Reuter}}, \bibinfo
  {author} {\bibfnamefont {A.~D.}\ \bibnamefont {Wieck}}, \ and\ \bibinfo
  {author} {\bibfnamefont {R.~J.}\ \bibnamefont {Warburton}},\ }\href {\doibase
  10.1103/PhysRevLett.108.107401} {\bibfield  {journal} {\bibinfo  {journal}
  {Phys. Rev. Lett.}\ }\textbf {\bibinfo {volume} {108}},\ \bibinfo {pages}
  {107401} (\bibinfo {year} {2012})}\BibitemShut {NoStop}%
\bibitem [{\citenamefont {Stockill}\ \emph {et~al.}(2016)\citenamefont
  {Stockill}, \citenamefont {Le~Gall}, \citenamefont {Matthiesen},
  \citenamefont {Huthmacher}, \citenamefont {Clarke}, \citenamefont {Hugues},\
  and\ \citenamefont {Atat{\"u}re}}]{Stockill2016}%
  \BibitemOpen
  \bibfield  {author} {\bibinfo {author} {\bibfnamefont {R.}~\bibnamefont
  {Stockill}}, \bibinfo {author} {\bibfnamefont {C.}~\bibnamefont {Le~Gall}},
  \bibinfo {author} {\bibfnamefont {C.}~\bibnamefont {Matthiesen}}, \bibinfo
  {author} {\bibfnamefont {L.}~\bibnamefont {Huthmacher}}, \bibinfo {author}
  {\bibfnamefont {E.}~\bibnamefont {Clarke}}, \bibinfo {author} {\bibfnamefont
  {M.}~\bibnamefont {Hugues}}, \ and\ \bibinfo {author} {\bibfnamefont
  {M.}~\bibnamefont {Atat{\"u}re}},\ }\href {\doibase 10.1038/ncomms12745}
  {\bibfield  {journal} {\bibinfo  {journal} {Nature Communications}\ }\textbf
  {\bibinfo {volume} {7}},\ \bibinfo {pages} {12745} (\bibinfo {year}
  {2016})}\BibitemShut {NoStop}%
\bibitem [{\citenamefont {Chekhovich}\ \emph {et~al.}(2017)\citenamefont
  {Chekhovich}, \citenamefont {Ulhaq}, \citenamefont {Zallo}, \citenamefont
  {Ding}, \citenamefont {Schmidt},\ and\ \citenamefont
  {Skolnick}}]{Chekhovich2017}%
  \BibitemOpen
  \bibfield  {author} {\bibinfo {author} {\bibfnamefont {E.~A.}\ \bibnamefont
  {Chekhovich}}, \bibinfo {author} {\bibfnamefont {A.}~\bibnamefont {Ulhaq}},
  \bibinfo {author} {\bibfnamefont {E.}~\bibnamefont {Zallo}}, \bibinfo
  {author} {\bibfnamefont {F.}~\bibnamefont {Ding}}, \bibinfo {author}
  {\bibfnamefont {O.~G.}\ \bibnamefont {Schmidt}}, \ and\ \bibinfo {author}
  {\bibfnamefont {M.~S.}\ \bibnamefont {Skolnick}},\ }\href {\doibase
  10.1038/nmat4959} {\bibfield  {journal} {\bibinfo  {journal} {Nature
  Materials}\ }\textbf {\bibinfo {volume} {16}},\ \bibinfo {pages} {982}
  (\bibinfo {year} {2017})}\BibitemShut {NoStop}%
\bibitem [{\citenamefont {Bulutay}\ \emph {et~al.}(2014)\citenamefont
  {Bulutay}, \citenamefont {Chekhovich},\ and\ \citenamefont
  {Tartakovskii}}]{PhysRevB.90.205425}%
  \BibitemOpen
  \bibfield  {author} {\bibinfo {author} {\bibfnamefont {C.}~\bibnamefont
  {Bulutay}}, \bibinfo {author} {\bibfnamefont {E.~A.}\ \bibnamefont
  {Chekhovich}}, \ and\ \bibinfo {author} {\bibfnamefont {A.~I.}\ \bibnamefont
  {Tartakovskii}},\ }\href {\doibase 10.1103/PhysRevB.90.205425} {\bibfield
  {journal} {\bibinfo  {journal} {Phys. Rev. B}\ }\textbf {\bibinfo {volume}
  {90}},\ \bibinfo {pages} {205425} (\bibinfo {year} {2014})}\BibitemShut
  {NoStop}%
\bibitem [{\citenamefont {Ragunathan}(2019)}]{ragunathan_2019}%
  \BibitemOpen
  \bibfield  {author} {\bibinfo {author} {\bibfnamefont {G.}~\bibnamefont
  {Ragunathan}},\ }\href {https://etheses.whiterose.ac.uk/25466/} {Ph.D.
  thesis} (\bibinfo {year} {2019})\BibitemShut {NoStop}%
\bibitem [{\citenamefont {Chekhovich}\ \emph {et~al.}(2018)\citenamefont
  {Chekhovich}, \citenamefont {Griffiths}, \citenamefont {Skolnick},
  \citenamefont {Huang}, \citenamefont {da~Silva}, \citenamefont {Yuan},\ and\
  \citenamefont {Rastelli}}]{PhysRevB.97.235311}%
  \BibitemOpen
  \bibfield  {author} {\bibinfo {author} {\bibfnamefont {E.~A.}\ \bibnamefont
  {Chekhovich}}, \bibinfo {author} {\bibfnamefont {I.~M.}\ \bibnamefont
  {Griffiths}}, \bibinfo {author} {\bibfnamefont {M.~S.}\ \bibnamefont
  {Skolnick}}, \bibinfo {author} {\bibfnamefont {H.}~\bibnamefont {Huang}},
  \bibinfo {author} {\bibfnamefont {S.~F.~C.}\ \bibnamefont {da~Silva}},
  \bibinfo {author} {\bibfnamefont {X.}~\bibnamefont {Yuan}}, \ and\ \bibinfo
  {author} {\bibfnamefont {A.}~\bibnamefont {Rastelli}},\ }\href {\doibase
  10.1103/PhysRevB.97.235311} {\bibfield  {journal} {\bibinfo  {journal} {Phys.
  Rev. B}\ }\textbf {\bibinfo {volume} {97}},\ \bibinfo {pages} {235311}
  (\bibinfo {year} {2018})}\BibitemShut {NoStop}%
\bibitem [{\citenamefont {Virtanen}\ \emph {et~al.}(2020)\citenamefont
  {Virtanen}, \citenamefont {Gommers}, \citenamefont {Oliphant}, \citenamefont
  {Haberland}, \citenamefont {Reddy}, \citenamefont {Cournapeau}, \citenamefont
  {Burovski}, \citenamefont {Peterson}, \citenamefont {Weckesser},
  \citenamefont {Bright}, \citenamefont {{van der Walt}}, \citenamefont
  {Brett}, \citenamefont {Wilson}, \citenamefont {Millman}, \citenamefont
  {Mayorov}, \citenamefont {Nelson}, \citenamefont {Jones}, \citenamefont
  {Kern}, \citenamefont {Larson}, \citenamefont {Carey}, \citenamefont {Polat},
  \citenamefont {Feng}, \citenamefont {Moore}, \citenamefont {{VanderPlas}},
  \citenamefont {Laxalde}, \citenamefont {Perktold}, \citenamefont {Cimrman},
  \citenamefont {Henriksen}, \citenamefont {Quintero}, \citenamefont {Harris},
  \citenamefont {Archibald}, \citenamefont {Ribeiro}, \citenamefont
  {Pedregosa}, \citenamefont {{van Mulbregt}},\ and\ \citenamefont {{SciPy 1.0
  Contributors}}}]{2020SciPy-NMeth}%
  \BibitemOpen
  \bibfield  {author} {\bibinfo {author} {\bibfnamefont {P.}~\bibnamefont
  {Virtanen}}, \bibinfo {author} {\bibfnamefont {R.}~\bibnamefont {Gommers}},
  \bibinfo {author} {\bibfnamefont {T.~E.}\ \bibnamefont {Oliphant}}, \bibinfo
  {author} {\bibfnamefont {M.}~\bibnamefont {Haberland}}, \bibinfo {author}
  {\bibfnamefont {T.}~\bibnamefont {Reddy}}, \bibinfo {author} {\bibfnamefont
  {D.}~\bibnamefont {Cournapeau}}, \bibinfo {author} {\bibfnamefont
  {E.}~\bibnamefont {Burovski}}, \bibinfo {author} {\bibfnamefont
  {P.}~\bibnamefont {Peterson}}, \bibinfo {author} {\bibfnamefont
  {W.}~\bibnamefont {Weckesser}}, \bibinfo {author} {\bibfnamefont
  {J.}~\bibnamefont {Bright}}, \bibinfo {author} {\bibfnamefont {S.~J.}\
  \bibnamefont {{van der Walt}}}, \bibinfo {author} {\bibfnamefont
  {M.}~\bibnamefont {Brett}}, \bibinfo {author} {\bibfnamefont
  {J.}~\bibnamefont {Wilson}}, \bibinfo {author} {\bibfnamefont {K.~J.}\
  \bibnamefont {Millman}}, \bibinfo {author} {\bibfnamefont {N.}~\bibnamefont
  {Mayorov}}, \bibinfo {author} {\bibfnamefont {A.~R.~J.}\ \bibnamefont
  {Nelson}}, \bibinfo {author} {\bibfnamefont {E.}~\bibnamefont {Jones}},
  \bibinfo {author} {\bibfnamefont {R.}~\bibnamefont {Kern}}, \bibinfo {author}
  {\bibfnamefont {E.}~\bibnamefont {Larson}}, \bibinfo {author} {\bibfnamefont
  {C.~J.}\ \bibnamefont {Carey}}, \bibinfo {author} {\bibfnamefont
  {{\.I}.}~\bibnamefont {Polat}}, \bibinfo {author} {\bibfnamefont
  {Y.}~\bibnamefont {Feng}}, \bibinfo {author} {\bibfnamefont {E.~W.}\
  \bibnamefont {Moore}}, \bibinfo {author} {\bibfnamefont {J.}~\bibnamefont
  {{VanderPlas}}}, \bibinfo {author} {\bibfnamefont {D.}~\bibnamefont
  {Laxalde}}, \bibinfo {author} {\bibfnamefont {J.}~\bibnamefont {Perktold}},
  \bibinfo {author} {\bibfnamefont {R.}~\bibnamefont {Cimrman}}, \bibinfo
  {author} {\bibfnamefont {I.}~\bibnamefont {Henriksen}}, \bibinfo {author}
  {\bibfnamefont {E.~A.}\ \bibnamefont {Quintero}}, \bibinfo {author}
  {\bibfnamefont {C.~R.}\ \bibnamefont {Harris}}, \bibinfo {author}
  {\bibfnamefont {A.~M.}\ \bibnamefont {Archibald}}, \bibinfo {author}
  {\bibfnamefont {A.~H.}\ \bibnamefont {Ribeiro}}, \bibinfo {author}
  {\bibfnamefont {F.}~\bibnamefont {Pedregosa}}, \bibinfo {author}
  {\bibfnamefont {P.}~\bibnamefont {{van Mulbregt}}}, \ and\ \bibinfo {author}
  {\bibnamefont {{SciPy 1.0 Contributors}}},\ }\href {\doibase
  10.1038/s41592-019-0686-2} {\bibfield  {journal} {\bibinfo  {journal} {Nature
  Methods}\ }\textbf {\bibinfo {volume} {17}},\ \bibinfo {pages} {261}
  (\bibinfo {year} {2020})}\BibitemShut {NoStop}%
\bibitem [{\citenamefont {Abragam}(1961)}]{abragam1961principles}%
  \BibitemOpen
  \bibfield  {author} {\bibinfo {author} {\bibfnamefont {A.}~\bibnamefont
  {Abragam}},\ }\href {https://books.google.co.uk/books?id=9M8U\_JK7K54C}
  {\emph {\bibinfo {title} {The Principles of Nuclear Magnetism}}},\
  International series of monographs on physics\ (\bibinfo  {publisher}
  {Clarendon Press},\ \bibinfo {year} {1961})\BibitemShut {NoStop}%
\bibitem [{\citenamefont {Cywi\ifmmode~\acute{n}\else \'{n}\fi{}ski}\ \emph
  {et~al.}(2009)\citenamefont {Cywi\ifmmode~\acute{n}\else \'{n}\fi{}ski},
  \citenamefont {Witzel},\ and\ \citenamefont
  {Das~Sarma}}]{PhysRevB.79.245314}%
  \BibitemOpen
  \bibfield  {author} {\bibinfo {author} {\bibfnamefont {L.}~\bibnamefont
  {Cywi\ifmmode~\acute{n}\else \'{n}\fi{}ski}}, \bibinfo {author}
  {\bibfnamefont {W.~M.}\ \bibnamefont {Witzel}}, \ and\ \bibinfo {author}
  {\bibfnamefont {S.}~\bibnamefont {Das~Sarma}},\ }\href {\doibase
  10.1103/PhysRevB.79.245314} {\bibfield  {journal} {\bibinfo  {journal} {Phys.
  Rev. B}\ }\textbf {\bibinfo {volume} {79}},\ \bibinfo {pages} {245314}
  (\bibinfo {year} {2009})}\BibitemShut {NoStop}%
\bibitem [{\citenamefont {H\"ogele}\ \emph {et~al.}(2012)\citenamefont
  {H\"ogele}, \citenamefont {Kroner}, \citenamefont {Latta}, \citenamefont
  {Claassen}, \citenamefont {Carusotto}, \citenamefont {Bulutay},\ and\
  \citenamefont {Imamoglu}}]{PhysRevLett.108.197403}%
  \BibitemOpen
  \bibfield  {author} {\bibinfo {author} {\bibfnamefont {A.}~\bibnamefont
  {H\"ogele}}, \bibinfo {author} {\bibfnamefont {M.}~\bibnamefont {Kroner}},
  \bibinfo {author} {\bibfnamefont {C.}~\bibnamefont {Latta}}, \bibinfo
  {author} {\bibfnamefont {M.}~\bibnamefont {Claassen}}, \bibinfo {author}
  {\bibfnamefont {I.}~\bibnamefont {Carusotto}}, \bibinfo {author}
  {\bibfnamefont {C.}~\bibnamefont {Bulutay}}, \ and\ \bibinfo {author}
  {\bibfnamefont {A.}~\bibnamefont {Imamoglu}},\ }\href {\doibase
  10.1103/PhysRevLett.108.197403} {\bibfield  {journal} {\bibinfo  {journal}
  {Phys. Rev. Lett.}\ }\textbf {\bibinfo {volume} {108}},\ \bibinfo {pages}
  {197403} (\bibinfo {year} {2012})}\BibitemShut {NoStop}%
\bibitem [{\citenamefont {Botzem}\ \emph {et~al.}(2016)\citenamefont {Botzem},
  \citenamefont {McNeil}, \citenamefont {Mol}, \citenamefont {Schuh},
  \citenamefont {Bougeard},\ and\ \citenamefont {Bluhm}}]{Botzem2016}%
  \BibitemOpen
  \bibfield  {author} {\bibinfo {author} {\bibfnamefont {T.}~\bibnamefont
  {Botzem}}, \bibinfo {author} {\bibfnamefont {R.~P.~G.}\ \bibnamefont
  {McNeil}}, \bibinfo {author} {\bibfnamefont {J.-M.}\ \bibnamefont {Mol}},
  \bibinfo {author} {\bibfnamefont {D.}~\bibnamefont {Schuh}}, \bibinfo
  {author} {\bibfnamefont {D.}~\bibnamefont {Bougeard}}, \ and\ \bibinfo
  {author} {\bibfnamefont {H.}~\bibnamefont {Bluhm}},\ }\href {\doibase
  10.1038/ncomms11170} {\bibfield  {journal} {\bibinfo  {journal} {Nature
  Communications}\ }\textbf {\bibinfo {volume} {7}},\ \bibinfo {pages} {11170}
  (\bibinfo {year} {2016})}\BibitemShut {NoStop}%
\bibitem [{\citenamefont {Neder}\ \emph {et~al.}(2011)\citenamefont {Neder},
  \citenamefont {Rudner}, \citenamefont {Bluhm}, \citenamefont {Foletti},
  \citenamefont {Halperin},\ and\ \citenamefont {Yacoby}}]{PhysRevB.84.035441}%
  \BibitemOpen
  \bibfield  {author} {\bibinfo {author} {\bibfnamefont {I.}~\bibnamefont
  {Neder}}, \bibinfo {author} {\bibfnamefont {M.~S.}\ \bibnamefont {Rudner}},
  \bibinfo {author} {\bibfnamefont {H.}~\bibnamefont {Bluhm}}, \bibinfo
  {author} {\bibfnamefont {S.}~\bibnamefont {Foletti}}, \bibinfo {author}
  {\bibfnamefont {B.~I.}\ \bibnamefont {Halperin}}, \ and\ \bibinfo {author}
  {\bibfnamefont {A.}~\bibnamefont {Yacoby}},\ }\href {\doibase
  10.1103/PhysRevB.84.035441} {\bibfield  {journal} {\bibinfo  {journal} {Phys.
  Rev. B}\ }\textbf {\bibinfo {volume} {84}},\ \bibinfo {pages} {035441}
  (\bibinfo {year} {2011})}\BibitemShut {NoStop}%
\bibitem [{\citenamefont {Cywi\ifmmode~\acute{n}\else \'{n}\fi{}ski}\ \emph
  {et~al.}(2008)\citenamefont {Cywi\ifmmode~\acute{n}\else \'{n}\fi{}ski},
  \citenamefont {Lutchyn}, \citenamefont {Nave},\ and\ \citenamefont
  {Das~Sarma}}]{PhysRevB.77.174509}%
  \BibitemOpen
  \bibfield  {author} {\bibinfo {author} {\bibfnamefont {L.}~\bibnamefont
  {Cywi\ifmmode~\acute{n}\else \'{n}\fi{}ski}}, \bibinfo {author}
  {\bibfnamefont {R.~M.}\ \bibnamefont {Lutchyn}}, \bibinfo {author}
  {\bibfnamefont {C.~P.}\ \bibnamefont {Nave}}, \ and\ \bibinfo {author}
  {\bibfnamefont {S.}~\bibnamefont {Das~Sarma}},\ }\href {\doibase
  10.1103/PhysRevB.77.174509} {\bibfield  {journal} {\bibinfo  {journal} {Phys.
  Rev. B}\ }\textbf {\bibinfo {volume} {77}},\ \bibinfo {pages} {174509}
  (\bibinfo {year} {2008})}\BibitemShut {NoStop}%
\bibitem [{\citenamefont {Cywi\ifmmode~\acute{n}\else
  \'{n}\fi{}ski}(2014)}]{PhysRevA.90.042307}%
  \BibitemOpen
  \bibfield  {author} {\bibinfo {author} {\bibfnamefont {L.}~\bibnamefont
  {Cywi\ifmmode~\acute{n}\else \'{n}\fi{}ski}},\ }\href {\doibase
  10.1103/PhysRevA.90.042307} {\bibfield  {journal} {\bibinfo  {journal} {Phys.
  Rev. A}\ }\textbf {\bibinfo {volume} {90}},\ \bibinfo {pages} {042307}
  (\bibinfo {year} {2014})}\BibitemShut {NoStop}%
\bibitem [{\citenamefont {Malinowski}\ \emph {et~al.}(2017)\citenamefont
  {Malinowski}, \citenamefont {Martins}, \citenamefont {Nissen}, \citenamefont
  {Barnes}, \citenamefont {Cywi{\'{n}}ski}, \citenamefont {Rudner},
  \citenamefont {Fallahi}, \citenamefont {Gardner}, \citenamefont {Manfra},
  \citenamefont {Marcus},\ and\ \citenamefont {Kuemmeth}}]{Malinowski2017}%
  \BibitemOpen
  \bibfield  {author} {\bibinfo {author} {\bibfnamefont {F.~K.}\ \bibnamefont
  {Malinowski}}, \bibinfo {author} {\bibfnamefont {F.}~\bibnamefont {Martins}},
  \bibinfo {author} {\bibfnamefont {P.~D.}\ \bibnamefont {Nissen}}, \bibinfo
  {author} {\bibfnamefont {E.}~\bibnamefont {Barnes}}, \bibinfo {author}
  {\bibfnamefont {{\L}.}~\bibnamefont {Cywi{\'{n}}ski}}, \bibinfo {author}
  {\bibfnamefont {M.~S.}\ \bibnamefont {Rudner}}, \bibinfo {author}
  {\bibfnamefont {S.}~\bibnamefont {Fallahi}}, \bibinfo {author} {\bibfnamefont
  {G.~C.}\ \bibnamefont {Gardner}}, \bibinfo {author} {\bibfnamefont {M.~J.}\
  \bibnamefont {Manfra}}, \bibinfo {author} {\bibfnamefont {C.~M.}\
  \bibnamefont {Marcus}}, \ and\ \bibinfo {author} {\bibfnamefont
  {F.}~\bibnamefont {Kuemmeth}},\ }\href {\doibase 10.1038/nnano.2016.170}
  {\bibfield  {journal} {\bibinfo  {journal} {Nature Nanotechnology}\ }\textbf
  {\bibinfo {volume} {12}},\ \bibinfo {pages} {16} (\bibinfo {year}
  {2017})}\BibitemShut {NoStop}%
\bibitem [{\citenamefont {Chekhovich}\ \emph {et~al.}(2015)\citenamefont
  {Chekhovich}, \citenamefont {Hopkinson}, \citenamefont {Skolnick},\ and\
  \citenamefont {Tartakovskii}}]{Chekhovich2015}%
  \BibitemOpen
  \bibfield  {author} {\bibinfo {author} {\bibfnamefont {E.~A.}\ \bibnamefont
  {Chekhovich}}, \bibinfo {author} {\bibfnamefont {M.}~\bibnamefont
  {Hopkinson}}, \bibinfo {author} {\bibfnamefont {M.~S.}\ \bibnamefont
  {Skolnick}}, \ and\ \bibinfo {author} {\bibfnamefont {A.~I.}\ \bibnamefont
  {Tartakovskii}},\ }\href {\doibase 10.1038/ncomms7348} {\bibfield  {journal}
  {\bibinfo  {journal} {Nature Communications}\ }\textbf {\bibinfo {volume}
  {6}},\ \bibinfo {pages} {6348} (\bibinfo {year} {2015})}\BibitemShut
  {NoStop}%
\end{thebibliography}%

\section{Methods}

\subsection{Device Growth \& Processing}
The structure was grown via molecular beam epitaxy (MBE). Individual QDs were grown via nanohole etching and infilling of the AlGaAs layer with GaAs\cite{Heyn2009,Atkinson2012}. During growth, the QD-containing layer was embedded between heavily n-doped and heavily p-doped layers of AlGaAs and AlGaAs-GaAs, respectively. The device was processed into an n-i-p diode using standard photolitographic steps outlined in \onlinecite{Supplementary}.

\subsection{Experimental Techniques}
Resonant fluorescence is implemented using polarisation rejection in a confocal microscope. For the Spin control, a 76 MHz Ti:Sapphire pulsed-laser source is split into two, with one path going through a motorised delay line. The light is subsequently pulse-picked by single-passing acousto-optic modulators with risetime of 5 ns. The 600 GHz-detuned control gates are spectrally filtered out from the photons emitted by the QD device using a holographic grating. A more detailed description of the experimental set-up and schematics are presented in the \onlinecite{Supplementary}.
In the NMR experiment, our QD sample is strained along the [110] axis to offset the quadrupolar sidebands ($\nu_{\pm\frac{1}{2}\leftrightarrow\pm\frac{3}{2}}$) by $\nu_\text{Q}\approx 300$\,kHz from the central transition frequency ($\nu_{+\frac{1}{2}\leftrightarrow-\frac{1}{2}}$). The QD nuclear spins are polarised optically, and the subsequent depolarisation arising from the radiofrequency pulse is read optically as a change in the Overhauser shift $\sum_i A_i \hat{I}_{z,i}$. To resolve both the `narrow' Arsenic ensemble (identified by previous NMR studies\cite{Chekhovich2020}), and the `broader' Arsenic ensemble (identified in this study), together with their relative weight we perform \emph{integral} NMR spectroscopy, which we then differentiate numerically \cite{Supplementary}.

\subsection{Data Analysis}
In order to deduce the electron spin state from the readout photon counts we introduce a \textit{visibility}:
\begin{equation}\label{visexp}
v=\frac{\mathrm{cts}_\uparrow-\mathrm{cts}_\downarrow}{\mathrm{cts}_\uparrow+\mathrm{cts}_\downarrow}
\end{equation}
where $\mathrm{cts}_\uparrow$ and $\mathrm{cts}_\downarrow$ are the readout counts collected in the experiments preceded by initialisation of the electron to $\ket{\uparrow}$ and $\ket{\downarrow}$ states, respectively. Since those counts are proportional to the electronic populations of $\ket{\uparrow}$ and $\ket{\downarrow}$ states we can identify (up to a sign dependent on the experimental sequence):
\begin{equation}
    v=S_z/S_{\mathrm{max}}
\end{equation}
where $S_z$ is the spin projection on the $z$-axis and $S_{\mathrm{max}}\leq 1$ is the maximal length of the electronic Bloch vector, limited by the preparation fidelity. 

For data points the error bars correspond to the standard deviation set by shot noise. For fitted values the errors represent $67\%$ confidence intervals. 

\subsection{Decoherence Model}
To model the decoherence of dynamically decoupled electron spin under strong hyperfine interactions with nuclear spin ensemble we employ an adapted ring diagram formalism. Our theory -- presented in the \onlinecite{Supplementary} -- extends previous work\cite{Cywinski2009,Neder2011a} by treating the nuclear quadrupolar effects from first principles.


\end{document}


	
	\title{Supplementary Materials for: \\ Ideal refocusing of an optically active spin qubit under strong hyperfine interactions}
	
\author{Leon Zaporski\textsuperscript{1,$\dagger$}}
\author{Noah Shofer\textsuperscript{1,*}}
\author{Jonathan H.\,Bodey\textsuperscript{1,*}}
\author{Santanu Manna\textsuperscript{2,*}}
\author{George Gillard\textsuperscript{3}}
\author{Daniel M.\,Jackson\textsuperscript{1}}
\author{Martin Hayhurst Appel\textsuperscript{1}}
\author{Christian Schimpf\textsuperscript{2}}
\author{Saimon Covre da Silva\textsuperscript{2}}
\author{John Jarman\textsuperscript{1}}
\author{Geoffroy Delamare\textsuperscript{1}}
\author{Gunhee Park\textsuperscript{1}}
\author{Urs Haeusler\textsuperscript{1}}
\author{Evgeny A.\,Chekhovich\textsuperscript{3}}
\author{Armando Rastelli\textsuperscript{2}}
\author{Dorian A.\,Gangloff\textsuperscript{1,4}}
\author{Mete Atat\"ure\textsuperscript{1,$\dagger$}}
\author{Claire Le Gall\textsuperscript{1,$\dagger$}}

\noaffiliation
\affiliation{Cavendish Laboratory, University of Cambridge, JJ Thomson Avenue, Cambridge, CB3 0HE, United Kingdom}
\affiliation{Institute of Semiconductor and Solid State Physics, Johannes Kepler University, Altenbergerstraße 69, Linz 4040, Austria
}\affiliation{Department of Physics and Astronomy, University of Sheffield, Sheffield, S3 7RH, United Kingdom}
\affiliation{Department of Engineering Science, University of Oxford, Parks Road, Oxford, OX1 3PJ
	\\ \ \\
	\textsuperscript{*}\,These authors contributed equally to this work.
	\\
	\textsuperscript{$\dagger$}\,Correspondence should be addressed to: lz412@cam.ac.uk; ma424@cam.ac.uk; cl538@cam.ac.uk.}
	
	\maketitle
	
	\tableofcontents

\section{Sample structure}

\begin{figure}[t!]
    \centering
    \includegraphics[width=\linewidth]{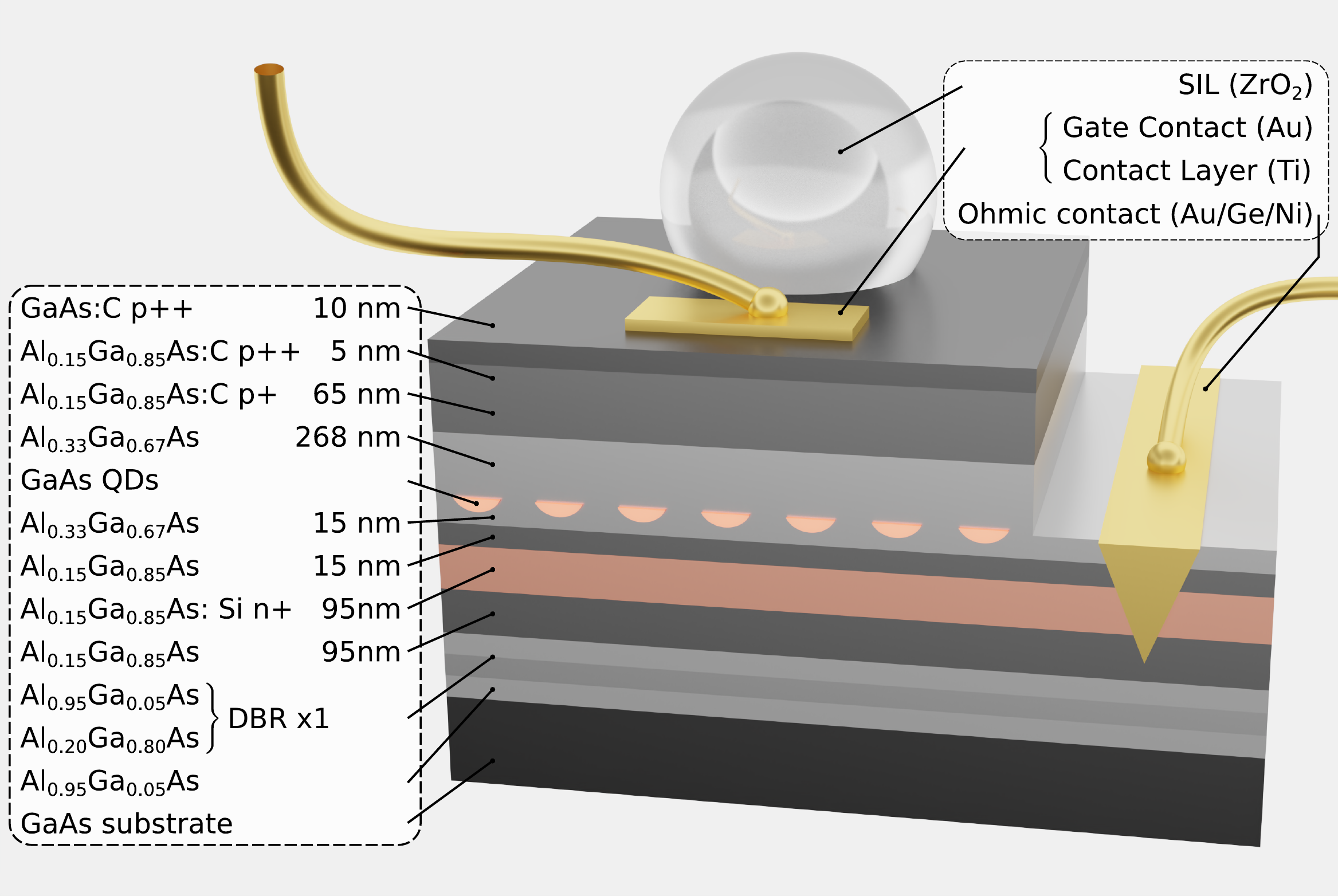}
    \caption{\textbf{QD device sample structure design (Linz Ref. SA0553)}} 
\label{struct}
\end{figure}

Fig.~\ref{struct} shows the heterostructure containing the quantum dots used in this work. The structure is grown via molecular beam epitaxy (MBE). Quantum dots are grown via nanohole etching with $\mathrm{Al}$-droplets and infilling of the $\mathrm{Al_{0.33}Ga_{0.67}As}$ layer with GaAs \cite{doi:10.1063/5.0057070,Heyn2009,Atkinson2012}. During growth, the quantum dot containing layer is embedded between a heavily n-doped (n\textsuperscript{+}) and heavily p-doped (p\textsuperscript{+}, p\textsuperscript{++}) layers of $\mathrm{Al_{0.15}Ga_{0.85}As}$ and $\mathrm{Al_{0.15}Ga_{0.85}As}/\mathrm{GaAs}$, respectively. To make the n-i-p-diode, contacts are fabricated onto the n\textsuperscript{+} and the top GaAs p\textsuperscript{++} layers. The back n-contact is fabricated by first etching down roughly $300$ nm using a sulfuric acid/hydrogen peroxide solution (1 $\mathrm{H}_{2}\mathrm{SO}_{4}$ : 8 $\mathrm{H}_{2}\mathrm{O}_{2}$ : 80 $\mathrm{H}_{2}\mathrm{O}$). About $100$ nm of AuGeNi alloy is then evaporated onto the surface of the sample to form the n-contact. The sample is then annealed at $220{}^\circ \mathrm{C}$ for $30$ seconds, followed by another step at $430{}^\circ \mathrm{C}$ for $160$ seconds. The top p-contact is fabricated by evaporating $10$ nm of Ti onto the top GaAs layer, followed by $100$ nm of Au. Prior to metal evaporation the surface of the sample is dipped in a $\sim10-50\%$ HCl solution to remove any native surface oxides that negatively affect the conductance of the top contact. The contacts are then wirebonded to a chip carrier to provide electrical connections.\par 

\begin{figure*}[t]
    \centering
    \includegraphics[width=\linewidth]{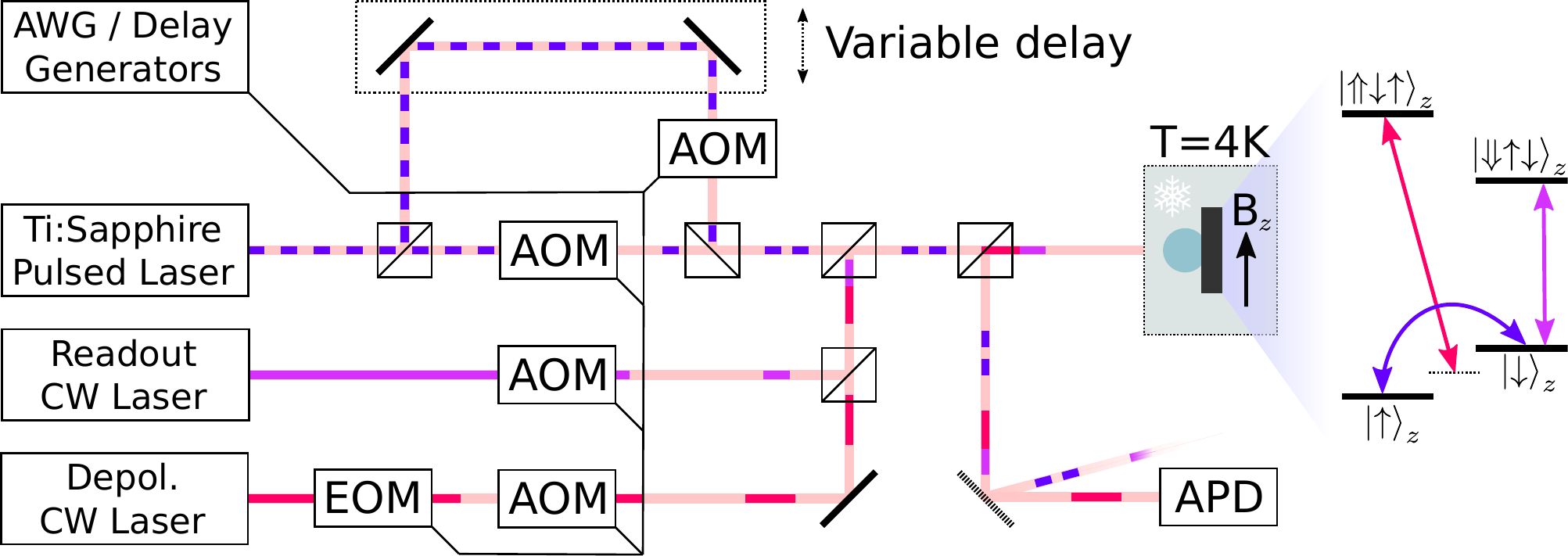}
    \caption{\textbf{Experimental set-up:} AOM $=$ acousto optic modulator; EOM $=$ electro-optic modulator; APD $=$ avalanche photodiode; AWG $=$ arbitrary wave generator. Colors of the laser beams correspond to the driven processes (readout, depolarization and spin control) outlined in the energy level diagram.} 
\label{exp-setup}
\end{figure*}

\section{Experimental Set-up for spin-control}

The experimental setup is outlined in Fig. \ref{exp-setup}, together with the energy level diagram showing three processes: readout, depolarization and spin control - each driven by one of the three, color-coded lasers.

Spin control is done via a Ti:Sapphire pulsed laser (Mira Optima 900-P) in a pico-second mode detuned by $500-600$ $\mathrm{GHz}$ from the resonance\cite{Berezovsky2008, Press2008}. Pulses are split into two optical paths, one of which contains a delay stage which allows to set an arbitrary time-offset between pulses in both paths. Readout and depolarization are effectuated via two continuous-wave diode lasers. 

All the lasers are fed through acousto-optic modulators (AOMs) to enable pulse picking essential for programming the pulse sequences. To improve the suppression of a powerful depolarization laser, it is also passed through an electro-optical modulator (EOM). 

The modulators are locked to $76$-$\mathrm{MHz}$ repetition rate of the pulsed laser and programmed using a combination of delay generators with $8$ $\mathrm{ps}$ jitter as well as an arbitrary wave generator (AWG). 

The optical paths are combined in a confocal microscope and sent down the bath cryostat to the sample held at $T=4$ $\mathrm{K}$. A $B$-field perpendicular to the growth axis is generated by a superconducting magnet. 

To suppress the resonant laser background we use cross-polarized linear polarizers with a quarter wave-plate in between to convert all laser polarizations arriving to the sample from linear to circular. 
Following the polarization suppresion, emission from the QD is spectrally filtered from the spin control pulses on a diffraction grating with a $20$ $\mathrm{GHz}$ bandwidth, and then detected by an avalanche photo diode (APD).

\section{Supplementary Measurements}
\subsection{Saturation curve}

In order to express the laser powers (stated in the units of the voltage measured on a photodiode in the confocal microscope) via system-specific parameters, we measure the saturation curve of counts scattered when resonantly driving the transition between the crystal ground state ($\ket{\mathrm{c.g.s.}}$) and $\ket{\Uparrow \downarrow}$ state at $B=0$ $\mathrm{T}$. 
We fit the two-level system model:
\begin{equation}
I(P)=\frac{I_{\mathrm{max}}}{1+(P/P_{\mathrm{sat}})^{-1}} 
\end{equation}
to the measured countrates $I$ for all of the considered laser powers to find $P_{\mathrm{sat}}=2.0(2) \times 10^{-7}$ $\mathrm{V}$ and $I_{\mathrm{max}}= 1.03(3)$ $\mathrm{MCounts/s}$.


\begin{figure}[h!] 
	\centering
	\includegraphics[width=\linewidth]{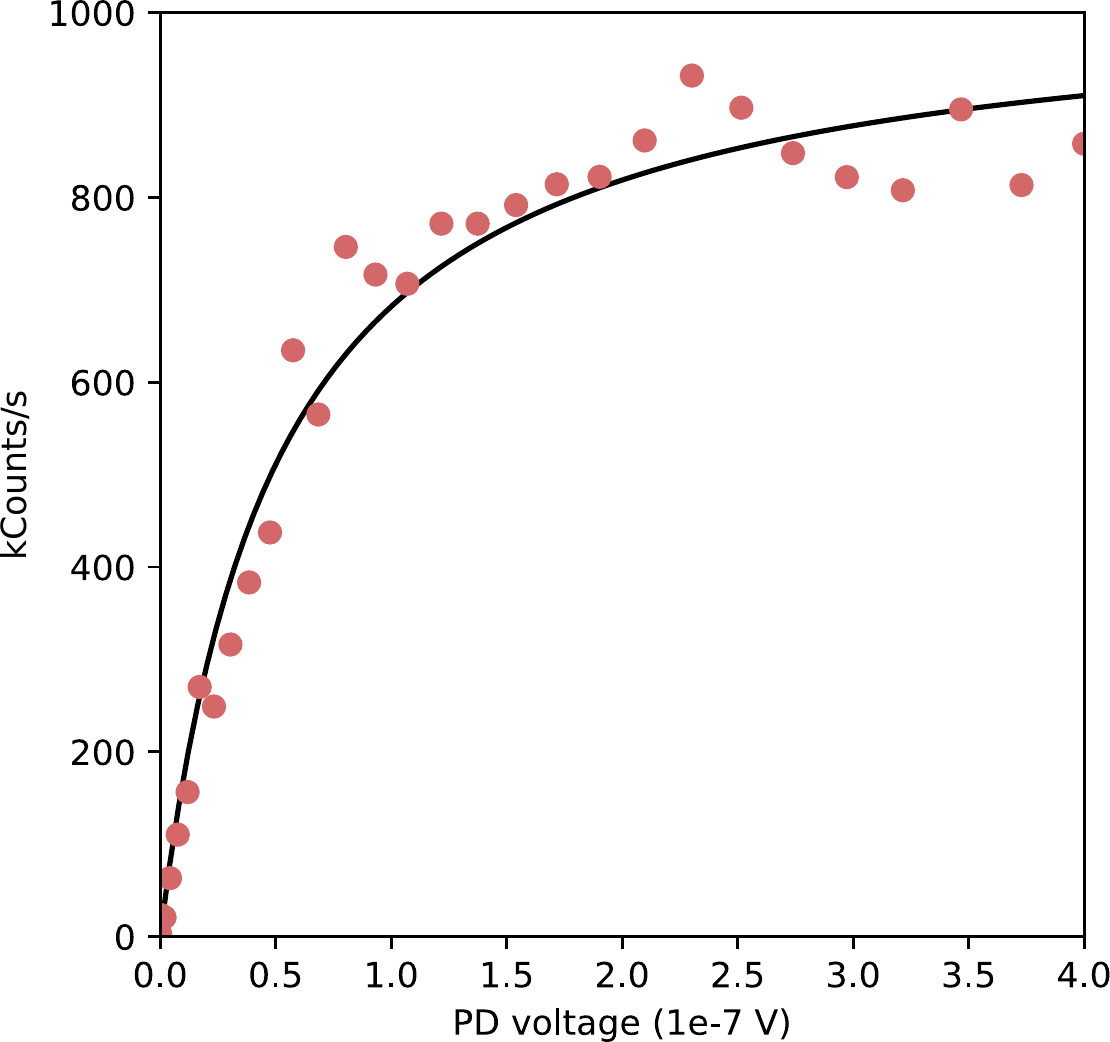}
	\caption{Saturation curve of counts scattered when driving the $\ket{\mathrm{c.g.s.}} \to \ket{\Uparrow \downarrow}$ transition resonantly. The red points are the countrates measured for a range of fixed laser powers, whereas the solid black curve is a fit to the data. }
	\label{fig:sat_curve}
\end{figure}

\subsection{Electron spin $T_1$}

\begin{figure}[b!] 
	\centering
	\includegraphics[width=\linewidth]{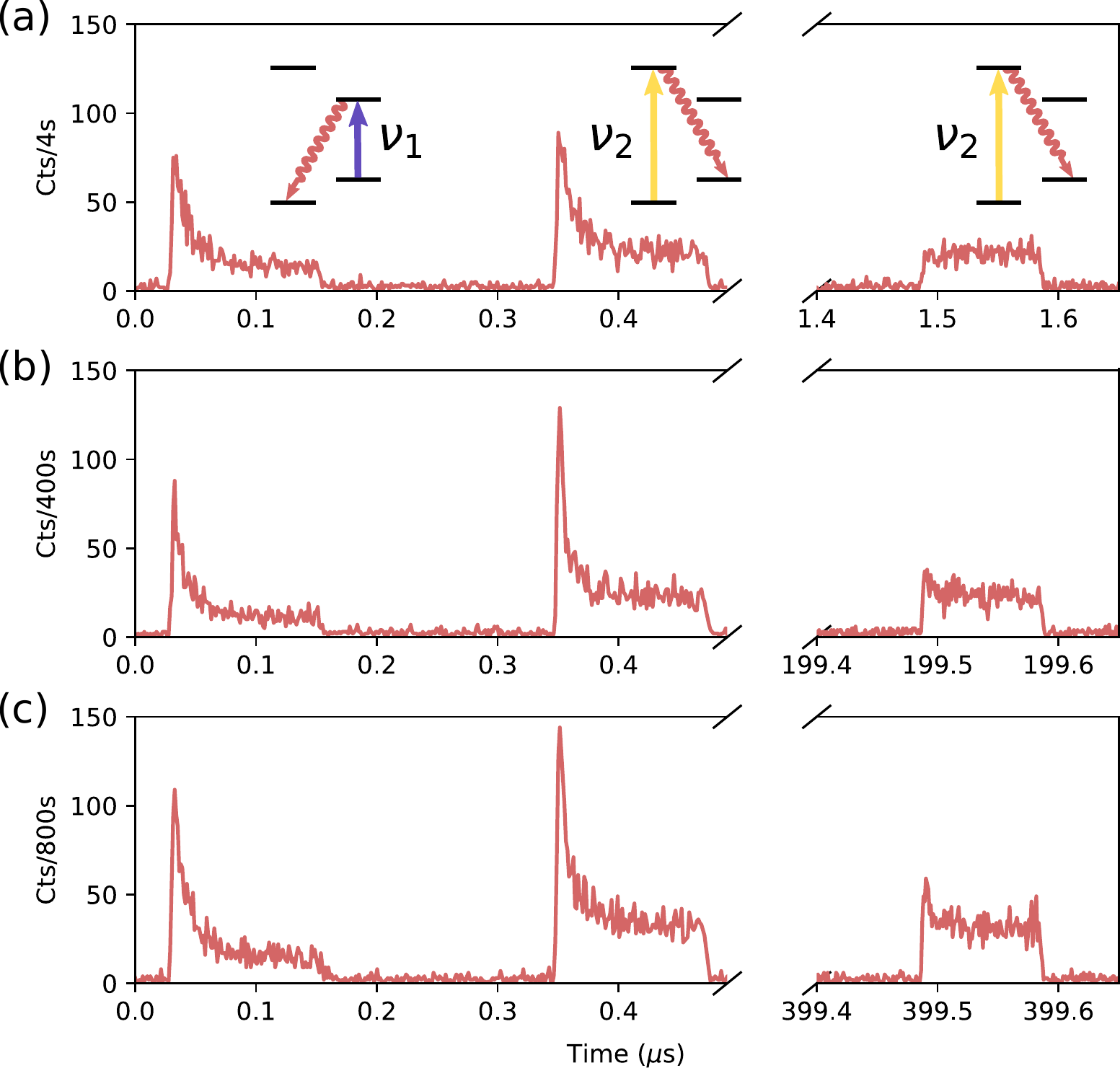}
	\caption{\textbf{a} Measurement of electron spin relaxation after delay of $1$ $\mu\mathrm{s}$; insets correspond to transitions driven during subsequent parts of the experiment (consistent across all panels). \textbf{b} Measurement of electron spin relaxation after delay of $199$ $\mu\mathrm{s}$. \textbf{c} Measurement of electron spin relaxation after delay of $399$ $\mu\mathrm{s}$.}
	\label{fig:spinT1}
\end{figure}

To place a lower bound on the electron spin $T_1$ we perform a population-relaxation measurement using two resonant lasers set to frequencies $\nu_1=380767$ $\mathrm{GHz}$ and $\nu_2=380793$ $\mathrm{GHz}$ (Fig. 1e of the main text), at the magnetic field of $6.5$ $\mathrm{T}$ (electron Zeeman splitting $\nu_{\mathrm{e}}= 4.453(6)$ $\mathrm{GHz}$). At the beginning of the experimental sequence, the first laser initialises the electron spin in the $\ket{\downarrow}$ state, whereas the second laser re-initialises it in $\ket{\uparrow}$ simultaneously revealing the (reference) population of $\ket{\downarrow}$. Following the delays of $1$, $199$ and $399$ $\mu s$, the second laser is pulsed again to measure the population of $\ket{\downarrow}$ again; photon counts recorded during the experiments with said delays are plotted in the panels a, b and c of the Fig. \ref{fig:spinT1}, respectively.

Small electron spin relaxation is observed over $399$ $\mu\mathrm{s}$, however, measured pick-up in $\ket{\downarrow}$-population following this delay is negligibly small to the reference. This allows us to place a weak lower bound $T_1>400$ $\mu\mathrm{s}$.

\subsection{Gate Fidelity}
Increasing the number of $\pi$-pulses - $N_\pi$ - in the CPMG sequence exposes the electron spin to increased amount of laser-induced dephasing\cite{Bodey2019}. During each pulse the Bloch vector's magnitude scales by a fraction $0\%\leq \mathcal{F} \leq 100\%$, i.e. \emph{gate fidelity}. In this simple model, visibility recorded at the beginning of the CPMG sequence ($v_{\mathrm{max}}=v(\tau \to 0)$ - see Eq. 2 of the main text) with $N_\pi$ pulses is given simply by $\mathcal{F}^{N_\pi}$. 

To measure the $\pi$-gate fidelity $\mathcal{F}$ we run a short-time CPMG experiment with $N_\pi=162$ pulses and a total time of $\tau=4.26$ $\mu\mathrm{s}$. Keeping the preparation-to-readout delay fixed to $6.50$ $\mu\mathrm{s}$, i.e. much shorter than the electron spin $T_1$, we find $v_{\mathrm{max}}=0.32(3)$.

\begin{figure}[h!]
	\centering
	\includegraphics[width=\linewidth]{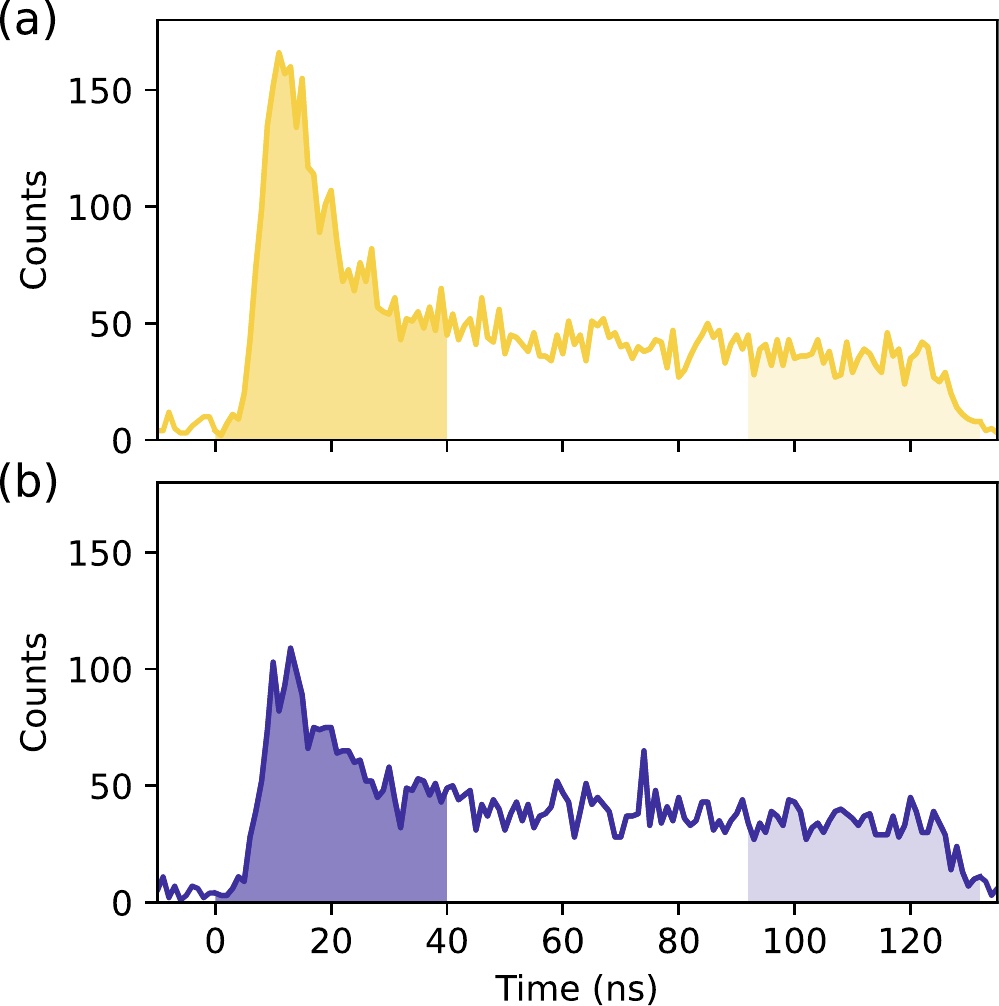}
	\caption{\text{(a)} Readout signal following CPMG sequence with $N_\pi=162$. Integrated counts under dark-shaded area were equal to 3062, and 1309 under light-shaded area (background).
	\textbf{(b)} Readout signal following the same pulse sequence with an additional $\pi$-pulse added after preparation. Integrated counts under dark-shaded area were equal to 2206, and 1289 under light-shaded area (background). }
	\label{fig:gatefid2}
\end{figure}

To do this we first integrate the photon counts within $41$ $\mathrm{ns}$ wide windows of the histograms presented in panels a and b of the Fig. \ref{fig:gatefid2} (shaded areas). Integrated counts of the light-shaded areas are a measure of background, and they are subtracted from the integrated counts of the corresponding dark-shaded areas in order to isolate the effect of dephasing from that of the imperfect state preparation. The signal in panel b has been obtained after CPMG sequence preceded by an additional $\pi$-pulse. Consequently, our measurement reveals the visibility as defined in Eq. 2 of the main text. The errors on the integrated counts are assumed to follow shot noise statistics, and are subsequently propagated into the expression for visibility. This analysis allows to quantify the best-achieved $\pi$-gate fidelity:
\begin{equation}
\mathcal{F}=99.30(5)\%
\end{equation}

This fidelity exceeds the best fidelities obtained in InGaAs/GaAs QDs structures \cite{Bodey2019}. The QD device structure used here does not contain a blocking-barrier -- an interface \cite{Houel2012} which may be responsible for limiting gate fidelities in other devices \cite{Stockill2016,Berezovsky2008,Press2008,Bodey2019}.

\begin{figure}[h!]
	\centering
	\includegraphics[width=\linewidth]{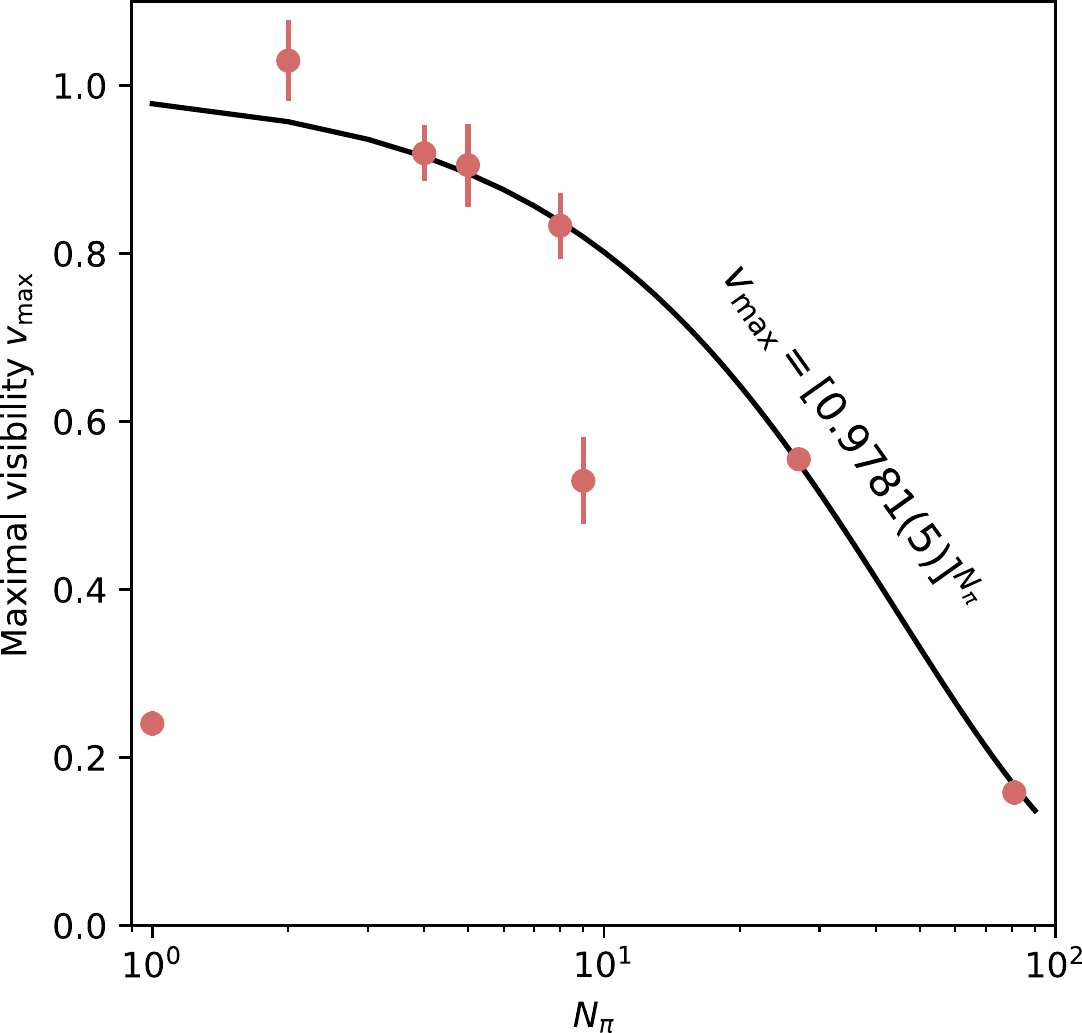}
	\caption{Maximal visibility as a function of the number of $\pi$-pulses in the CPMG sequence. Datapoints were obtained from stretched-exponential fits ($f(\tau)=v_{\mathrm{max}}\times\exp{-(\tau/T_2)^\alpha}+c$) to the unnormalized dynamical decoupling data; these are plotted in Fig. \ref{fig:best_fit} following normalization.}
	\label{fig:gatefid}
\end{figure}

Measuring the coherence of dynamically decoupled electron spin requires hours of integration time. Mechanical stability of our current set-up on hour-long timescales bounds the achieved fidelity to $\bar{\mathcal{F}}<\mathcal{F}$, which we now constrain, for completeness. In this analysis, for every single recorded CPMG data set (plotted, following normalisation, in the Fig. \ref{fig:best_fit}) we find $v_{\mathrm{max}}$ from the stretched exponential fit ($f(\tau)=v_{\mathrm{max}}\times \exp{-(\tau/T_2)^\alpha}+c$). We plot the fitted values of $v_{\mathrm{max}}$ as a function of $N_\pi$ in the Fig. \ref{fig:gatefid}, and find the average gate fidelity as $\bar{\mathcal{F}}=97.81(5)\%$ from a fit of $v_{\mathrm{max}}(N_\pi)=\bar{\mathcal{F}}^{N_\pi}$ to the data (except for $N_\pi=1,3$ and $9$, due to experimental errors or lack of a visibility measurement). 

\section{Free Induction Decay: data analysis}\label{sec:FID}
\subsection{Experimental details}
The experiment was performed at $B_z=3.5$ $\mathrm{T}$ and applied gate voltage of $517$ $\mathrm{mV}$ (c.f. Fig. 1e of the main text). The readout laser frequency was set to that of the highest-frequency transition - $380754.5$ $\mathrm{GHz}$ - and its power was stabilised around $0.9P_{\mathrm{sat}}$. The depolarisation laser was set to drive the two highest-energy transitions (frequency of $380751$ $\mathrm{GHz}$) with a power of $7P_{\mathrm{sat}}$. To compensate for the small offset of depolarisation laser's polarization from circular, its frequency did not coincide with the arithmetic mean of the two frequencies, exactly. Visibility constructed according to Eq. 2 of the main text (which coincided with normalized spin projection $S_y/S_{\mathrm{max}}$ in this instance - see the main text) was taken as a signal in further parts of data analysis. Preparation of the opposite spin states relied on insertion of the additional $\pi$-pulse following the first resonant laser pulse.

\begin{figure}[b!]
	\centering
	\includegraphics[width=\linewidth]{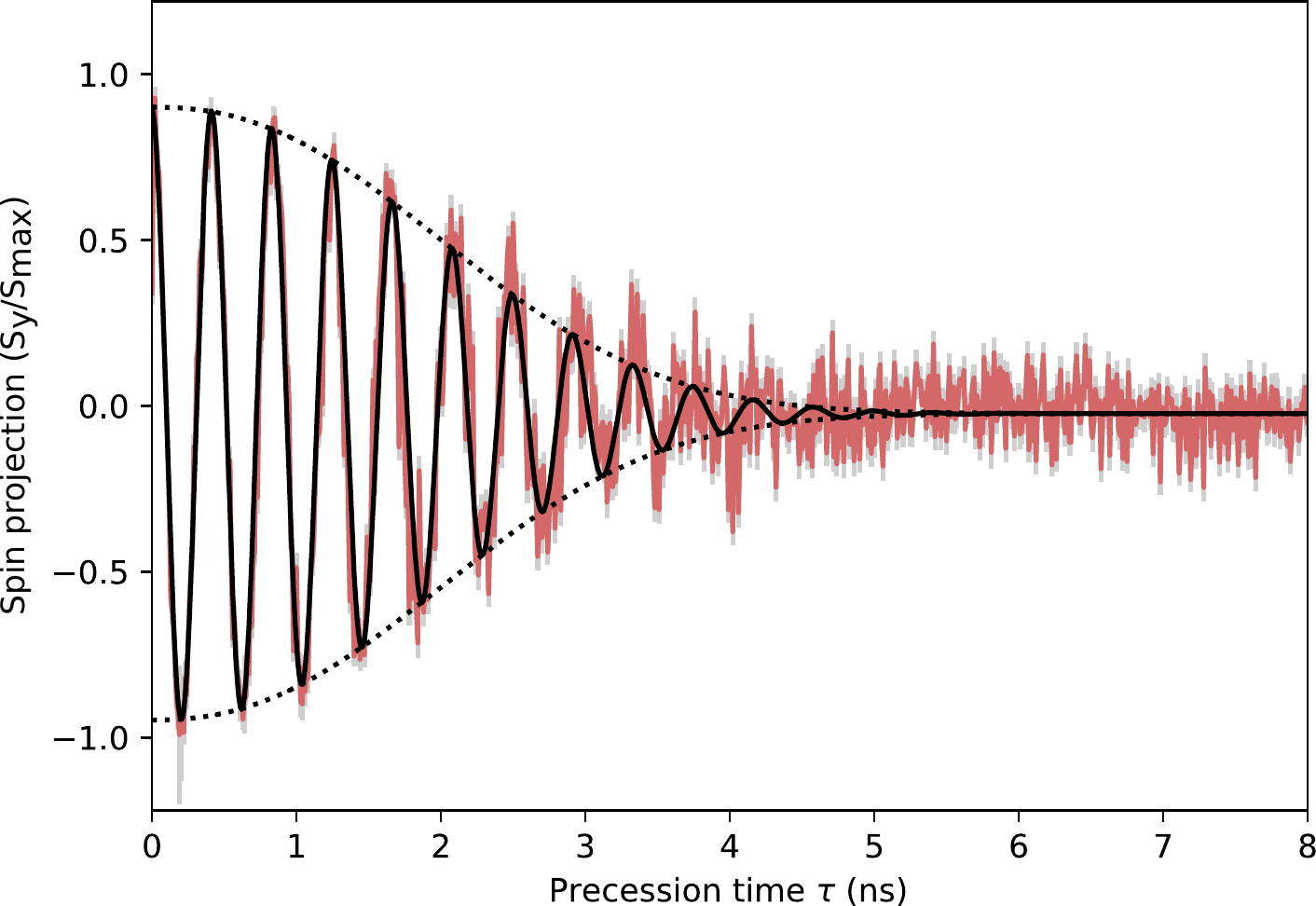}
	\caption{FID signal (solid red curve with faint grey error bars), overlaid with the damped sinusoidal fit $f(\tau)=A\sin(2\pi \nu_{\mathrm{L}} \tau +\phi)e^{-(\tau/T_2^*)^\alpha}+C$ (solid black curve). Dotted curves represent the fit envelopes, $Ae^{-(\tau/T_2^*)^\alpha}+C$ .} 
	\label{fig:Ramsey2}
\end{figure}

\subsubsection{Fit function and parameters}
The signal was fitted with $f(\tau)=A\sin(2\pi \nu_{\mathrm{L}} \tau +\phi)e^{-(\tau/T_2^*)^\alpha}+C$,
where:
\begin{itemize}
	\item $A=0.93(2)$
	\item $\nu_{\mathrm{L}}=2.398(3)$ $\mathrm{GHz}$
	\item $\phi =1.65(2)$
	\item $T_2^*=2.55(5)$ $\mathrm{ns}$
	\item $\alpha=2.3(1)$
	\item $C=-0.0234(3)$.
\end{itemize}
The envelope and fit (offset by the mean normalized countrate) are presented in Fig. \ref{fig:Ramsey2}, overlaid on top of the signal. Fitted frequency of Larmor precession $\nu_{\mathrm{L}}$ corresponds to the $g$-factor of magnitude $|g_{\mathrm{e}}|=0.04895(6)$ - in a good agreement with $|g_e|=0.0495(9)$ extracted from the RF map of $X^-$ in co-tunnelling regime (c.f. Fig. 1e of the main text).

\subsection{Constraining the total number of nuclei}
The drop of coherence in Free Induction Decay (FID) by a factor $e$ is observed over a timescale $T_2^*$. This characterizes the magnitude of shot-to-shot fluctuations of the Overhauser field $B_{\mathrm{OH}}$, and the relationship is expressed via:
\begin{equation}
T_2^*=\frac{\sqrt{6}\hbar}{g_{\mathrm{e}}\mu_{\mathrm{B}} \sqrt{\Delta^2 B_{\mathrm{OH}}}}
\end{equation}
Calculating the variance of the Overhauser field from the infinite temperature distribution of nuclear spin projections yields: 
\begin{equation}
\Delta^2 B_{\mathrm{OH}}=\frac{2}{(g_{\mathrm{e}}\mu_{\mathrm{B}}\sqrt{N})^2}\sum_k \eta_k A_k^2 I_k(I_k+1)
\end{equation}
where the sum runs over distinct nuclear species with hyperfine constants, $A_k$, and concentrations, $\eta_k$, equal to those stated in the Table \ref{table:material_constants}. 
This allows us to express the total number of nuclei through:
\begin{equation}
N=\frac{5}{4\hbar^2} \sum_k \eta_k A_k^2 (T_2^*)^2
\end{equation}
Using the measured value of $T_2^*=2.55(5) \, \mathrm{ns}$, we constrain the total number of nuclei to $N=6.5(3)\times 10^{4}$.


\section{Reconstructing distribution of quadrupolar shifts from NMR data}\label{NMR_section}

\begin{figure}[b!]
	\centering
	\includegraphics[width=\linewidth]{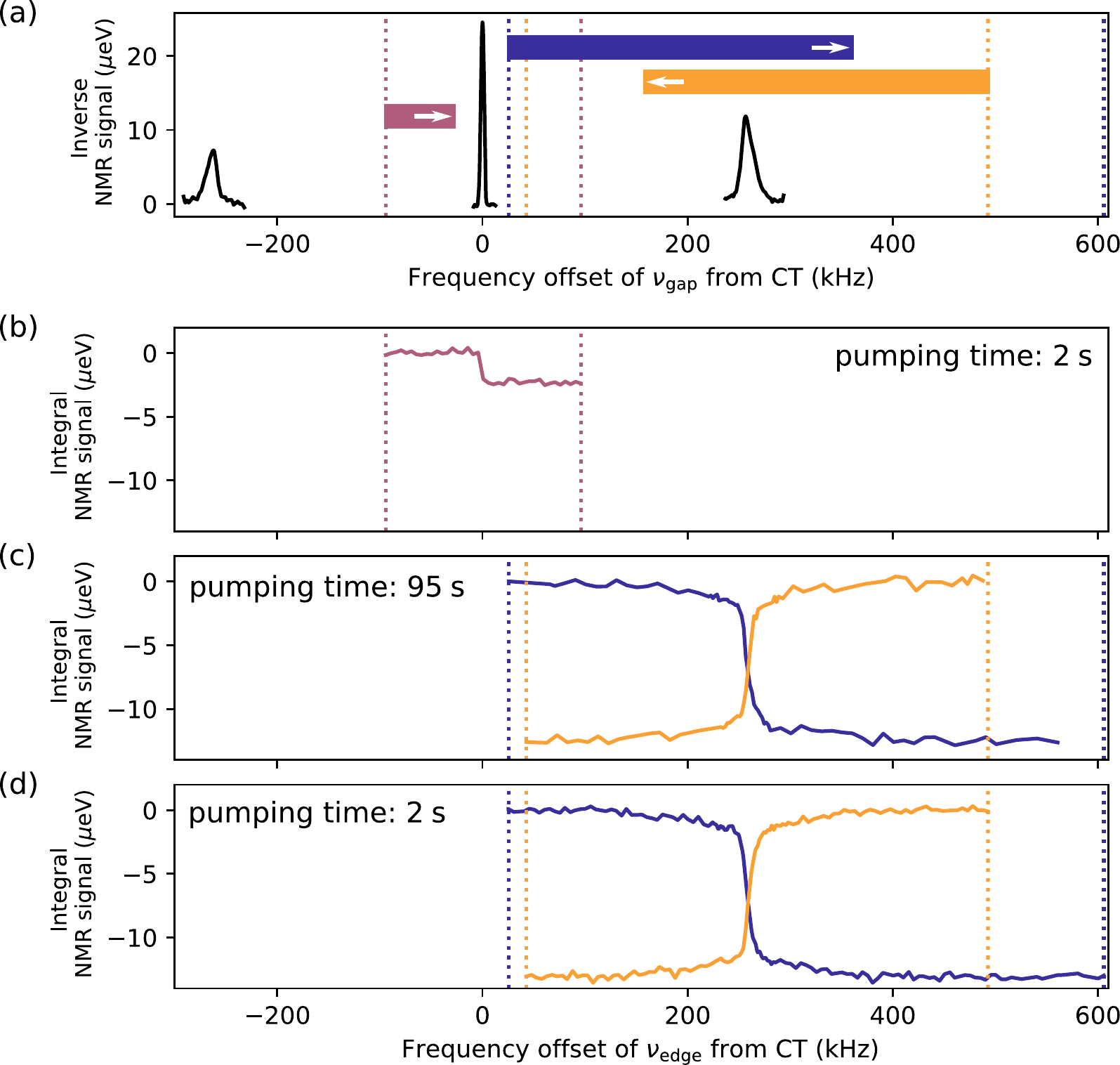}
	\caption{\textbf{(a)} Inverse NMR spectrum showing a narrow central transition and broadened satellite transitions. The x-axis states the frequency offset of applied $\nu_{\mathrm{gap}}$ from central transition. The horizontal bars with arrows indicate the positions of the fixed edges and directions of the integral NMR scans shown in the panels below.  \textbf{(b)} Integral NMR signal from the Central Transition. \textbf{(c)} Integral NMR signal from the higher-frequency Satellite Transition taken with a long NMR pumping time of $95$ $\mathrm{s}$. \textbf{(d)} Integral NMR signal from the higher-frequency Satellite Transition taken with a short NMR pumping time of $2$ $\mathrm{s}$} 
	\label{fig:NMR_principles}
\end{figure}

In order to reconstruct the distribution of quadrupolar shifts, $P_{\mathrm{As}}(\nu)$, we perform NMR measured optically via photoluminescence detection on another piece of the same wafer (i.e. a different QD device). 

Following the optical preparation of a low-temperature nuclear spin state \cite{Chekhovich2017}, a burst of radiofrequency oscillating magnetic field is sent to the QD for a time sufficiently long to equilibrate the populations of addressed 
subsets of nuclear spin states. Regardless of whether we apply `Inverse NMR' or `Integral NMR' excitation schemes, the direct observables are the hyperfine splittings of $X^-$ transitions that reveal $\sum_i A^{\mathrm{(e)}}_i\langle \hat{I}^i_z \rangle$ and $\sum_i A^{(\mathrm{h})}_i\langle \hat{I}^i_z \rangle$, where hyperfine constants $A^{(\mathrm{e})}_i$ and $A^{(\mathrm{h})}_i$ are those of the electron and the hole, respectively. Under radiofrequency-induced relaxation the splittings change, informing about the spectral overlap of nuclear spin transitions with the radio-frequency comb. Such measurement is therefore sensitive to Zeeman splitting of different nuclear species, together with the quadrupolar shifts. 

In all these experiments, a magnetic field is applied parallel to the growth axis (Faraday geometry). External strain causing a constant quadrupolar shift of $\sim250$ $\mathrm{kHz}$ (for Arsenic nuclei) is applied along the $[110]$ crystal axis, and perpendicular to the growth axis in order to detect the broad spectral features of the $+\frac{1}{2}\to+\frac{3}{2}$ satellite transition, unobstructed by the other two NMR transitions of the spin-$3/2$ nuclei.

\subsection{Inverse NMR}
In the Inverse NMR excitation scheme, a broad frequency comb with a $6$ $\mathrm{kHz}$-wide gap centered around the variable frequency $\nu_{\mathrm{gap}}$ is sent to the QD \cite{PhysRevB.90.205425}. The pulse drives all the allowed NMR transitions except for those in the frequency gap.

Panel a of Fig. \ref{fig:NMR_principles} shows the complete Inverse NMR spectrum of the Arsenic-75 nuclei, featuring a sharp central transition and the two broadened transitions.

\subsection{Integral NMR}
In the Integral NMR excitation scheme, a frequency comb of increasing width (from 0 $\mathrm{kHz}$ to 600 $\mathrm{kHz}$) is sent to the QD\cite{ragunathan_2019}. Throughout the measurement either the lower frequency edge ($38.45$ $\mathrm{MHz}$) or higher frequency edge ($38.92$ $\mathrm{MHz}$) of the comb is fixed (see Fig. \ref{fig:NMR_principles}a). Increasing the width of the comb (i.e. sweeping the frequency of a moving edge: $\nu_{\mathrm{edge}}$) relaxes the populations of an increasing number of nuclei giving rise to a signal propotional to the integral of the spectral distribution of NMR transitions. Such measurement is more sensitive to broad, weak features which are much harder to resolve under the Inverse NMR scheme. 

Panel b of Fig. \ref{fig:NMR_principles} shows the integral NMR signal obtained via sweeping the comb edge over the central transition. Panels c and d of Fig. \ref{fig:NMR_principles} depict the corresponding signal for a high-frequency (quadrupolar-shifted) satellite transition with long and short NMR pumping times, respectively. Regardless of the pumping time and direction of sweep the integral NMR signal picks up a $200$ $\mathrm{kHz}$-broad spectral feature. 

\subsection{Fits to the data} \label{NMR_fits}

\begin{figure}[t!]
	\centering
	\includegraphics[width=\linewidth]{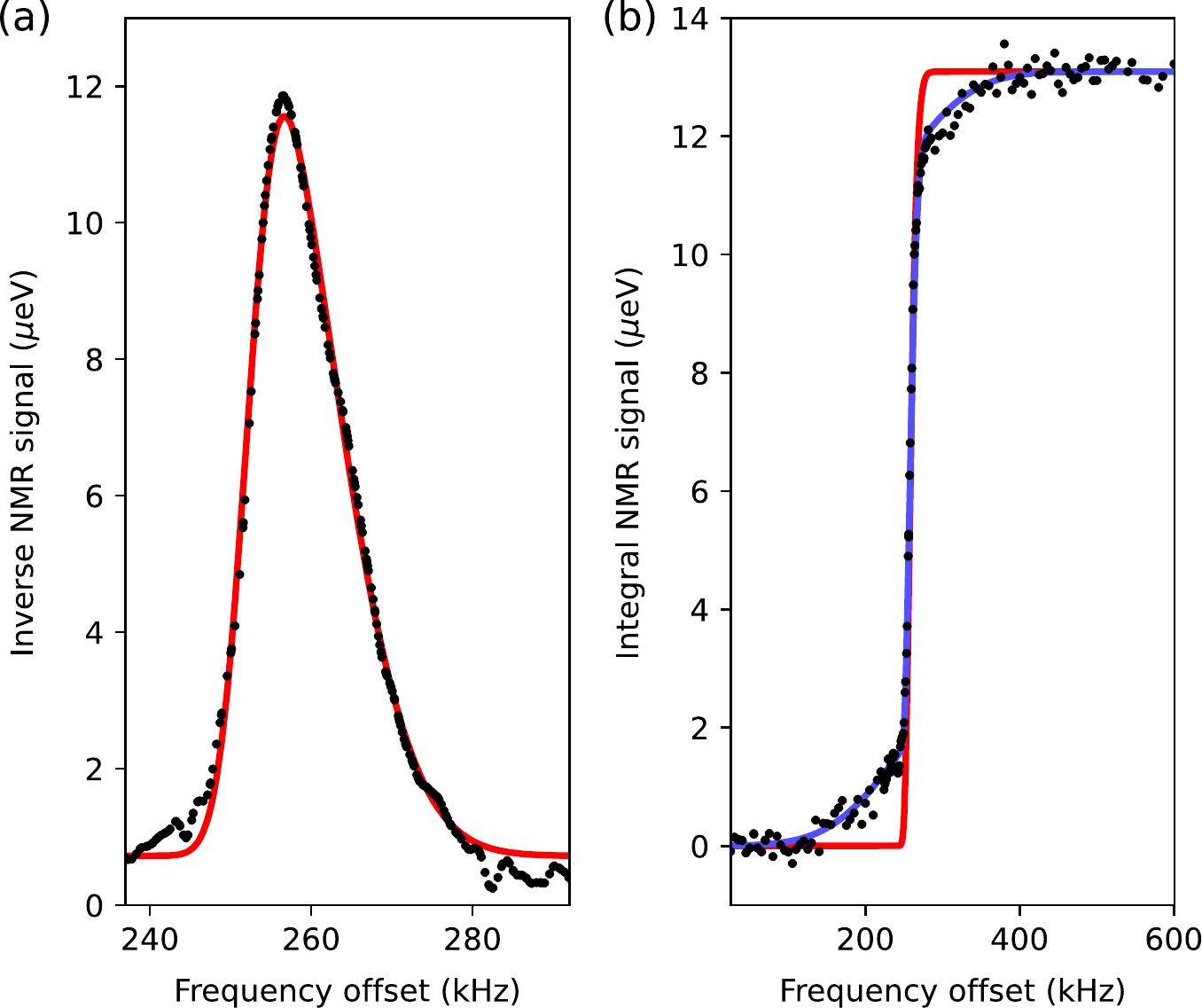}
	\caption{\textbf{(a)} Inverse NMR signal (black dots) alongside the skew normal fit (solid red curve). \textbf{(b)} Integral NMR signal (black dots) alongside scaled $\mathrm{CDF}_{(\mathrm{A})}(\nu)$ (solid red curve), and fitted $\mathrm{CDF}(\nu)$ (solid blue curve) which accommodates a residual Gaussian feature.
	} 
	\label{fig:NMR}
\end{figure}

First, we fit the spectral lineshape of the high-frequency satellite transition addressed via Inverse NMR with a scaled skew normal distribution:
\begin{equation}
S_{(\mathrm{A})}(\nu)=A\frac{2}{\sigma_{(\mathrm{A})}}\phi\bigg(\frac{\nu-\mu_{(\mathrm{A})}}{\sigma_{(\mathrm{A})}}\bigg)\Phi\bigg(\zeta_{(\mathrm{A})}\bigg(\frac{\nu-\mu_{(\mathrm{A})}}{\sigma_{(\mathrm{A})}}\bigg)\bigg)+C
\end{equation}
where:
\begin{equation}
\begin{split}
\phi(x)&=\frac{1}{\sqrt{2\pi}}e^{-x^2/2}\\
\Phi(x)&=\int_{-\infty}^{x}dx \, \phi(x)=\frac{1}{2}\bigg(1+\erf \bigg(\frac{x}{\sqrt 2} \bigg)\bigg)
\end{split}
\end{equation}
The best fit (shown in Fig. \ref{fig:NMR}a alongside the data) gives:
\begin{itemize}
\item $A=158(1)$ $\mu\mathrm{eV}\times \mathrm{kHz}$
\item $\mu_{(\mathrm{A})}=252.21(4)$ $\mathrm{kHz}$
\item $\sigma_{(\mathrm{A})}=9.74(8)$ $\mathrm{kHz}$
\item $\zeta_{(\mathrm{A})}=3.22(8)$ 
\item $C=0.72(3)$ $\mu\mathrm{eV}$
\end{itemize}
The cumulative distribution function corresponding to the fitted skew normal distribution is:
\begin{equation}
\mathrm{CDF}_{(\mathrm{A})}(\nu) = \Phi\bigg(\frac{\nu-\mu_{(\mathrm{A})}}{\sigma_{(\mathrm{A})}}\bigg)-2T\bigg(\frac{\nu-\mu_{(\mathrm{A})}}{\sigma_{(\mathrm{A})}}, \zeta_{(\mathrm{A})} \bigg)
\end{equation}
where $T(x,y)$ is an Owen's $T$ function.

We juxtapose the $\mathrm{CDF}_{(\mathrm{A})}(\nu)$ (the red curve in Fig. \ref{fig:NMR}b) with the normalized Integral NMR data to arrive at a clear mismatch, suggestive of the presence of two sub-ensembles ($\mathrm{A}$) and ($\mathrm{B}$) featuring lower and higher degrees of broadening, respectively.

This inconsistency is best resolved by
fitting the integral NMR data with:
\begin{equation}
\mathrm{CDF}(\nu)=\beta\mathrm{CDF}_{(\mathrm{B})}(\nu)+(1-\beta)\mathrm{CDF}_{(A)}(\nu)
\end{equation}
where $\mathrm{CDF}_{(\mathrm{A})}(\nu)$ is fixed by the inverse NMR fit and the added weighted residual $\mathrm{CDF}_{(\mathrm{B})}(\nu)$ corresponds to a broad Gaussian feature:
\begin{equation}
\mathrm{CDF}_{(\mathrm{B})}(\nu)=\Phi\bigg(\frac{\nu-\mu_{(\mathrm{B})}}{\sigma_{(\mathrm{B})}}\bigg)
\end{equation}
Best fit is found for the following set of parameters:
\begin{itemize}
\item $\mu_{(\mathrm{B})}=246(2)$ $\mathrm{kHz}$
\item $\sigma_{(\mathrm{B})}=73(4)$ $\mathrm{kHz}$
\item $\beta=0.244(7)$  
\end{itemize}
The (scaled) fit is presented in the Fig. \ref{fig:NMR}b as the solid blue curve. 

Next, by differentiating the fitted $\mathrm{CDF}(\nu)$ we reconstruct the probability distribution of Arsenic quadrupolar shifts:
\begin{equation}\label{fit_to_nmr_dist}
\begin{split}
P_{\mathrm{As}}(\nu)&=\beta\frac{1}{\sigma_{(\mathrm{B})}}\phi\bigg(\frac{\nu-\mu_{(\mathrm{B})}}{\sigma_{(\mathrm{B})}}\bigg)\\&+(1-\beta)\frac{2}{\sigma_{(\mathrm{A})}}\phi\bigg(\frac{\nu-\mu_{(\mathrm{A})}}{\sigma_{(\mathrm{A})}}\bigg)\Phi\bigg(\zeta_{(\mathrm{A})}\bigg(\frac{\nu-\mu_{(\mathrm{A})}}{\sigma_{(\mathrm{A})}}\bigg)\bigg)
\end{split}
\end{equation}

Finally, to use this distribution in our CPMG model (see section \ref{sec:modelCPMG}), we must account for the fact that the NMR experiments were realized in Faraday geometry ($B$-field parallel to the growth axis) and under external applied strain, while the all-optical readout and control of the electron spin required operation in Voigt geometry ($B$-field perpendicular to the growth axis) and were performed on unstrained piece of the same wafer. We assume that the broadening of the sub-ensemble ($\mathrm{B}$) results from random alloying in $\mathrm{Al}_x\mathrm{Ga}_{1-x}\mathrm{As}$, which leads to the same contribution to $P_\mathrm{As}(\nu)$ regardless of which geometry is considered, due to the rotation symmetry of zincblende structure.
To factor out the effect of the externally applied strain, we further centralize the entire distribution via translating it by $\mu_{(\mathrm{B})}$. The sub-ensemble $(\mathrm{A})$ is assumed to feature quadrupolar broadening coming purely from the residual strain, which is pinned to the growth axis. This necessitates a transformation of the quadrupolar transition frequencies as $\nu \to -\nu/2$ when transforming the distribution from Faraday to Voigt geometry \cite{PhysRevB.97.235311}. The above considerations lead us to:
\begin{widetext}
\begin{equation}\label{reconstructed_distribution}
\begin{split}
P_{\mathrm{As}}(\nu)=\beta\frac{1}{\sigma_{(\mathrm{B})}}\phi\bigg(\frac{\nu}{\sigma_{(\mathrm{B})}}\bigg)+(1-\beta)\frac{4}{\sigma_{(\mathrm{A})}}\phi\bigg(\frac{\nu-(\mu_{(\mathrm{B})}-\mu_{(\mathrm{A})})}{\sigma_{(\mathrm{A})}/2}\bigg)\Phi\bigg(-\zeta_{(\mathrm{A})}\bigg(\frac{\nu-(\mu_{(\mathrm{B})}-\mu_{(\mathrm{A})})}{\sigma_{(\mathrm{A})}/2}\bigg)\bigg)
\end{split}  
\end{equation}
\end{widetext}

In the process of fitting we made use of the \textit{SciPy} implementation of the error function $\erf(x)$ and the Owen's $T$ function $T(x,y)$ \cite{2020SciPy-NMeth}.

\section{Modelling of a dynamically-decoupled electron spin}\label{sec:modelCPMG}

\subsection{Hamiltonian of the electron-nuclear system}
In the absence of a magnetic field the Hamiltonian for electron-nuclear system is given by:
\begin{equation}
\hat{H}(B=0)= \underbrace{ \sum_i A_i \hat{\mathbf{S}}\cdot \hat{\mathbf{I}}_i}_{\hat{H}_{\mathrm{hf}}} +  \underbrace{\sum_i \omega^i_{\mathrm{Q}} ( \hat{\mathbf{n}}_i \cdot \hat{\mathbf{I}}_i )^2 }_{\hat{H}_{\mathrm{Q}}} 
\end{equation}
The term denoted by $\hat{H}_{\mathrm{hf}}$ stands for the hyperfine interaction between the electron and the nuclei, whereas $\hat{H}_{\mathrm{Q}}$ captures the quadrupolar interaction of high-spin ($I_i=\tfrac{3}{2}$) nuclei originating from strain-induced and alloying-induced electric field gradients. The unit vector $\hat{\mathbf{n}}_i$ defines the unique principal axis of an electric field gradient tensor \cite{abragam1961principles}. 

In the regime of high magnetic fields ($B>1$\,T) aligned with the $z$-axis, the Zeeman interaction ($\hat{H}_{\mathrm{z}}$) splits the electron spin states, as well as the nuclear spin states, modifying the system Hamiltonian in the following way:

\begin{equation}
\hat{H}(B\ne 0)=\hat{H}(B=0)+ \underbrace{\omega_{\mathrm{e}} \hat{S}_z + \sum_i \omega^i_{\mathrm{n}}\hat{I}^i_z}_{\hat{H}_{\mathrm{z}}} 
\end{equation}
The energy scales involved in the dynamics are:
\begin{equation}
\begin{split}
\omega_{\mathrm{e}} &\sim 1 \, \mathrm{GHz}, \quad \omega^i_{\mathrm{n}} \sim 10 \, \mathrm{MHz} \\ A_i &\sim 100 \, \mathrm{kHz}, \quad \omega^i_{\mathrm{Q}} \sim 10 \, \mathrm{kHz} 
\end{split} 
\end{equation}
which dictates the direction of quantisation axis along the $z$-axis.
Consequently, the terms in the Hamiltonian that are non-diagonal in the $\ket{S_z,\{m_i\}_{i=1,..,N}}$ basis should be treated perturbatively.

Projecting the electron-nuclear flip-flop terms $\propto \hat{S}_+ \hat{I}^i_-+\hat{S}_-\hat{I}^i_+$ present in $\hat{H}_{\mathrm{hf}}$ onto a low-energy subspace (and ignoring small corrections to the electron and nuclear Zeeman interaction) via a canonical transformation one arrives at \cite{PhysRevB.79.245314}:
\begin{equation}
\hat{H}_{\mathrm{hf}}\approx \hat{S}_z \sum_i A_i \hat{I}^i_z+\hat{S}_z\sum_{i\ne j} \frac{A_iA_j}{2\omega_{\mathrm{e}}}\hat{I}^i_+\hat{I}^j_-
\end{equation}
where the second term introduces the electron-mediated nuclear spin flip-flops. 

Turning the attention to the quadrupolar Hamiltonian $\hat{H}_{\mathrm{Q}}$ one can isolate the diagonal part:
\begin{equation}\label{eqn:H_Q_D}
\hat{H}^{\mathrm{D}}_{\mathrm{Q}}=\sum_i \sum_{m_i} \omega^i_{\mathrm{Q}} \bra{m_i}(\hat{\mathbf{n}}_i \cdot \hat{\mathbf{I}}_i )^2\ket{m_i} \ketbra{m_i}
\end{equation} 
which simply changes the nuclear Zeeman transition frequencies to $\omega^i_{\mathrm{n}}+\Delta^i_{\mathrm{Q}}$, $\omega^i_{\mathrm{n}}$ and $\omega^i_{\mathrm{n}}-\Delta^i_{\mathrm{Q}}$, giving rise to two satellite transitions on top of a central transition.  

In the particular case of $\hat{\mathbf{n}}_i\cdot\hat{\mathbf{z}}=1$ one arrives at $\Delta^i_{\mathrm{Q}} = 2 \omega^i_{\mathrm{Q}}$, whereas for $\hat{\mathbf{n}}_i\cdot\hat{\mathbf{z}}=0$ the quadrupolar shift is $\Delta^i_{\mathrm{Q}} =  -\omega^i_{\mathrm{Q}}$.

The non-diagonal part of quadrupolar Hamiltonian $\hat{H}_{\mathrm{Q}}$ leads to a non-collinear interaction captured by: 
\begin{equation}
\hat{H}_{\mathrm{nc}}\approx \sum_i A_{\mathrm{nc}}^i \hat{S}_z \hat{I}^i_x,
\end{equation} 
where $A^i_{\mathrm{nc}} \propto \omega^i_{\mathrm{Q}} A_i/\omega^i_{\mathrm{n}}$.
This term leads to a number of rich phenomena such as dynamic nuclear polarization studied in Ref.
\cite{PhysRevLett.108.197403}, 
or coherence modulation observed in Ref. \cite{Botzem2016}. 

In our system, the rate of the processes induced via the non-collinear interaction is $\sqrt{N} A^i_{\mathrm{nc}} \sim 10 \, \mathrm{kHz}$. In contrast, the hyperfine interaction leads to dynamics at rates of $\sqrt{N} A_i \sim 10 \, \mathrm{MHz}$, for the collinear ($\propto \hat{I}_z^i$) part, and $(\sqrt{N})^2\frac{A_iAj}{2\omega_{\mathrm{e}}} \sim 1 \, \mathrm{MHz}$ for the electron-mediated nuclear spin flip-flops. Consequently, the non-collinear interaction can be ignored when modelling the electron spin decoherence.

\subsection{Hamiltonian of the dynamically-decoupled electron}\label{sec:dd_hamiltonian}
A particularly useful picture of the electron-nuclear interaction comes from transforming the system Hamiltonian into the frame evolving due to $\hat{H}_z +\hat{H}^{\mathrm{D}}_{\mathrm{Q}}$ (c.f. `interaction picture'). Following this transformation, the Hamiltonian can be recast as:
\begin{equation}
\hat{H}_{I}(t)=g\mu_{\mathrm{B}}\Big(\underbrace{\sum_i \frac{A_i}{g \mu_{\mathrm{B}}}\hat{I}_z^i(t)}_{B_{\mathrm{OH}}^\parallel(t)}+\underbrace{\sum_{i\ne j} \frac{A_iA_j}{2g \mu_{\mathrm{B}} \omega_{\mathrm{e}}}\hat{I}_+^i(t)\hat{I}_-^j(t)}_{(B_{\mathrm{OH}}^\perp(t))^2/2B_{\mathrm{ext}}}\Big)\hat{S}_z
\end{equation}
where $B_{\mathrm{OH}}^\parallel$ and $B_{\mathrm{OH}}^\perp$ are the time-dependent longitudinal and transverse components of the nuclear Overhauser field, respectively. In this picture, species-specific transverse components of the Overhauser field precess under the combined effect of the external $B$-field and the quadrupolar shifts (Eq. \ref{eqn:H_Q_D}). While the external $B$-field is uniform across the nuclei, their quadrupolar shifts follow a $10-100$ $\mathrm{kHz}$ wide distribution (see section \ref{NMR_section}) which results in the electronic dephasing.



Since the spin-control pulses are much shorter than any system-specific timescale, including the period of the electronic Larmor precession $2\pi/\omega_{\mathrm{e}}$ and the inhomogeneous dephasing time $T_2^*$, they are assumed to be instantaneous, which allows to write the interaction Hamiltonian as:

\begin{equation}
\hat{H}_{\mathrm{I}}(t)=f(\tau;t)g\mu_{\mathrm{B}}\Big(B_{\mathrm{OH}}^\parallel(t)+\frac{(B_{\mathrm{OH}}^{\perp}(t))^2}{2B_{\mathrm{ext}}}\Big)\hat{S}_z
\end{equation}
where the filter function $f(\tau;t)$ encodes the action of ultra-fast $\pi$-pulses that flip the electron spin during the CPMG sequence of total duration $\tau$. The filter function is a piecewise constant function $f:\mathbb{R}\to\{-1,0,+1\}$ that changes sign during pulse sequence every time the $\pi$-pulse is applied, and stays zero outside of the
pulse sequence. 


Importantly, at the time $t=0$ the state of the system is given by:
\begin{equation}
\hat{\rho}(t=0)=\hat{\rho}_{\mathrm{e}}(t=0)\otimes\hat{\rho}_{\mathrm{n}}
\end{equation}
with $\hat{\rho}_{\mathrm{e}}(t=0) =\tfrac{1}{2}(1+\sigma_x)$ and $\hat{\rho}_{\mathrm{n}}\propto \mathds{1}$, since at $T=4$ $\mathrm{K}$ the nuclear energy scales are much smaller that the magnitude of a thermal fluctuation ($\hbar \omega_{\mathrm{N}}^i \ll k_{\mathrm{B}}T$).

To capture the electronic coherence that remains after the CPMG pulse sequence, it is convenient to use the decoherence function defined as follows:
\begin{equation}\label{eqn:decoherence_fn}
W(\tau)= \frac{\bra{\uparrow}\Tr_{\mathrm{n}}\hat{\rho}(\tau)\ket{\downarrow}}{\bra{\uparrow}\Tr_{\mathrm{n}}\hat{\rho}(0)\ket{\downarrow}}
\end{equation}

The quasi-static noise component of $B_{\mathrm{OH}}^\parallel$ originating from the shot-to-shot fluctuation of the total nuclear polarization is filtered out via successive electron spin-inversions, and is only relevant to the decoherence process in the absence of decoupling (see section \ref{sec:FID}). The remaining spectral-diffusion processes that contribute to the evolution of longitudinal component of the nuclear noise (like dipolar nuclear interactions \cite{PhysRevB.79.245314}) are very slow and estimated to proceed on tens of microsecond timescales \cite{PhysRevB.84.035441}.To a good approximation the longitudinal noise component $B_{\mathrm{OH}}^\parallel$ is uncorrelated with the transverse noise component  $B_{\mathrm{OH}}^\perp$, leading to \cite{PhysRevB.84.035441}:
\begin{equation}
W(\tau)\approx W_{\perp}(\tau)\underbrace{W_{\parallel}(\tau)}_{\sim 1}
\end{equation}

The interaction picture Hamiltonian is further simplified in this limit. By transforming it into the frame evolving due to:
\begin{equation}\label{eq:int_pict_diag}
\hat{H}_0(t)=\hat{H}_z + \hat{H}^{\mathrm{D}}_{\mathrm{Q}} +f(\tau;t)\hat{S}_z \sum_i A_i \hat{I}^i_z,
\end{equation}
finally, one arrives at:
\begin{equation}\label{eq:final_hamiltonian}
\hat{H}_{\mathrm{I}}(t)\approx f(\tau;t)g\mu_{\mathrm{B}}\frac{(B_{\mathrm{OH}}^{\perp}(t))^2}{2B_{\mathrm{ext}}}\hat{S}_z  
\end{equation}
\newline

\subsection{Overview of existing models of dynamical decoupling} 




Due to the strong hyperfine interaction of the electron with the nuclear spins ($\omega_{\mathrm{e}}<\mathcal{A}$), as well as high nuclear homogeneity, the environmental noise felt by the electron spin qubit
is non-Gaussian. Therefore, modeling the electron dephasing requires a treatment more advanced than the usual Filter Function formalism\cite{PhysRevB.77.174509,PhysRevA.90.042307,Stockill2016}. Instead, it should be based on casting the decoherence of the electron as a fully unitary process involving interaction with nuclear ensemble with infinitely long bath correlation time and at an infinite temperature. In such a model, the exact quantum evolution of the system is calculated and averaged over all equiprobable initial conditions of the ensemble (as also done in this work).

Models applicable here, both of which lead to equivalent results, have been developed in Ref. \cite{PhysRevB.79.245314} (using ring diagram formalism, briefly explained in subsection \ref{sec:ringdiagram}) and Ref. \cite{PhysRevB.84.035441} (using semi-classical model). The key conclusion presented in both references was that the decoherence function in presence of dynamical decoupling could be straightforwardly related to a sequence-specific $\mathbf{T}$-matrix via:
\begin{equation}
W_\perp(\tau)=\frac{1}{\det(1+i\mathbf{T}(\tau))}
\end{equation} 
In particular, for a CPMG sequence with $n$ $\pi$-pulses, a $\mathbf{T}$-matrix was found as \cite{Malinowski2017}:
\begin{widetext}
\begin{equation}\label{eqn:originalT}
\mathbf{T}_{i,j}(\tau)= \frac{5A_iA_j}{2g_{\mathrm{e}}\mu_B B}\frac{\omega_{i,j}}{\omega^2_{i,j}-A^2_{i,j}}\Bigg[ 1-\frac{\cos{\frac{A_{i,j}\tau}{2n}}}{\cos{\frac{\omega_{i,j}\tau}{2n}}} \Bigg] \sin{\frac{\omega_{i,j}\tau+n\pi}{2}}e^{i\frac{\omega_{i,j}\tau+n\pi}{2}}
\end{equation}
\end{widetext}
where indices $i$ and $j$ run over all nuclei, and:
\begin{equation}
\omega_{i,j}=\omega^i_{\mathrm{n}}-\omega^j_{\mathrm{n}}, \quad A_{i,j}=A_i-A_j
\end{equation}
Since the ensemble consists of $\sim10^5$ nuclei, the size of the $\mathbf{T}$-matrix is prohibitively large to perform any simulation, which motivates coarse graining the nuclei into $K$ groups per nuclear species, and multiplying $(k,l)$-th component of the matrix by factor $\sqrt{N_k N_l}$ - where $N_k$ is a number of the nuclei in the $k$-th group - in order to account for collective enhancement of dynamics of a resulting `macrospin'. 

It is readily seen that at Tesla-strong $B$-fields where $\omega_{\mathrm{n}}^i \gg A_i$, the dominant terms in the $\mathbf{T}$-matrix come from the nuclei of similar Zeeman frequency. This has motivated a phenomenological treatment of quadrupolar effects in the Ref. \cite{PhysRevB.79.245314} and Ref. \cite{PhysRevB.84.035441}, where their presence has been simply emulated by the broadening of single-species Larmor precession frequencies $\omega_{\mathrm{n}}^i$. Non-trivial broadening of the nuclear spectra measured in our QD devices (see section \ref{NMR_section}) resulting in emergence of resolvable satellite transitions in the NMR, invites a more careful treatment. We develop an extension of the model which treats the quadrupolar effects from first principles, and therefore allows us to constrain the simulation with complimentary NMR data that reveals the exact distribution of quadrupolar shifts $\omega_\mathrm{Q}^i$.
\textit{A posteriori}, we verify that whilst the effect of quadrupolar Hamiltonian $\hat{H}_{\mathrm{Q}}$ on the state-mixing (i.e. non-collinear interactions) can be ignored, the quadrupolar shifts to the nuclear spin transitions are essential in capturing the decoherence of the electron spin.

Development of our theory extension follows closely the reasoning from Ref. \cite{PhysRevB.79.245314}. For that reason, we stick to the notation and conventions from that work, and we briefly outline them in the next two subsections.

%
\subsection{Conventions}
\subsubsection{Ensemble averaging}
In the sections to follow, we write
\begin{equation}
\langle ... \rangle = \Tr_{\mathrm{n}}...
\end{equation}
as $\hat{\rho}_{\mathrm{n}}(t=0)\propto \mathds{1}$.

\subsubsection{Keldysh contour}
When the evolution of the system is governed by a time-dependent Hamiltonian $\hat{H}(t)$, the time evolution of pure quantum states is calculated via the action of a unitary operator $\hat{U}(t)$:
\begin{equation}
\ket{\psi(t)}=\underbrace{\mathcal{T}e^{-i\int^t_0\hat{H}(t^\prime)dt^\prime}}_{\hat{U}(t)}\ket{\psi(0)} 
\end{equation}
where $\mathcal{T}$ is the time ordering operator.

Correspondingly, action of the propagator on the density operator looks as follows:
\begin{equation}
\hat{\rho}(t)=\hat{U}^\dagger(t) \hat{\rho}(0) \hat{U}(t) 
\end{equation}
where the hermitian adjoint of the unitary operator $\hat{U}^\dagger(t)$ involves the reverse time ordering operator $\tilde{\mathcal{T}}$ and the flipped sign of the interaction:
\begin{equation}
\hat{U}^\dagger(t)=\tilde{\mathcal{T}}e^{i\int^t_0\hat{H}(t^\prime)dt^\prime}
\end{equation}
Naturally, one expects that in such case, any calculation that involves propagating the density operator $\hat{\rho}(t)$ in time is a tedious task.  

A compact shortcut is introduced via Keldysh contour path-ordering \cite{PhysRevB.79.245314} imposed by the operator $\mathcal{T}_C$. All the operators following $\mathcal{T}_C$ have to follow a time contour $0 \to \infty \to 0$ (also called a Keldysh contour). In addition, an integral of an arbitrary function along the Keldysh contour is defined via:
\begin{equation}\label{eqn:keldyshintegral}
\int_C F(t,c)dt_c = \int_0^\infty F(t,1)dt + \int_\infty^0F(t,-1)dt
\end{equation} 
where $c$ has been introduced as a sign of interaction, which changes from positive along $0\to \infty$, to negative along $\infty \to 0$.

\begin{widetext}
\subsection{Ring diagram formalism}\label{sec:ringdiagram}
In solving the problem of decoherence of the electron interacting with a nuclear ensemble, the decoherence function (defined in the Eq. \ref{eqn:decoherence_fn}) has been shown to satisfy \cite{PhysRevB.79.245314}:
\begin{equation}\label{eq:startingpoint_Tmatrix}
W_\perp (\tau) = \Big \langle \mathcal{T}_C \exp(-i\int_C \hat{\mathcal{V}}(t_c)d t_c) \Big \rangle
\end{equation}
i.e. to be an expectation value of the Keldysh-contour ordered propagator, generated by the following nuclear operator:
\begin{equation}\label{eq:interaction}
\hat{\mathcal{V}}(t_c)= c \bra{\uparrow} \hat{H}_{\mathrm{I}}(t) \ket{\uparrow}
\end{equation}
where $\hat{H}_{\mathrm{I}}(t)$ is the interaction picture Hamiltonian derived at the end of the section \ref{sec:dd_hamiltonian} (Eq. \ref{eq:final_hamiltonian}) and $c$ is the sign of interaction (see Eq. \ref{eqn:keldyshintegral}). 

Explicitly:
\begin{equation}\label{eq:interaction_explicitly}
\hat{\mathcal{V}}(t_c)= c f(\tau;t)\hat{U}_0^\dagger(t_c) \Big(\sum_{i \ne j} \frac{A_iA_j}{4\omega_{\mathrm{e}}}\hat{I}_i^+\hat{I}_j^-\Big)\hat{U}_0(t_c)  
\end{equation}
where:
\begin{equation}
\hat{U}_0(t_c)= \exp(-i\sum_j \omega^j_{\mathrm{n}}\hat{I}_z^jt - ic\sum_j\frac{A_j}{2}\hat{I}_z^j\int_0^t f(\tau;t^\prime)dt^\prime-i\hat{H}^{\mathrm{D}}_{\mathrm{Q}}t)
\end{equation}
is simply a unitary transformation generated by $\hat{H}_0$ from Eq. \ref{eq:int_pict_diag}.


By expanding the Keldysh-contour ordered exponential from Eq. \ref{eq:startingpoint_Tmatrix} into Dyson series, one arrives at $W_\perp (\tau)=1+\sum_{k=1}^\infty W_\perp^{(2k)}(\tau)$ where $2k$-th order terms in the expansion are related to Keldysh-contour ordered $2k$-point auto-correlators of $\hat{\mathcal{V}}(t_c)$, i.e.:
\begin{equation}\label{eqn:Wperp_higher_orders}
W_\perp^{(2k)}(\tau) \sim \int_C \prod^{2k}_{i=1} dt_{c_i} \bigg \langle \mathcal{T}_C \prod^{2k}_{i=1}\hat{\mathcal{V}}(t_{c_i}) \bigg \rangle
\end{equation}
Authors of Ref. \cite{PhysRevB.79.245314} recognise an algebraic structure behind these correlators which allows expressing $W_\perp^{(2k)}(\tau)$, for arbitrary $k$, via weighted sums of \emph{ring diagrams}:
\begin{equation}\label{eq:ring_diagrams}
R_{2k}(\tau)=\Tr_{\mathrm{n}}\mathbf{T}^{2k}\Big(1 +\mathcal{O}\Big(\frac{1}{N}\Big)\Big),
\end{equation}
where for the case of $\hat{H}^{\mathrm{D}}_{\mathrm{Q}}=0$ the $\mathbf{T}$-matrix is the same as that from the equation \ref{eqn:originalT}.

Central result of the Ref. \cite{PhysRevB.79.245314} shows that resummation of all the ring diagrams in the expansion of $W_\perp(\tau)$ results in:
\begin{equation}\label{eqn:central_result}
W_\perp(\tau)\approx \exp(\sum_{k=1}^\infty \frac{(-i)^{2k}}{2k} R_{2k}(\tau))\approx \exp(\sum_{k=1}^\infty \frac{(-i)^{2k}}{2k} \Tr_{\mathrm{n}}\mathbf{T}^{2k}(\tau))
\end{equation}
which, following a straight-forward calculation involving the diagonalization of $\mathbf{T}$-matrix, results in a compact relationship:
\begin{equation}\label{eqn_visibility}
W_\perp(\tau)\approx\frac{1}{\det(1+i\mathbf{T})}
\end{equation}

The approximate nature of the expression from Eq. \ref{eqn:central_result} comes from $\mathcal{O}(N^{-1})$ relative error in the relation from Eq. \ref{eq:ring_diagrams}, and allowing the sum in the exponent to run up to $\infty$, rather than $N$. Both approximations can be safely invoked in the regime of $N\sim 10^{5-6}$.


	\subsection{Generalised $\mathbf{T}$-matrix}
	
%
%
	The Ring diagram formalism applied to the case $\hat{H}^{\mathrm{D}}_{\mathrm{Q}}=0$ has been clearly laid out in a great detail in Ref. \cite{PhysRevB.79.245314}. Here we generalise this result to the case of $\hat{H}^{\mathrm{D}}_{\mathrm{Q}} \ne 0$ and we show that the algebraic Ring diagram structure of the problem survives, albeit with a generalised expression for a $\mathbf{T}$-matrix. 
	
	
	As in the Ref. \cite{PhysRevB.79.245314}, we begin by calculating the lowest order contribution to the decoherence function ($k=1$ term from the Eq. \ref{eqn:Wperp_higher_orders}):
	\begin{equation}
	W^{(2)}_\perp (\tau) = -\frac{1}{2}\int_C d t_{c_1}\int_C d t_{c_2} \langle \mathcal{T}_C \hat{\mathcal{V}}( t_{c_1}) \hat{\mathcal{V}}( t_{c_2}) \rangle
	\end{equation}
	where $\hat{\mathcal{V}}(t_c)$ is given by Eq. \ref{eq:interaction_explicitly}.
	The calculation starts from expanding the $+$-rotating component of the $\mathbf{B}_{\mathrm{OH}}(t)$ in the nuclear product basis:
	\begin{equation}
	\begin{split}
	\hat{U}_0^\dagger(t_c) \Big(\sum_j A_j \hat{I}^j_+ \Big) \hat{U}_0(t_c)
	&=\sum_j A_j \Big(\sum_{m_j} \sqrt{I_j(I_j+1)- m_j(m_j+1)}(\hat{U}_0^j)^\dagger(t_c)\ketbra{m_j+1}{m_j} \hat{U}_0^j(t_c)\Big)
	\end{split}
	\end{equation}
	The action of unitaries on $\ketbra{m_j+1}{m_j}$ operators simply brings state-dependent oscillating factors:
	\begin{equation}
	\exp(i(\omega_{m_j+1}-\omega_{m_j})t + ic \frac{A_j}{2}\int_0^t f(\tau;t^\prime)dt^\prime)\equiv e_{m_j}(t_c)
	\end{equation}
	Where $\omega_{m_j+1}-\omega_{m_j}$ are equal to $\ket{m_j+1}\to \ket{m_j}$ nuclear transition frequencies; here $\hat{H}_{\mathrm{Q}} \ne 0$ introduces quadrupolar shifts, which split these transition frequencies into $\omega^j_{\mathrm{n}}+\Delta^j_{\mathrm{Q}}$, $\omega^j_{\mathrm{n}}$ and $\omega^j_{\mathrm{n}}-\Delta^j_{\mathrm{Q}}$ for $m_j=1/2$, $m_j=-1/2$, and $m_j=-3/2$, respectively.
	
	For convenience, we define the set of operators:
	\begin{equation}
	\hat{\mathcal{O}}^j_{m_j} \equiv \sqrt{ 2I_j+1} \ketbra{m_j+1}{m_j}
	\end{equation}
	and turn our attention to the following correlator:
	
	\begin{equation}\label{correlator}
	\begin{split}
	\langle \mathcal{T}_C \hat{\mathcal{V}}( t_{c_1}) \hat{\mathcal{V}}( t_{c_2}) \rangle &=\sum_{k\ne l} \sum_{p\ne q} \sum_{m_k,m_l,m^\prime_p,m^\prime_q} \mathcal{V}_{k,l}^{m_k,m_l}(t_{c_1})\mathcal{V}_{p,q}^{m^\prime_p,m^\prime_q}(t_{c_2}) \langle \mathcal{T}_C \hat{\mathcal{O}}^k_{m_k} \hat{\mathcal{O}}^{l,\dagger}_{m_l} \hat{\mathcal{O}}^p_{m^\prime_p} \hat{\mathcal{O}}^{q,\dagger}_{m^\prime_q} \rangle
	\end{split}
	\end{equation}
	With:
	\begin{equation}
	\mathcal{V}_{k,l}^{m_k,m_l}(t_c)=cf(\tau;t)e_{m_k}(t_c)e^*_{m_l}(t_c)\underbrace{\sqrt{\frac{I_k(I_k+1)-m_k(m_k+1)}{2I_k+1}}}_{\equiv \sqrt{a(I_k,m_k)}}\underbrace{\sqrt{\frac{I_l(I_l+1)-m_l(m_l+1)}{2I_l+1}}}_{\equiv \sqrt{a(I_l,m_l)}}\frac{A_kA_l}{4\omega_{\mathrm{e}}}
	\end{equation}
	The remaining correlator is further simplified by considering the only possible non-zero contractions of nuclear indices:
	\begin{equation}
	\begin{split}
	\langle \mathcal{T}_C \hat{\mathcal{O}}^k_{m_k} \hat{\mathcal{O}}^{l,\dagger}_{m_l} \hat{\mathcal{O}}^p_{m^\prime_p} \hat{\mathcal{O}}^{q,\dagger}_{m^\prime_q} \rangle &= \delta_{k,q}\delta_{l,p}\langle \mathcal{T}_C\hat{\mathcal{O}}^k_{m_k}
	\hat{\mathcal{O}}^{k,\dagger}_{m^\prime_k}\rangle
	\langle \mathcal{T}_C \hat{\mathcal{O}}^{l,\dagger}_{m_l} \hat{\mathcal{O}}^l_{m_l^\prime}  \rangle \\
	&= \delta_{k,q}\delta_{l,p}\delta_{m_k,m_k^\prime}\delta_{m_l,m_l^\prime}\langle \mathcal{T}_C\hat{\mathcal{O}}^k_{m_k}
	\hat{\mathcal{O}}^{k,\dagger}_{m_k}\rangle
	\langle \mathcal{T}_C \hat{\mathcal{O}}^{l,\dagger}_{m_l} \hat{\mathcal{O}}^l_{m_l}  \rangle
	\end{split}
	\end{equation}
	The first line follows from the fact that $k \ne l$ and $p \ne q$, as well as the lack of inter-nuclear coherences in a thermal state. Similarly, the second line results from the lack of inter-state coherences for a single nucleus in a thermal state.
	
	At this stage it becomes clear that the Keldysh-contour ordering operator can be removed. For a $k$-th nucleus, permutation of the two operators will bring either $(2I_k+1)\langle\ketbra{m_k+1}\rangle$ or $(2I_k+1)\langle\ketbra{m_k}\rangle$; for an infinite temperature bath the populations of all nuclear spin states are equal:
	\begin{equation}
	\langle\ketbra{m_k}\rangle=1/(2I_k+1)
	\end{equation}
	rendering the two results identical. 
	
	Finally:
	\begin{equation}
	\begin{split}
	\langle \mathcal{T}_C \hat{\mathcal{O}}^k_{m_k} \hat{\mathcal{O}}^{l,\dagger}_{m_l} \hat{\mathcal{O}}^p_{m^\prime_p} \hat{\mathcal{O}}^{q,\dagger}_{m^\prime_q} \rangle &=\delta_{k,q}\delta_{l,p}\delta_{m_k,m_k^\prime}\delta_{m_l,m_l^\prime} 
	\end{split}
	\end{equation}
	so the correlator from the equation \ref{correlator} simplifies significantly:
	\begin{equation}
	\langle \mathcal{T}_C \hat{\mathcal{V}}( t_{c_1}) \hat{\mathcal{V}}( t_{c_2}) \rangle = \sum_{k\ne l} \sum_{m_k,m_l} \mathcal{V}_{k,l}^{m_k,m_l}(t_{c_1})\mathcal{V}_{l,k}^{m_l,m_k}(t_{c_2})
	\end{equation}
	For convenience we free the sums to run over all indices, and we re-arrange the $4$ dimensional tensors into matrices by introducing combined indices $\alpha=(k,m_k)$ and $\beta=(l,m_l)$, such that:
	\begin{equation}
	\langle \mathcal{T}_C \hat{\mathcal{V}}( t_{c_1}) \hat{\mathcal{V}}( t_{c_2}) \rangle = \sum_{\alpha,\beta}  (1-\delta_{k,l})\mathcal{V}_{\alpha,\beta}(t_{c_1})(1-\delta_{l,k})\mathcal{V}_{\beta,\alpha}(t_{c_2})
	\end{equation}
	Finally, we write:
	\begin{equation}
	W^{(2)}_\perp (\tau)=-\frac{1}{2}\sum_{\alpha,\beta} \mathbf{T}_{\alpha,\beta} \mathbf{T}_{\beta,\alpha} = -\frac{1}{2} \Tr_{\mathrm{n}} \mathbf{T}^2
	\end{equation}
	having arrived at a generalised $\mathbf{T}$-matrix of the following form:
	\begin{equation}
	\mathbf{T}_{\alpha,\beta}= (1-\delta_{k,l})\int_C d t_c \mathcal{V}^{m_k,m_l}_{k,l}(t_c)
	\end{equation}
	Where $\alpha$ and $\beta$ are combined indices.
	
	Higher-order contributions ($k>1$) to the decoherence function will also follow exactly the ring diagram structure of the $\hat{H}_{\mathrm{Q}}=0$ problem (see Ref. \cite{PhysRevB.79.245314}), since the combinatorics of contractions of the $k$-point correlators here is analogous, as it also relies on:
	\begin{itemize}
		\item lack of inter-nuclear correlations,
		\item lack of inter-state correlations,
		\item equal populations of nuclear spin-states,
	\end{itemize}
	all resulting from the bath having an infinite temperature. Again, this leads to the identity from Eq. \ref{eqn_visibility}, where now the $\mathbf{T}$-matrix has a generalized form which we identified.
	
	For spin-$\frac{3}{2}$ species our theory extension requires working with $12K\times 12K$ matrices rather than $3K\times 3K$ matrices, but it straight-forwardly captures the effect of quadrupolar broadening from first principles. Performing the remaining integrals can be done exactly like in Ref. \cite{PhysRevB.79.245314} and \cite{Malinowski2017}. In particular, for a CPMG sequence with $n$ $\pi$-pulses we recover:
	\begin{equation}
	\begin{split}
	\mathbf{T}_{\alpha,\beta}=& (1-\delta_{k,l})\sqrt{a(I_k,m_k)a(I_l,m_l)}\frac{A_kA_l}{\omega_{\mathrm{E}}}\frac{\omega^{m_k,m_l}_{k,l}}{(\omega^{m_k,m_l}_{k,l})^2 -A^2_{k,l}}\\
	&\times \Bigg\{1-\frac{\cos(\tfrac{A_{k,l}\tau}{2n})}{\cos(\tfrac{\omega^{m_k,m_l}_{k,l}\tau}{2n})}\Bigg\}\sin(\frac{\omega^{m_k,m_l}_{k,l}\tau+n\pi}{2})\exp(i\frac{\omega^{m_k,m_l}_{k,l}\tau+n\pi}{2})
	\end{split}
	\end{equation}
	where:
	\begin{equation}
	\begin{split}
	A_{k,l}&\equiv A_k-A_l\\
	\omega^{m_k,m_l}_{k,l}&\equiv (\omega_{m_k+1}-\omega_{m_k})-(\omega_{m_l+1}-\omega_{m_l})
	\end{split}
	\end{equation}
	are differences of single-nucleus hyperfine constants and transition frequencies. Coarse-graining the $\mathbf{T}$-matrix to describe the interactions between groups of nuclei again results in the appearance of additional factors $\sqrt{N_kN_l}$ where $N_k$ and $N_l$ are numbers of nuclei in $k$-th and $l$-th groups, respectively. 
\end{widetext}

\section{Numerical simulation}

We use a simple python code to assemble $\mathbf{T}$-matrix and calculate the determinant of $1+i\mathbf{T}$ for delays $\tau$ between two $\pi/2$-puses in CPMG sequence. This is straightforwardly related to the electronic coherence according to Eq. \ref{eqn_visibility}.

We set species-dependent hyperfine constants, concentrations and Zeeman splittings present in the model to the material constants presented in the table \ref{table:material_constants}. The dot-specific parameters, i.e. the total number of nuclei and the g-factor, are constrained through independent measurements, and summarised in the table \ref{table:constrained_params}.  

The fit of the model to the data relies on the distribution of quadrupolar broadening, $P_{\mathrm{As}}(\nu)$, which is reconstructed in the section \ref{NMR_section} for another QD device made from the same wafer. In order to account for the possible dot-to-dot variations, our model has two free parameters: the electron's wavefunction overlap with an inhomogeneous nuclear sub-ensemble affected by random alloying, $\beta$, and a `scaling factor', $\kappa$, that accounts for the variation in the magnitude of quadrupolar shifts. The `scaling factor' simply scales the width of distribution via $P_{\mathrm{As}}(\nu)\to \kappa^{-1}P_{\mathrm{As}}(\nu/\kappa)$ transformation. 

In order to sample the distribution $P_{\mathrm{As}}(\nu)$ reconstructed via the Integral and Inverse NMR spectra (see section \ref{NMR_section}), we partition nuclear spins into $K=200$ groups of unequal size, each containing $N_k$ nuclear spins with a quadrupolar shift $\nu_{k}$. We consider $K$ equally spaced quadrupolar shifts $\nu_{k}$ within $[-5\sigma_{(\mathrm{B})},5\sigma_{(\mathrm{B})}]$ interval (where $\sigma_{(\mathrm{B})}$ is the width of the detected broad Gaussian distribution - see section~\ref{NMR_section}), and we assign each shift to the following number of nuclei:
\begin{equation}
N_k=N \frac{c_{\mathrm{As}}}{2}\int_{\nu_{k}-\Delta\nu/2}^{\nu_{k}+\Delta\nu/2} d\nu \, P_{\mathrm{As}}(\nu)
\end{equation}
where $c_{\mathrm{As}}$ is the concentration of Arsenic, $N$ is a total number of nuclei and the bin width $\Delta \nu=10\sigma_{(\mathrm{B})}/K$.


In order to capture the residual hetero-nuclear effects we include both Gallium isotopes ($^{71}\mathrm{Ga}$ and $^{69}\mathrm{Ga}$) in the simulation, under the assumption that the effect of random alloying in $\mathrm{AlGaAs}$ on their spectra of quadrupolar inhomogeneities is negligible\cite{Chekhovich2015}. For each of the Gallium isotopes we constrain the distribution of quadrupolar shifts to: 
\begin{equation}
P_{\mathrm{Ga}}(\nu) \propto P^{(\mathrm{A})}_{\mathrm{As}}(-2\nu)
\end{equation}
where $P^{(\mathrm{A})}_{\mathrm{As}}(\nu)$ is the narrow mode of the Arsenic's distribution (see section \ref{NMR_section}). The relative scaling of width follows from the measurement on ${}^{75}\mathrm{As}$ and ${}^{69}\mathrm{Ga}$ nuclei performed in Ref.~\cite{PhysRevB.97.235311}, according to which:
\begin{equation}
\frac{Q^{\mathrm{As}}S^{\mathrm{As}}_{11}}{Q^{\mathrm{Ga}}S^{\mathrm{Ga}}_{11}} \approx -2
\end{equation}
where $Q^{\alpha}$ is the nuclear quadrupolar moment of species $\alpha$, and $S^\alpha_{11}$ is a diagonal component of its gradient-elastic tensor. 

To represent the distribution of quadrupolar inhomogeneities for both Gallium isotopes - $P_{\mathrm{Ga}}(\nu)$ - we arrange the Gallium nuclei into bins of the same width as those of Arsenic ($\Delta \nu=10\sigma_{(\mathrm{B})}/K$), and retain only the bins that have no fewer nuclei than in any of Arsenic's bins. Such cut-off prevents the simulation from crashing due to the presence of empty bins resulting from rounding errors.


\begin{figure}[t!]
    \centering
    \includegraphics[width=\linewidth]{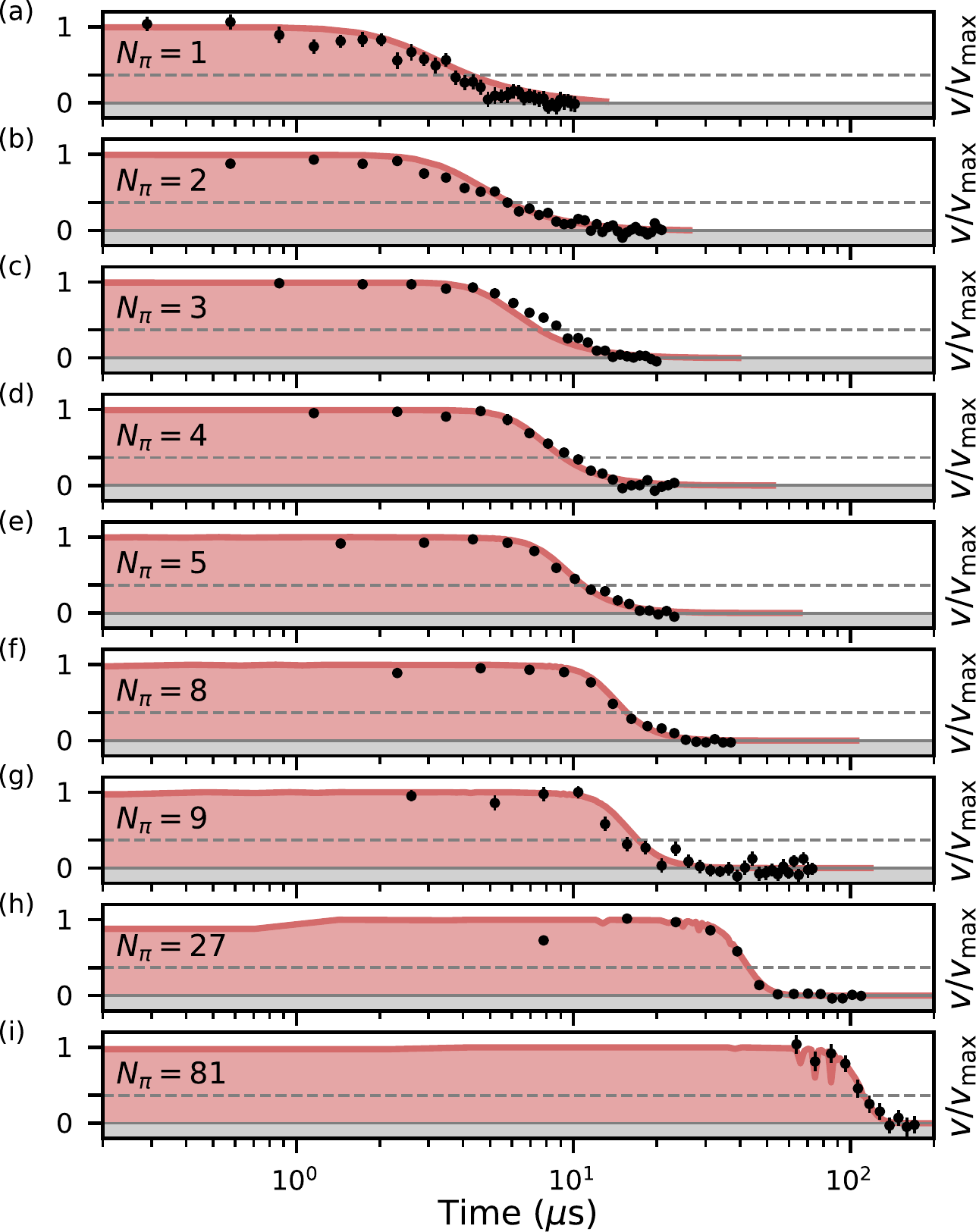}
    \caption{\textbf{(a)-(i)} CPMG datasets (black dots) for $N_\pi=1,2,3,4,5,8,9,27$ and $81$, plotted together with best simultaneous fit of the theory curves (red solid curve).} 
\label{fig:best_fit}
\end{figure}
 
\begin{figure}[h!]
	\centering
	\includegraphics[width=\linewidth]{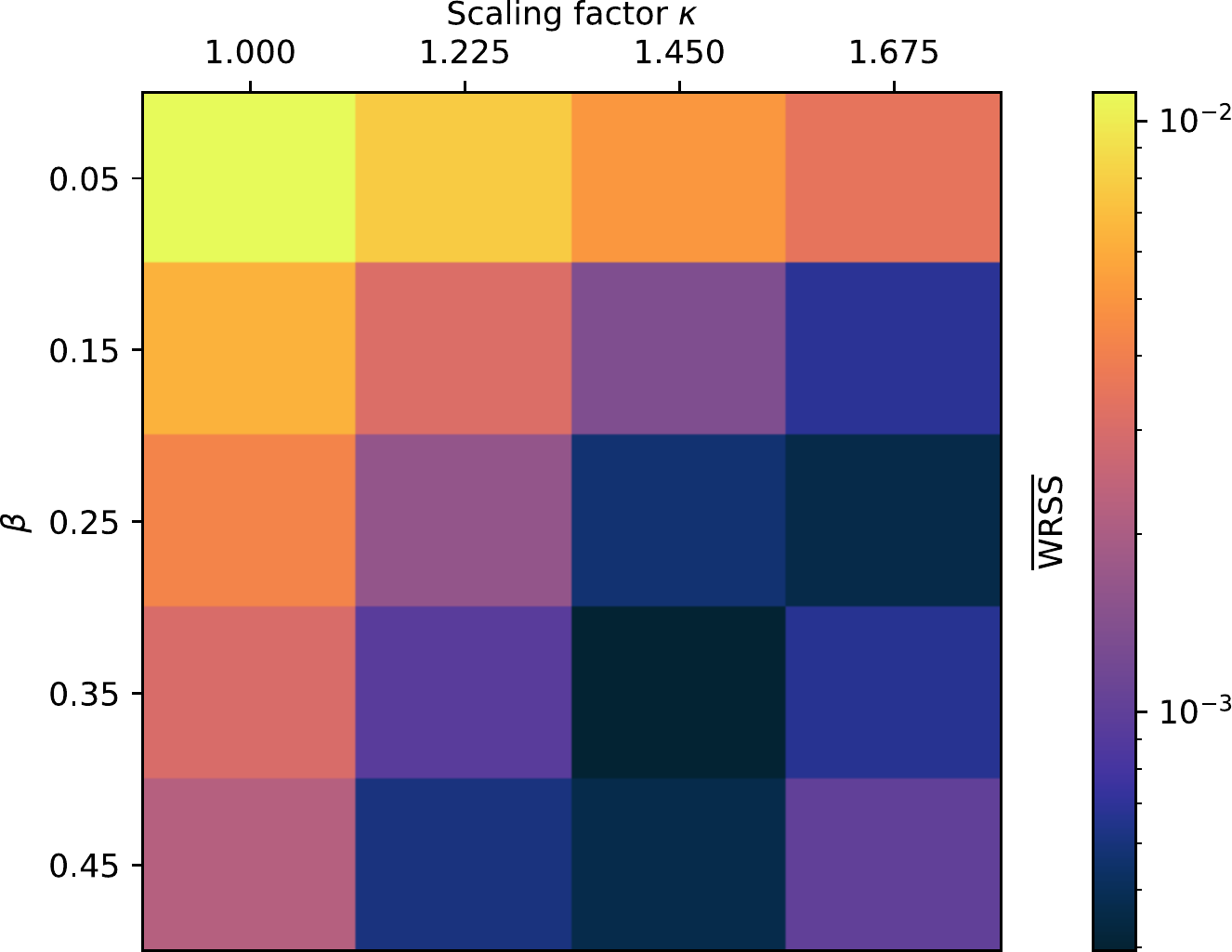}
	\caption{Weighted residual sum of squares averaged over all the CPMG datasets, as a function of two free parameters in a numerical simulation: $\beta$ and scaling factor $\kappa$. Best fit is found for $\beta=0.35$ and $\kappa=1.45$.}  
	\label{fig:wrss}
\end{figure}

\subsection{Simultaneous fit to the data}
We use our model with two free parameters, $\beta$ and $\kappa$, to simultaneously fit the coherence decay in all the CPMG datasets presented in Fig. \ref{fig:best_fit}. The goodness of the simultaneous fit is characterized by the weighted residual sum of squares per data point ($\mathrm{WRSS}$) averaged over fits to distinct $N_\pi$ datasets. For each CPMG dataset containing $M$ datapoints $y_i$ with errors $\Delta_i$ we calculate:
\begin{equation}
    \mathrm{WRSS(N_\pi)}=\Big(\sum^M_i (y_i-\hat{y}_i)^2\tfrac{1}{\Delta_i^{2}}\Big)/\Big(M\sum^M_i \tfrac{1}{\Delta_i^{2}}\Big),
\end{equation}
where $\hat{y}_i$ are the corresponding model values,
and we subsequently average the expression over different $N_\pi$ datasets to arrive at a robust measure: $\overline{\mathrm{WRSS}}$. 

Due to the size of the $\mathbf{T}$-matrix, the simulation time is too long to allow for an automated least-squares optimization. We thus perform a raster scan of the $\overline{\mathrm{WRSS}}$ over models with $(\beta,\kappa) \in \{0.05,0.15,0.25,0.35,0.45\}\times \{1.000,1.225,1.450,1.675\}$. The calculated $\overline{\mathrm{WRSS}}(\beta,\kappa)$ has a clear local minimum on that set, corresponding to $\beta=0.35$ and $\kappa=1.450$ (see Fig. \ref{fig:wrss}). For our best fit $\overline{\mathrm{WRSS}}=3.9\times10^{-4}$. Theory curves of the best simultaneous fit are plotted on top of the data in Fig. \ref{fig:best_fit}.

\section{Material constants and constrained parameters}

\begin{table}[h!]
\begin{tabular}{|c||c|c|c|}
	\hline
	Species $\alpha$&${}^{75}\mathrm{As}$&${}^{71}\mathrm{Ga}$&${}^{69}\mathrm{Ga}$\\
	\hline
	Hyperfine interaction constant, $A_\alpha$ ($\mathrm{GHz}$)& $65.3$ & $69.9$ & $54.7$\\ 
    \hline
	Concentration, $c_{\alpha}$ & $1$ & $0.396$ & $0.604$\\ 
	\hline 
	
	Zeeman splitting, $\omega_{\alpha}/B$ $(\mathrm{MHz}/\mathrm{T})$ & $7.22$&$12.98$&$10.22$\\ 
	\hline
		
\end{tabular}

\caption{\label{table:material_constants}Material constants consistent with Ref. \cite{Malinowski2017}.}
\end{table}

%

\begin{table}[h!]
\begin{tabular}{|c||c|}
    \hline
	External magnetic field, $B$   & $6.5~\mathrm{T}$\\ 
	\hline
	Total number of nuclei, N & $6.5(3)\times 10^4$\\ 
	\hline
	Electron $g$-factor, $|g_\mathrm{e}|$ & $0.04895(6)$\\ 
	\hline
	$\pi$-gate fidelity, $\mathcal{F}$ & $99.30(5)\%$\\ 
	\hline
	
\end{tabular}

\caption{\label{table:constrained_params}Parameters constrained through independent measurement or set experimentally.}
\end{table}

\bibliography{SI_bib}